\documentclass[a4paper, 11 pt]{article}
\usepackage[a4paper, hmargin=1.25in,vmargin=1.25in]{geometry}
\usepackage{amsmath,amssymb,amsthm,mathrsfs,amsfonts,mathtools,dsfont,bm} % Essential packages for math
\usepackage[dvipsnames,svgnames, x11names]{xcolor}% Must before tikz
\usepackage{natbib}
\usepackage{bibunits}
\usepackage{marginnote} % Add notes at margin
\usepackage[bottom,flushmargin]{footmisc} % Force footnotes to say on the corresponding pages
\usepackage{enumitem} % Properly numbering exercises in lecture notes
\newlist{problems}{enumerate}{1}
\setlist[problems]{label={\arabic*.}, ref={\arabic{part}.\thechapter.\arabic*}}
\usepackage{pdfpages}
\usepackage[hyphens]{url}
\usepackage[labelsep=period]{caption} % Replace colon with period in the captions
\usepackage{subcaption}
\usepackage{environ,trimspaces}
\usepackage[breakable,most]{tcolorbox}
\tcbuselibrary{breakable}
\usepackage[titletoc]{appendix}
\usepackage{fancyhdr,setspace, titlesec,titling} % Space and title formatting
\usepackage{booktabs,threeparttable,multirow,makecell} % Table specification
\usepackage{pdflscape,selectp,float,rotating,tabularx}
\usepackage{tikz,pgfplots} % Graph packages
\pgfplotsset{compat=1.11}
\usetikzlibrary{arrows,intersections}
\usetikzlibrary{backgrounds,positioning,fit,petri}
\usetikzlibrary{mindmap,trees,calendar,external,shapes}
\usetikzlibrary{decorations.fractals,decorations.pathmorphing,shadows,shadings,patterns}
\usetikzlibrary{spy,calc} % Magnifying Parts of Pictures
\usepackage[bookmarksnumbered,bookmarksopen,psdextra,unicode]{hyperref}
\makeatletter
\renewcommand\paragraph{\@startsection{paragraph}{4}{\z@}%
            {-2.5ex\@plus -1ex \@minus -.25ex}%
            {1.25ex \@plus .25ex}%
            {\normalfont\normalsize\bfseries}}
\makeatother
\newtheorem{thm}{Theorem}[section]

\newtheorem{lem}{Lemma}[section]
\newtheorem{pro}{Proposition}[section]
\newtheorem{ass}{Assumption}[section]

\theoremstyle{definition}
\newtheorem{defn}{Definition}[section]

\newtheorem{rem}{Remark}[section]

\theoremstyle{definition}
\newtheorem{ex}{Example}[section]
\newtheorem*{ex*}{Example}

\newtheoremstyle{exctd}
{\topsep} {\topsep}%
{\upshape}% Body font
{}% Indent amount (empty = no indent, \parindent = para indent)
{\bfseries\scshape}% Thm head font
{.}% Punctuation after thm head
{1em}% Space after thm head (\newline = linebreak)
{\thmname{#1} \thmnumber{ #2}\thmnote{#3} (Continued)}% Thm head spec
\theoremstyle{exctd}

\newenvironment{assprime}[1]
  {\renewcommand{\theass}{\ref*{#1}\!'}%
   \addtocounter{ass}{-1}%
   \begin{ass}}
  {\end{ass}}

\newcommand{\transpose}{\text{\scalebox{0.7}{$\intercal$}}}

\newcommand{\amin}{\operatornamewithlimits{arg\,min}}

\newcommand{\convas}{\overset{as}{\to}}
\newcommand{\convl}{\xrightarrow{L}}

\newcommand{\convp}{\xrightarrow{p}}

\newcommand{\ul}{\underline}

\def\be{\begin{enumerate}} % Begin Enumerate
\def\ee{\end{enumerate}} % End Enumerate
\def\bi{\begin{itemize}} % Begin Itemize
\def\ei{\end{itemize}} % End Itemize

\defaultbibliographystyle{ims}
\defaultbibliography{mybibliography}

\begin{document}

\begin{bibunit}
\pdfbookmark[1]{Title}{title}
\title{Improved Inference on the Rank of a Matrix}
\author{
Qihui Chen\\ School of Management and Economics,\\ CUHK Business School,\\ Shenzhen Finance Institute,\\ The \hspace{-0.03cm}Chinese\hspace{-0.03cm} University\hspace{-0.03cm} of\hspace{-0.03cm} Hong\hspace{-0.03cm} Kong, \hspace{-0.03cm}Shenzhen\\ qihuichen@cuhk.edu.cn
\and
Zheng Fang\thanks{We are indebted to Frank Schorfheide and three anonymous referees for valuable and constructive comments that helped greatly improve this paper. We would also like to thank Yonghong An, Brendan Beare, Jiaying Gu, Yingyao Hu, Ivana Komunjer, Qi Li,
Aureo de Paula, Peter Phillips, Andres Santos, Yixiao Sun, Xun Tang and seminar participants for helpful discussions and comments, and Frank Kleibergen for generously sharing the data with us. Qihui Chen gratefully acknowledges financial support by the National Natural Science Foundation of China (Grant No.\ 71803169).} \\ Department of Economics \\ Texas A\&M University\\ zfang@tamu.edu}
\date{\today}
\maketitle

\begin{abstract}

This paper develops a general framework for conducting inference on the rank of an unknown matrix $\Pi_0$. A defining feature of our setup is the null hypothesis of the form $\mathrm H_0: \mathrm{rank}(\Pi_0)\le r$. The problem is of first order importance because the previous literature focuses on $\mathrm H_0': \mathrm{rank}(\Pi_0)= r$ by implicitly assuming away $\mathrm{rank}(\Pi_0)<r$, which may lead to invalid rank tests due to over-rejections. In particular, we show that limiting distributions of test statistics under $\mathrm H_0'$ may not stochastically dominate those under $\mathrm{rank}(\Pi_0)<r$. A multiple test on the nulls $\mathrm{rank}(\Pi_0)=0,\ldots,r$, though valid, may be substantially conservative. We employ a testing statistic whose limiting distributions under $\mathrm H_0$ are highly nonstandard due to the inherent irregular natures of the problem, and then construct bootstrap critical values that deliver size control and improved power. Since our procedure relies on a tuning parameter, a two-step procedure is designed to mitigate concerns on this nuisance. We additionally argue that our setup is also important for estimation. We illustrate the empirical relevance of our results through testing identification in linear IV models that allows for clustered data and inference on sorting dimensions in a two-sided matching model with transferrable utility.
\end{abstract}

\begin{center}
\end{center}

\newpage
\section{Introduction}\label{Sec: 1}

The rank of a matrix plays a number of fundamental roles in economics, not just as crucial technical identification conditions \citep{Fisher1966IDBook}, but also of central empirical relevance in numerous settings such as inference on cointegration rank \citep{Engle_Granger1987Co-In,Johansen1991CoIntegration}, specification of finite mixture models \citep{McLachlanPeel2004Mixture,KasaharaShimotsu2009NPIDmixture} and estimation of matching dimensions \citep{DupuyGalichon2014Personality} -- more can be found in Supplemental Appendix \ref{Sec: B}. These problems reduce to examining the hypotheses: for an unknown matrix $\Pi_0$ of size $m\times k$ with $m\ge k$,
\begin{align}\label{Eqn: hypothesis, intro}
\mathrm H_0: \mathrm{rank}(\Pi_0)\le r \qquad \text{v.s.} \qquad  \mathrm H_1: \mathrm{rank}(\Pi_0)> r~,
\end{align}
where $r\in\{0,\ldots,k-1\}$ is some prespecified value and $\mathrm{rank}(\Pi_0)$ denotes the rank of $\Pi_{0}$. If $r=k-1$, then \eqref{Eqn: hypothesis, intro} is concerned with whether $\Pi_0$ has full rank.

Despite a rich set of results in the literature, previous studies instead focus on
\begin{align}\label{Eqn: hypothesis, literature, intro}
\mathrm H_0': \mathrm{rank}(\Pi_0)= r \qquad \text{v.s.} \qquad  \mathrm H_1: \mathrm{rank}(\Pi_0)> r~.
\end{align}
In effect, the testing problem \eqref{Eqn: hypothesis, literature, intro} assumes away the possibility $\mathrm{rank}(\Pi_0)< r$, which is often unrealistic to be excluded. This, unfortunately, has drastic consequences. As elaborated through an analytic example in Section \ref{Sec: 2}, a number of popular tests, including \citet{Robin_Smith2000rank} and \citet{Kleibergen_Paap2006rank}, may over-reject for some data generating processes and under-reject for others, both having $\mathrm{rank}(\Pi_0)< r$. In particular, contrary to what appears to have been conjectured in the literature (\citeauthor{Cragg_Donald1993TestID}, \citeyear{Cragg_Donald1993TestID}, p.225; \citeauthor{Johansen1995likelihood}, \citeyear{Johansen1995likelihood}, p.168), our analysis suggests that {\it limiting distributions of tests obtained under $\mathrm H_0'$ may not first order stochastically dominate those under $\mathrm{rank}(\Pi_0)< r$}. Hence, ignoring the possibility $\mathrm{rank}(\Pi_0)< r$ may lead to tests that are not even first order valid.

One may nonetheless justify the setup \eqref{Eqn: hypothesis, literature, intro} for two reasons. First, the problem \eqref{Eqn: hypothesis, intro} may be studied by a multiple test on the nulls $\mathrm{rank}(\Pi_0)=0,1,\ldots,r$. Our simulations show, however, that such a procedure, though valid, may be substantially conservative and have trivial power against local alternatives that are close to matrices whose rank is strictly less than $r$. Second, the setup \eqref{Eqn: hypothesis, literature, intro} suits well for estimation by sequentially testing $\mathrm{rank}(\Pi_0)=j$ for $j=0,1,\ldots,k-1$. Crucially, however, all steps except for $j=0$ ignore type I errors (false rejection) potentially made in previous steps, and may have limited capability of controlling type II errors (false acceptance) -- see Supplemental Appendix \ref{Sec: 4-3} for more details. Hence, the setup \eqref{Eqn: hypothesis, intro} is desirable for estimation as well.

We thus conclude that developing a valid and powerful test for \eqref{Eqn: hypothesis, intro} is of first order importance. To the best of our knowledge, no direct tests to date exist in this regard. Our objective in this paper is therefore to develop an inferential framework under the setup \eqref{Eqn: hypothesis, intro}. A key insight we exploit to this end is that \eqref{Eqn: hypothesis, intro} is equivalent to
\begin{align}\label{Eqn: hypothesis, phi, intro}
\mathrm H_0: \phi_r(\Pi_0)=0 \qquad \text{v.s.} \qquad  \mathrm H_1: \phi_r(\Pi_0)>0~,
\end{align}
where $\phi_r(\Pi_0)\equiv\sum_{j=r+1}^k\sigma_j^{2}(\Pi_0)$ is the sum of the $k-r$ smallest squared singular values $\sigma_j^{2}(\Pi_0)$ of $\Pi_0$  -- see Supplemental Appendix for a review on singular values. Such a reformulation is attractive because it converts an unwieldy inference problem on an integer-valued parameter (i.e., rank) into a more tractable one on a real-valued functional (i.e., a sum of singular values). Given an estimator $\hat\Pi_n$ of $\Pi_0$, it is thus natural to base the testing statistic on the plug-in estimator $\phi_r(\hat\Pi_n)$ and then invoke the Delta method. As it turns out, the formulation \eqref{Eqn: hypothesis, phi, intro} reveals two crucial irregular natures involved, namely, $\phi_r$ admits a zero first order derivative under $\mathrm{H}_0$ and is second order nondifferentiable precisely when $\mathrm{rank}(\Pi_0)< r$ -- see Proposition \ref{Pro: phi, differentiability} and Lemma \ref{Lem: phi, socHD}. While the null limiting distributions of $\phi_r(\hat\Pi_n)$ can nonetheless be derived by existing generalizations of the Delta method \citep{Shapiro2000inference}, constructions of critical values are nontrivial because the limits are non-pivotal and highly nonstandard. In particular, they depend on the true rank (among other things), upholding the importance of taking into account the possibility $\mathrm{rank}(\Pi_0)<r$. For this, we appeal to modified bootstrap schemes recently developed by \citet{Fang_Santos2014HDD} and \citet{Chen_Fang2015FOD}, which yield tests for \eqref{Eqn: hypothesis, intro} that have asymptotically pointwise exact size control and are consistent. We further characterize analytically classes of local perturbations of the data generating processes under which our tests enjoy size control and nontrivial power.

A common feature of our tests is their dependence on tuning parameters, although we stress that this is only in line with the irregular natures of nonstandard problems \citep{CHT2007,AndrewsandSoares2010,Linton2010}. While we are unable to offer a general theory guiding their choices, a two-step procedure similar to \citet{RomanoShaikhWolf2014TwoStep} is proposed to mitigate potential concerns. The intuition is as follows. First, the appearance of $r_0\equiv \mathrm{rank}(\Pi_0)$ in the limits  suggests the need of a consistent rank estimator $\hat r_n$, which may be achieved by a sequential testing procedure coupled with a significance level $\alpha_n$ (serving as the tuning parameter) that tends to zero suitably. Although the estimation error of $\hat r_n$, i.e., the probability of false selection, is asymptotically negligible (as $\alpha_n\to 0$), that probability is positive in any finite samples. Thus, we account for false selection by fixing $\alpha_n=\beta$ rather than letting it tend to zero. Given an estimator $\hat r_n$ with $\liminf_{n\to\infty}P(\hat r_n=r_0)\ge 1-\beta$, the two-step procedure at a significance level $\alpha$ is: reject $\mathrm H_0$ if $\hat r_n>r$ in the first step; otherwise in the second step incorporate $\hat r_n$ into our bootstrap and conduct the test at the adjusted significance level $\alpha-\beta>0$. We show in a number of simulation designs that the procedure is quite insensitive to our choices of $\beta$, even for small sample sizes.

% The simulation results are particularly encouraging because we fix crude choices of tuning parameters throughout the paper, and yet the performance of our methods is consistently better.

The marked size and power properties rest with several attractive features. First, since we rely on the Delta method, the theory is conceptually simple and requires mild assumptions. Essentially, all we need are a matrix estimator $\hat\Pi_n$ that converges weakly and a consistent bootstrap analog. In particular, the data may be non-i.i.d.\ and non-stationary, the convergence rate may be non-$\sqrt n$ and even heterogeneous across entries of $\hat\Pi_n$ -- see Supplemental Appendix \ref{Sec: coint appendix}, the limit $\mathcal M$ of $\hat\Pi_n$ may be non-Gaussian, the bootstrap for $\mathcal M$ (a crucial ingredient of our method) may be virtually any consistent resampling scheme, and no side rank conditions are directly imposed beyond those entailed by the restrictions on the population quantiles. Second, computation of our testing statistic and the critical values are quite simple as both involve only calculations of singular value decompositions -- we reiterate that the need of resampling only reflects the irregular natures of the problem rather than because of an exclusive attribute of our treatment. Finally, the superior testing properties of our procedure translate to more accurate rank estimators through the aforementioned two channels, namely, reducing type I and type II errors. Simulations confirm that our methods work better when $\mathrm{rank}(\Pi_0)<r$ or when $\mathrm{\Pi_0}$ is close to a matrix whose rank is strictly less than $r$.

We illustrate the application of our framework by testing identification in linear IV models that accommodates clustered data. To draw further attention to the empirical relevance of our results, we study a two-sided bipartite matching model with transferrable utility, building upon the work of \citet{DupuyGalichon2014Personality}. A central question here is: how many attributes are relevant for the matching? Under a parametric specification of the surplus function, this number is equal to the rank of the so-called affinity matrix. We show that our procedure and \citet{Kleibergen_Paap2006rank} can produce quite different results with regards to several model specifications, in terms of both $p$-values of the tests and actual estimates of the matching dimension.

As mentioned previously, the literature has been mostly concerned with the hypotheses \eqref{Eqn: hypothesis, literature, intro}. In the context of multivariate regression, \citet{Anderson1951Estimating} develops a likelihood ratio test based on canonical correlations. This test is restrictive in that it crucially depends on the asymptotic variance $\Omega_0$ of $\mathrm{vec}(\hat\Pi_n)$ having a Kronecker product structure. Building upon \citet{Gill_Lewbel1992rank}, \citet{Cragg_Donald1996LDU} propose a test that requires nonsingularity of $\Omega_0$ and may be sensitive to the transformations involved. \citet{Cragg_Donald1997infer} provide a test based on a constrained minimum distance criterion, which, in addition to the nonsingularity requirement of $\Omega_0$, is in general computationally intensive. To relax the nonsingularity condition, \citet{Robin_Smith2000rank} employ a class of testing statistics which are asymptotically equivalent to ours, but their results only apply to the setup \eqref{Eqn: hypothesis, literature, intro}. \citet{Kleibergen_Paap2006rank} study a Wald-standardized version of our statistic in order to obtain pivotal asymptotic distributions (under $\mathrm H_0'$), but at the expense of a side rank condition. We refer the reader to \citet{Camba_Kapetanios2009rank}, \citet{Portier_Delyon2014} and \citet{Sadoon2017Rank} for further discussions.

There are a few exceptions that study \eqref{Eqn: hypothesis, intro}. \citet{Johansen1988CoInt,Johansen1991CoIntegration} obtains his likelihood ratio statistics under $ \mathrm H_0$ but only establishes their asymptotic distributions under $\mathrm H_0'$. Shortly after, \citet[p.157-8,168]{Johansen1995likelihood} presents the limits under $\mathrm H_0$, and essentially argues based on simulations that the asymptotic distributions under $\mathrm{rank}(\Pi_0)< r$ are first order stochastically dominated by those
under $\mathrm H_0'$ and ``hence not relevant for calculating the $p$-value''. However, the counterexample given in Section \ref{Sec: 2} disproves this conjecture. \citet[p.225]{Cragg_Donald1993TestID} recognize the importance of studying \eqref{Eqn: hypothesis, intro}, but do not derive the asymptotic distributions under $\mathrm{H}_0$. Instead, they show that their statistic has first order stochastically dominant limiting laws under $\mathrm H_0'$ with somewhat restrictive conditions. Our results suggest that may not be true in general.

We now introduce some notation. The space of $m\times k$ matrices is denoted by $\mathbf M^{m\times k}$. For a matrix $A$, we write its transpose by $A^\transpose$, its trace by $\mathrm{tr}(A)$ if it is square, its vectorization by $\text{vec}(A)$, and its Frobenius norm by $\|A\|\equiv\sqrt{\mathrm{tr}(A^\transpose A)}$. The identity matrix of size $k$ is denoted $I_k$, the $k\times 1$ vectors of zeros and ones are respectively denoted by $\mathbf 0_{k}$ and $\mathbf 1_k$, and  the $m\times k$ matrix of zeros is denoted $\mathbf 0_{m\times k}$. We let $\mathrm{diag}(a)$ denote the diagonal matrix whose diagonal entries compose $a$. The $j$th largest singular value of a matrix $A\in\mathbf M^{m\times k}$ is denoted $\sigma_j(A)$. We define the set $\mathbb S^{m\times k}=\{A\in\mathbf M^{m\times k}: A^\transpose A=I_k\}$ and let $\overset{d}{=}$ signify ``equal in distribution.'' Finally, $\lfloor a\rfloor$ is the integer part of $a\in\mathbf R$.

The remainder of the paper is organized as follows. Section \ref{Sec: 2} illustrates the consequences of ignoring $\mathrm{rank}(\Pi_0)<r$, and provides an overview of our tests, together with a step-by-step implementation guide. Section \ref{Sec: 3} develops our inferential framework. Section \ref{Sec: 4} presents Monte Carlo studies. Section \ref{Sec: mathching} further illustrates the empirical relevance of our results by studying a matching model. Section \ref{Sec: 5} briefly concludes. Proofs are collected in a Supplemental Appendix. We also study the estimation problem, but, due to space limitation, relegate the results to Supplemental Appendix \ref{Sec: 4-3}. Finally, we have developed a Stata command \texttt{bootranktest} to test whether a matrix of the form $E[VZ^\transpose]$ has full rank -- see the Supplemental Appendix for a brief description.

\section{Motivations, Overview and Implementation}\label{Sec: 2}
% This section is written mostly for applied people

In this section, we first motivate the development of our theory by illustrating how serious the issue can be if one ignores the possibility $\mathrm{rank}(\Pi_0)<r$ in conducting rank tests. This is accomplished by examining the influential test proposed by \citet{Kleibergen_Paap2006rank}, referred to as the KP test hereafter, and its multiple testing version. Then we provide an overview of our tests, together with a step-by-step implementation guide that applies to general settings.

To elucidate the consequences of ignoring $\mathrm{rank}(\Pi_0)<r$, consider an example where $\Pi_0=\mathbf 0_{2\times 2}$ and $r=1$ so that $\mathrm{rank}(\Pi_0)<r$. Suppose $\Pi_0$ admits an estimator $\hat\Pi_n$ such that $\sqrt{n}\hat\Pi_n\overset{d}{=}\mathcal M$ for all $n$ (rather than just asymptotically), where $\mathcal M\in\mathbf M^{2\times 2}$ satisfies $\mathrm{vec}(\mathcal M)\sim N(0,\Omega_0)$ with $\Omega_0$ nonsingular and {\it known}. In this case, the KP test for \eqref{Eqn: hypothesis, literature, intro} employs critical values from $\chi^2(1)$, while the actual distribution of the KP statistic is
\begin{align}\label{Eqn: KPTestStat: WeakLimit}
T_{n,\mathrm{kp}}\overset{d}{=} \frac{\sigma_2^2(\mathcal M)}{(\mathcal{Q}_{2}\otimes \mathcal{P}_{2})^{\transpose}\Omega_0(\mathcal{Q}_{2}\otimes \mathcal{P}_{2})}~,
\end{align}
where $\mathcal P_2$ and $\mathcal Q_2$ are the left and right singular vectors associated with $\sigma_2(\mathcal M)$, both having unit length. Note the distribution of $T_{n,\mathrm{kp}}$ depends only on $\Omega_0$. Figure \ref{Fig: KP limits} plots (based on simulations) two cdfs $F_1$ and $F_2$ of $T_{n,\mathrm{kp}}$ in \eqref{Eqn: KPTestStat: WeakLimit} respectively determined by
\begin{align}\label{Eqn: overreject, var}
\Omega_{1} = \left[
                   \begin{array}{rrrr}
                   1&0&0&0\\
                   0&1&0&0\\
                   0&0&1&0\\
                   0&0&0&1\\
                   \end{array}
                 \right]~\text{ and }~
\Omega_{2} = \left[
                   \begin{array}{rrrr}
                   1&0&0&-0.9\sqrt{5}\\
                   0&1&0.9\sqrt{5}&0\\
                   0&0.9\sqrt{5}&5&0\\
                   -0.9\sqrt{5}&0&0&5\\
                   \end{array}
                 \right]~,
\end{align}
together with the cdf $F_0$ of $\chi^2(1)$. Note that $F_0$ is stochastically dominated by $F_2$ but stochastically dominates $F_1$, both in the first order sense. Hence, the KP test is invalid due to over-rejection when $\Omega_0=\Omega_2$. We have thus {\it disproved} that the limits under $\mathrm{rank}(\Pi_0)=r$ are first order stochastically dominant in general, a conjecture by \citet{Cragg_Donald1993TestID} for their statistic which they show to hold under somewhat restrictive conditions. These erratic behaviors can also be expected for the test of \citet{Robin_Smith2000rank} in view of its relation to the KP test -- see Supplemental Appendix \ref{Sec: comparision with KP}.

 % -- see also Theorem \ref{thm: Weaklimit: Test} and remarks therein, and for \citet{Anderson1951Estimating} since it is a special case of the KP test if $\Omega_0$ admits a Kronecker product structure.

\pgfplotstableread{
X Y1 Y2 Y3
0	3.73507364540145e-09	2.43226576729300e-09	0
0.0500000000000000	0.00909715271568395	0.00160932765346150	0.00393214000001952
0.100000000000000	0.0365618586579883	0.00635285996713346	0.0157907740934312
0.150000000000000	0.0842008779330303	0.0144885301338314	0.0357657791558976
0.200000000000000	0.151732607950905	0.0259754396467885	0.0641847546673016
0.250000000000000	0.237603608717244	0.0414901201290033	0.101531044267622
0.300000000000000	0.339919101427980	0.0607538991724877	0.148471861832545
0.350000000000000	0.459730566363366	0.0846617946454281	0.205900125227766
0.400000000000000	0.596126853011560	0.113759309645023	0.274995897728455
0.450000000000000	0.751643108364612	0.147976474823992	0.357317168286320
0.500000000000000	0.922756852624857	0.189495334531488	0.454936423119572
0.550000000000000	1.12039054578462	0.237734592980602	0.570651862051189
0.600000000000000	1.34938018483665	0.295545944496175	0.708326300800793
0.650000000000000	1.60197335354828	0.365623241253882	0.873457142989230
0.700000000000000	1.89960737697793	0.451494584493166	1.07419417085759
0.750000000000000	2.25620670728667	0.558068021483900	1.32330369693147
0.800000000000000	2.69908578532006	0.695736867924860	1.64237441514982
0.850000000000000	3.26558877040370	0.884018476842553	2.07225085582223
0.900000000000000	4.09641366876075	1.16204775101438	2.70554345409541
0.950000000000000	5.49430410654744	1.66751998558478	3.84145882069413
}\datainvalid

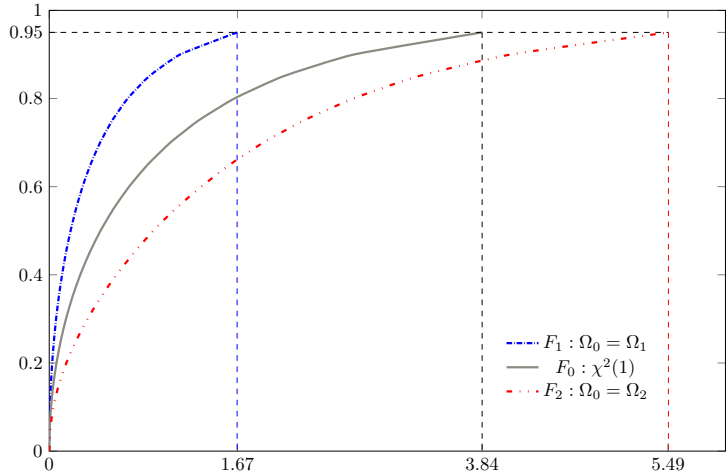
\begin{figure}[!htp]
\small\centering
\begin{tikzpicture}[scale=0.7]
\begin{axis}[scale = 1.6,
    legend style={draw=none},
    grid = minor,
    xmax=6,xmin=0,
    ymax=1,ymin=0,
    xtick={0,1.67,3.84,5.49},
    ytick={0,0.2,0.4,0.6,0.8,0.95,1},
    width=0.65\textwidth,
    height=6.8cm,
    legend style={at={(0.9,0.28)}}
    ]
\addplot[smooth,tension=0.3,densely dashdotted, color=blue, line width=1.2pt] table[x = Y2,y=X] from \datainvalid ;
\addlegendentry{$F_1: \Omega_0 = \Omega_1$}  ;
\addplot[smooth,tension=0.3, color=Ivory4, line width=1.2pt] table[x = Y3,y=X] from \datainvalid ;
\addlegendentry{$F_0: \chi^{2}(1)$}  ;
\addplot[smooth,tension=0.3,loosely dashdotdotted, color=red, line width=1.2pt] table[x = Y1,y=X] from \datainvalid ;
\addlegendentry{$F_2: \Omega_0 = \Omega_2$}  ;
\draw[dashed] (0,0.95)--(5.494,0.95);
\draw[dashed,blue] (1.667,0.95)--(1.667,0);
\draw[dashed] (3.841,0.95)--(3.841,0);
\draw[dashed,red] (5.494,0.95)--(5.494,0);
\end{axis}
\end{tikzpicture}
\caption{The cdfs of the KP statistic when $\Pi_0 = \mathbf{0}_{2\times 2}$ and $r=1$}
\end{figure}\label{Fig: KP limits}
% The distribution functions based on 100,000 simulation replications are plotted in Figure \ref{Graph: Invalid}.

Alternatively, one might aim to construct a valid test for \eqref{Eqn: hypothesis, intro} by a multiple test on $\mathrm{rank}(\Pi_0)=0,1,\ldots,r$. However, the validity is achieved at the expense of conservativeness -- see Supplemental Appendix \ref{Sec: comparision with KP}, which may generate substantial power loss. To illustrate, consider the following data generating process:
\begin{align}\label{Eqn: Motivation: DGPs}
Z =\Pi_{0}^\transpose V+u~,
\end{align}
where $V,u\in N(0,I_{6})$ are independent and, for $\delta\ge 0$ and $d\in\{1,\ldots,6\}$,
\begin{align}\label{Eqn: Motivation: DGPs1}
\Pi_{0}=\textrm{diag}(\mathbf{1}_{6-d},\mathbf{0}_{d})+\delta I_{6} ~.
\end{align}
We test the hypotheses in \eqref{Eqn: hypothesis, intro} with $r=5$ at the level $\alpha=5\%$, and note that $\mathrm{H}_{0}$ holds if and only if $\delta=0$. For an i.i.d.\ sample $\{V_i,Z_i\}_{i=1}^{1000}$ generated according to \eqref{Eqn: Motivation: DGPs}, we conduct tests based on the matrix estimator $\hat{\Pi}_{n}=\frac{1}{1000}\sum_{i=1}^{1000}V_{i}Z_{i}^{\transpose}$ for $\Pi_{0}$.

\pgfplotstableread{
delta alpha d1 d2 d3 d4 d5 d6
0	0.0500000000000000	0.0492000000000000	0.00410000000000000	0.000300000000000000	0	0	0
0.0200000000000000	0.0500000000000000	0.0984000000000000	0.00840000000000000	0.000500000000000000	0	0	0
0.0400000000000000	0.0500000000000000	0.246600000000000	0.0444000000000000	0.00560000000000000	0.000400000000000000	0.000100000000000000	0
0.0600000000000000	0.0500000000000000	0.471500000000000	0.158600000000000	0.0414000000000000	0.00740000000000000	0.00140000000000000	0.000200000000000000
0.0800000000000000	0.0500000000000000	0.717800000000000	0.399000000000000	0.179600000000000	0.0674000000000000	0.0215000000000000	0.00610000000000000
0.100000000000000	0.0500000000000000	0.887300000000000	0.678300000000000	0.444100000000000	0.257000000000000	0.126200000000000	0.0538000000000000
0.120000000000000	0.0500000000000000	0.963700000000000	0.876300000000000	0.732000000000000	0.567300000000000	0.396100000000000	0.243700000000000
0.140000000000000	0.0500000000000000	0.992000000000000	0.966400000000000	0.915300000000000	0.831000000000000	0.715300000000000	0.570400000000000
0.160000000000000	0.0500000000000000	0.998800000000000	0.995500000000000	0.983300000000000	0.958200000000000	0.917600000000000	0.843900000000000
0.180000000000000	0.0500000000000000	0.999800000000000	0.999500000000000	0.998300000000000	0.993300000000000	0.982400000000000	0.964400000000000
0.200000000000000	0.0500000000000000	1	1	1	0.999600000000000	0.998000000000000	0.994700000000000
}\motivation

\pgfplotstableread{
delta alpha d1 d2 d3 d4 d5 d6
0	0.0500000000000000	0.0502000000000000	0.0536000000000000	0.0516000000000000	0.0535000000000000	0.0529000000000000	0.0541000000000000
0.0200000000000000	0.0500000000000000	0.100900000000000	0.0832000000000000	0.0739000000000000	0.0668000000000000	0.0621000000000000	0.0636000000000000
0.0400000000000000	0.0500000000000000	0.248300000000000	0.197300000000000	0.150900000000000	0.124900000000000	0.110700000000000	0.0985000000000000
0.0600000000000000	0.0500000000000000	0.472100000000000	0.434500000000000	0.345800000000000	0.278500000000000	0.222000000000000	0.184400000000000
0.0800000000000000	0.0500000000000000	0.716400000000000	0.701300000000000	0.620600000000000	0.545100000000000	0.454300000000000	0.370200000000000
0.100000000000000	0.0500000000000000	0.883900000000000	0.885800000000000	0.848700000000000	0.795600000000000	0.723300000000000	0.632300000000000
0.120000000000000	0.0500000000000000	0.963700000000000	0.963200000000000	0.952700000000000	0.932000000000000	0.898300000000000	0.851100000000000
0.140000000000000	0.0500000000000000	0.991900000000000	0.990100000000000	0.988600000000000	0.982400000000000	0.970500000000000	0.956200000000000
0.160000000000000	0.0500000000000000	0.998600000000000	0.997400000000000	0.997300000000000	0.995700000000000	0.993900000000000	0.990700000000000
0.180000000000000	0.0500000000000000	0.999800000000000	0.999700000000000	0.999500000000000	0.999300000000000	0.998700000000000	0.998100000000000
0.200000000000000	0.0500000000000000	1	1	1	0.999800000000000	0.999800000000000	0.999500000000000
}\dominationcf
{
\begin{figure}[!h]
\begin{subfigure}[b]{0.48\textwidth}
\centering
\resizebox{\linewidth}{!}{
\begin{tikzpicture}
\begin{axis}[
    legend style={draw=none},
    grid = minor,
    xmax=0.2,xmin=0,
    ymax=1,ymin=0,
    xtick={0,0.1,0.2},
    ytick={0.5,1},
    tick label style={/pgf/number format/fixed},
legend style={at={(0.82,0.47)},anchor=north,
    row sep = 3pt}]
\addlegendimage{empty legend};
\addlegendentry{\footnotesize \hspace{-0.8cm}$d=1$}
\addplot[smooth,tension=0.5,color=SeaGreen4, line width=0.75pt] table[x = delta,y=d1] from \dominationcf ;
\addlegendentry{\footnotesize CF-A}  ;
\addplot[smooth,tension=0.5,color=black, line width=0.75pt, dashdotted] table[x = delta,y=d1] from \motivation ;
\addlegendentry{\footnotesize KP-M}  ;
\addplot[smooth,tension=0.5,no markers, color=NavyBlue, line width=0.75pt, densely dotted] table[x = delta,y=alpha] from \motivation ;
\addlegendentry{\footnotesize $5\%$ level}  ;
\end{axis}
\end{tikzpicture}}
\end{subfigure}
\begin{subfigure}[b]{0.48\textwidth}
\centering
\resizebox{\linewidth}{!}{
\begin{tikzpicture}
\begin{axis}[
    legend style={draw=none},
    grid = minor,
    xmax=0.2,xmin=0,
    ymax=1,ymin=0,
    xtick={0,0.1,0.2},
    ytick={0.5,1},
    tick label style={/pgf/number format/fixed},
legend style={at={(0.82,0.47)},anchor=north,
    row sep = 3pt}]
\addlegendimage{empty legend};
\addlegendentry{\footnotesize \hspace{-0.8cm}$d=2$}
\addplot[smooth,tension=0.5,color=SeaGreen4, line width=0.75pt] table[x = delta,y=d2] from \dominationcf ;
\addlegendentry{\footnotesize CF-A}  ;
\addplot[smooth,tension=0.5,color=black, line width=0.75pt,dashdotted] table[x = delta,y=d2] from \motivation ;
\addlegendentry{\footnotesize KP-M}  ;
\addplot[smooth,tension=0.5,no markers, color=NavyBlue, line width=0.75pt, densely dotted] table[x = delta,y=alpha] from \motivation ;
\addlegendentry{\footnotesize $5\%$ level}  ;
\end{axis}
\end{tikzpicture}}
\end{subfigure}

\begin{subfigure}[b]{0.48\textwidth}
\centering
\resizebox{\linewidth}{!}{
\begin{tikzpicture}
\begin{axis}[
    legend style={draw=none},
    grid = minor,
    xmax=0.2,xmin=0,
    ymax=1,ymin=0,
    xtick={0,0.1,0.2},
    ytick={0.5,1},
    tick label style={/pgf/number format/fixed},
legend style={at={(0.82,0.47)},anchor=north,
    row sep = 3pt}]
\addlegendimage{empty legend};
\addlegendentry{\footnotesize \hspace{-0.8cm}$d=3$}   ;
\addplot[smooth,tension=0.5,color=SeaGreen4, line width=0.75pt] table[x = delta,y=d3] from \dominationcf ;
\addlegendentry{\footnotesize CF-A}  ;
\addplot[smooth,tension=0.5,color=black, line width=0.75pt, dashdotted] table[x = delta,y=d3] from \motivation ;
\addlegendentry{\footnotesize KP-M}  ;
\addplot[smooth,tension=0.5,no markers, color=NavyBlue, line width=0.75pt, densely dotted] table[x = delta,y=alpha] from \motivation ;
\addlegendentry{\footnotesize $5\%$ level}  ;
\end{axis}
\end{tikzpicture}}
\end{subfigure}
\begin{subfigure}[b]{0.48\textwidth}
\centering
\resizebox{\linewidth}{!}{
\begin{tikzpicture}
\begin{axis}[
    legend style={draw=none},
    grid = minor,
    xmax=0.2,xmin=0,
    ymax=1,ymin=0,
    xtick={0,0.1,0.2},
    ytick={0.5,1},
    tick label style={/pgf/number format/fixed},
legend style={at={(0.82,0.47)},anchor=north,
    row sep = 3pt}]
\addlegendimage{empty legend};
\addlegendentry{\footnotesize \hspace{-0.8cm}$d=4$}  ;
\addplot[smooth,tension=0.5,color=SeaGreen4, line width=0.75pt] table[x = delta,y=d4] from \dominationcf ;
\addlegendentry{\footnotesize CF-A}  ;
\addplot[smooth,tension=0.5,color=black, line width=0.75pt,dashdotted] table[x = delta,y=d4] from \motivation ;
\addlegendentry{\footnotesize KP-M}  ;
\addplot[smooth,tension=0.5,no markers, color=NavyBlue, line width=0.75pt, densely dotted] table[x = delta,y=alpha] from \motivation ;
\addlegendentry{\footnotesize $5\%$ level}  ;
\end{axis}
\end{tikzpicture}}
\end{subfigure}

\begin{subfigure}[b]{0.48\textwidth}
\centering
\resizebox{\linewidth}{!}{
\begin{tikzpicture}
\begin{axis}[
    legend style={draw=none},
    grid = minor,
    xmax=0.2,xmin=0,
    ymax=1,ymin=0,
    xtick={0,0.1,0.2},
    ytick={0.5,1},
    tick label style={/pgf/number format/fixed},
legend style={at={(0.82,0.47)},anchor=north,
    row sep = 3pt}]
\addlegendimage{empty legend};
\addlegendentry{\footnotesize \hspace{-0.8cm}$d=5$}
\addplot[smooth,tension=0.5,color=SeaGreen4, line width=0.75pt] table[x = delta,y=d5] from \dominationcf ;
\addlegendentry{\footnotesize CF-A}  ;
\addplot[smooth,tension=0.5,color=black, line width=0.75pt, dashdotted] table[x = delta,y=d5] from \motivation ;
\addlegendentry{\footnotesize KP-M}  ;
\addplot[smooth,tension=0.5,no markers, color=NavyBlue, line width=0.75pt, densely dotted] table[x = delta,y=alpha] from \motivation ;
\addlegendentry{\footnotesize $5\%$ level}  ;
\end{axis}
\end{tikzpicture}}
\end{subfigure}
\begin{subfigure}[b]{0.48\textwidth}
\centering
\resizebox{\linewidth}{!}{
\begin{tikzpicture}
\begin{axis}[
    legend style={draw=none},
    grid = minor,
    xmax=0.2,xmin=0,
    ymax=1,ymin=0,
    xtick={0,0.1,0.2},
    ytick={0.5,1},
    tick label style={/pgf/number format/fixed},
legend style={at={(0.82,0.47)},anchor=north,
    row sep = 3pt}]
\addlegendimage{empty legend};
\addlegendentry{\footnotesize \hspace{-1cm}$d=6$}
\addplot[smooth,tension=0.5,color=SeaGreen4, line width=0.75pt] table[x = delta,y=d6] from \dominationcf ;
\addlegendentry{\footnotesize CF-A}  ;
\addplot[smooth,tension=0.5,color=black, line width=0.75pt,dashdotted] table[x = delta,y=d6] from \motivation ;
\addlegendentry{\footnotesize KP-M}  ;
\addplot[smooth,tension=0.5,no markers, color=NavyBlue, line width=0.75pt, densely dotted] table[x = delta,y=alpha] from \motivation ;
\addlegendentry{\footnotesize $5\%$ level}  ;
\end{axis}
\end{tikzpicture}}
\end{subfigure}
\caption{Conservativeness of the KP-M test. The number of Monte Carlo simulations is 10,000, the number of bootstrap repetitions is 500, and $\kappa_n=n^{-1/4}$ (for CF-A).}\label{Fig: CF vs KP}
\end{figure}
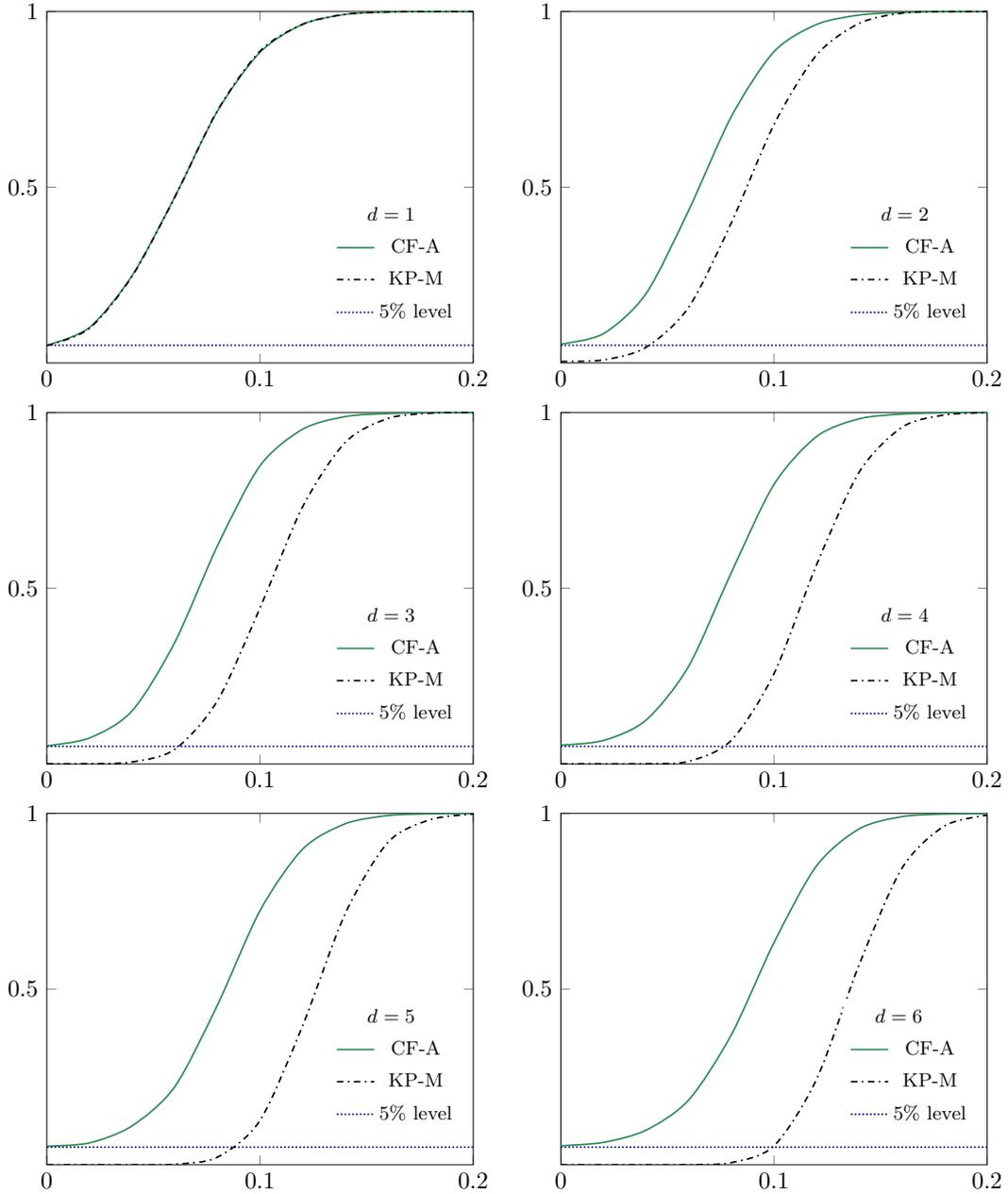
}

\pgfplotstableread{
delta alpha kp kpm cft cfa cfn
0	0.0500000000000000	0.00500000000000000	0.00460000000000000	0.0444000000000000	0.0514000000000000	0.0482000000000000
0.0200000000000000	0.0500000000000000	0.00990000000000000	0.00900000000000000	0.0629000000000000	0.0733000000000000	0.0693000000000000
0.0400000000000000	0.0500000000000000	0.0386000000000000	0.0374000000000000	0.160300000000000	0.187200000000000	0.180600000000000
0.0600000000000000	0.0500000000000000	0.154900000000000	0.153000000000000	0.338200000000000	0.430900000000000	0.411500000000000
0.0800000000000000	0.0500000000000000	0.392300000000000	0.390500000000000	0.540400000000000	0.703500000000000	0.685500000000000
0.100000000000000	0.0500000000000000	0.678900000000000	0.678500000000000	0.734800000000000	0.890200000000000	0.883000000000000
0.120000000000000	0.0500000000000000	0.885000000000000	0.885000000000000	0.890300000000000	0.967900000000000	0.969000000000000
0.140000000000000	0.0500000000000000	0.972600000000000	0.972600000000000	0.969200000000000	0.991100000000000	0.994400000000000
0.160000000000000	0.0500000000000000	0.995800000000000	0.995800000000000	0.994800000000000	0.997800000000000	0.999200000000000
0.180000000000000	0.0500000000000000	0.999300000000000	0.999300000000000	0.999200000000000	0.999600000000000	0.999900000000000
0.200000000000000	0.0500000000000000	0.999900000000000	0.999900000000000	0.999900000000000	0.999900000000000	1
}\motivationa

\iffalse
\pgfplotstableread{
delta alpha kp kpm cft cfa cfn
0	0.0500000000000000	0.115100000000000	0.0290000000000000	0.0469000000000000	0.0501000000000000	0.0420000000000000
0.0400000000000000	0.0500000000000000	0.239800000000000	0.119700000000000	0.157700000000000	0.169600000000000	0.150100000000000
0.0800000000000000	0.0500000000000000	0.446800000000000	0.402200000000000	0.416900000000000	0.483200000000000	0.455800000000000
0.1200000000000000	0.0500000000000000	0.602000000000000	0.600600000000000	0.594300000000000	0.756700000000000	0.730700000000000
0.1600000000000000	0.0500000000000000	0.744600000000000	0.744600000000000	0.734100000000000	0.908100000000000	0.889600000000000
0.200000000000000	0.0500000000000000	0.873700000000000	0.873700000000000	0.866600000000000	0.953300000000000	0.962300000000000
0.240000000000000	0.0500000000000000	0.952400000000000	0.952400000000000	0.947800000000000	0.964200000000000	0.989100000000000
0.280000000000000	0.0500000000000000	0.986600000000000	0.986600000000000	0.984600000000000	0.987100000000000	0.997600000000000
0.320000000000000	0.0500000000000000	0.997200000000000	0.997200000000000	0.996800000000000	0.997100000000000	0.999700000000000
0.360000000000000	0.0500000000000000	0.999500000000000	0.999500000000000	0.999500000000000	0.999500000000000	0.999900000000000
0.400000000000000	0.0500000000000000	0.999900000000000	0.999900000000000	0.999900000000000	0.999900000000000	1
}\motivationb
\fi

\pgfplotstableread{
delta alpha kp kpm cft cfa cfn
0	0.0500000000000000	0.115100000000000	0.0290000000000000	0.0469000000000000	0.0501000000000000	0.0420000000000000
0.0400000000000000	0.0500000000000000	0.239800000000000	0.119700000000000	0.157700000000000	0.169600000000000	0.150100000000000
0.0800000000000000	0.0500000000000000	0.446800000000000	0.402200000000000	0.416900000000000	0.483200000000000	0.455800000000000
0.1200000000000000	0.0500000000000000	0.602000000000000	0.600600000000000	0.594300000000000	0.756700000000000	0.730700000000000
0.1600000000000000	0.0500000000000000	0.744600000000000	0.744600000000000	0.734100000000000	0.908100000000000	0.889600000000000
0.200000000000000	0.0500000000000000	0.873700000000000	0.873700000000000	0.866600000000000	0.953300000000000	0.962300000000000
0.240000000000000	0.0500000000000000	0.952400000000000	0.952400000000000	0.947800000000000	0.964200000000000	0.989100000000000
0.280000000000000	0.0500000000000000	0.986600000000000	0.986600000000000	0.984600000000000	0.987100000000000	0.997600000000000
0.320000000000000	0.0500000000000000	0.997200000000000	0.997200000000000	0.996800000000000	0.997100000000000	0.999700000000000
0.360000000000000	0.0500000000000000	0.999500000000000	0.999500000000000	0.999500000000000	0.999500000000000	0.999900000000000
0.400000000000000	0.0500000000000000	0.999900000000000	0.999900000000000	0.999900000000000	0.999900000000000	1
}\motivationb

{
\begin{figure}[!h]
\begin{subfigure}[b]{0.48\textwidth}
\centering
\resizebox{\linewidth}{!}{
\begin{tikzpicture}
\begin{axis}[
    legend style={draw=none},
    grid = minor,
    xmax=0.2,xmin=0,
    ymax=1,ymin=0,
    xtick={0,0.1,0.2},
    ytick={0.5,1},
    tick label style={/pgf/number format/fixed},
legend style={at={(0.98,0.6)}}]
\addlegendimage{empty legend};
\addlegendentry{\footnotesize \hspace{-1cm}$\Omega_0=\Omega_1$}
\addplot[smooth,tension=0.5,mark=*, color=blue, line width=0.75pt] table[x = delta,y=cfa] from \motivationa ;
\addlegendentry{\footnotesize CF-A}  ;
\addplot[smooth,tension=0.5,mark=pentagon, color=green, line width=0.75pt] table[x = delta,y=cfn] from \motivationa ;
\addlegendentry{\footnotesize CF-N}  ;
\addplot[smooth,tension=0.5,mark=triangle, color={rgb:red,4;green,2;yellow,1}, line width=0.75pt] table[x = delta,y=cft] from \motivationa ;
\addlegendentry{\footnotesize CF-T}  ;
\addplot[smooth,tension=0.5,mark =square , color=magenta, line width=0.75pt] table[x = delta,y=kp] from \motivationa ;
\addlegendentry{\footnotesize KP}  ;
\addplot[smooth,tension=0.5,mark=star, color=red, line width=0.75pt] table[x = delta,y=kpm] from \motivationa ;
\addlegendentry{\footnotesize KP-M}  ;
\addplot[smooth,tension=0.5,no markers, color=NavyBlue, line width=0.75pt, densely dotted] table[x = delta,y=alpha] from \motivationa ;
\addlegendentry{\footnotesize $5\%$ level}  ;
\end{axis}
\end{tikzpicture}}
\end{subfigure}
\begin{subfigure}[b]{0.48\textwidth}
\centering
\resizebox{\linewidth}{!}{
\begin{tikzpicture}
\begin{axis}[
    legend style={draw=none},
    grid = minor,
    xmax=0.4,xmin=0,
    ymax=1,ymin=0,
    xtick={0,0.2,0.4},
    ytick={0.5,1},
    tick label style={/pgf/number format/fixed},
legend style={at={(0.98,0.6)}}]
\addlegendimage{empty legend};
\addlegendentry{\footnotesize \hspace{-1cm}$\Omega_0=\Omega_2$}
\addplot[smooth,tension=0.5,mark=*, color=blue, line width=0.75pt] table[x = delta,y=cfa] from \motivationb ;
\addlegendentry{\footnotesize CF-A}  ;
\addplot[smooth,tension=0.5,mark=pentagon, color=green, line width=0.75pt] table[x = delta,y=cfn] from \motivationb ;
\addlegendentry{\footnotesize CF-N}  ;
\addplot[smooth,tension=0.5,mark=triangle, color={rgb:red,4;green,2;yellow,1}, line width=0.75pt] table[x = delta,y=cft] from \motivationb ;
\addlegendentry{\footnotesize CF-T}  ;
\addplot[smooth,tension=0.5,mark =square , color=magenta, line width=0.75pt] table[x = delta,y=kp] from \motivationb ;
\addlegendentry{\footnotesize KP}  ;
\addplot[smooth,tension=0.5,mark=star, color=red, line width=0.75pt] table[x = delta,y=kpm] from \motivationb ;
\addlegendentry{\footnotesize KP-M}  ;
\addplot[smooth,tension=0.5,no markers, color=NavyBlue, line width=0.75pt, densely dotted] table[x = delta,y=alpha] from \motivationb ;
\addlegendentry{\footnotesize $5\%$ level}  ;
\end{axis}
\end{tikzpicture}}
\end{subfigure}
\caption{Comparisons with the KP and the KP-M tests. The number of Monte Carlo simulations is 10,000, the number of bootstrap repetitions is 1000, $\kappa_n=n^{-1/4}$ (for both CF-A and CF-N), and $\beta=\alpha/10$ (for CF-T).}\label{Fig: CF vs KP 2}
\end{figure}
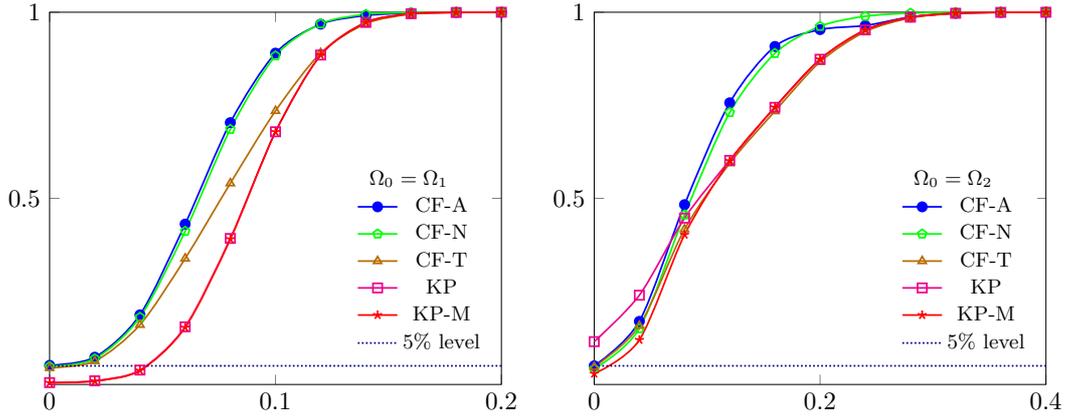
}

Figure \ref{Fig: CF vs KP} plots the power functions (against $\delta$) of the multiple KP test, labelled KP-M. For $d=1$ (and so $\mathrm{rank}(\Pi_0)=r$), the null rejection rate is 5\%, while the power increases to unity as $\delta$ increases. As soon as $d>1$ (so that $\mathrm{rank}(\Pi_0)<r$), the power curves shift downward dramatically: the null rejection rates are close to zero and the power is well below 5\% when $\delta$ is close to zero. Moreover, the power deteriorates as $\Pi_0$ becomes more degenerate in the sense that $\Pi_0$ is close to a matrix whose rank becomes smaller as $d$ increases. This reinforces the critical importance to accommodate $\mathrm{rank}(\Pi_0)<r$.

% This reinforces our belief that it is critical to develop a valid and powerful test that accommodates $\mathrm{rank}(\Pi_0)<r$.

% The results for other choices of $\kappa_n$, and those based on the numerical estimator and on the two-step test share similar patterns, and are available upon request.

% The number of simulation replications is set to be 10,000, with $500$ nonparametric i.i.d.\ bootstrap repetitions for each loop.

% Having illustrated the limitations of existing work,

To compare, we first show that three versions of our test -- CF-A, CF-N and CF-T (see below) -- control size even when the KP test does not. Let $\{Z_i\}_{i=1}^{1000}$ be an i.i.d.\ sample in $\mathbf M^{2\times 2}$ such that $\mathrm{vec}(Z_1)\sim N(\mathrm{vec}(\Pi_0), \Omega_0)$, where $\mathrm{vec}(\Pi_0) = \delta \Omega_0^{1/2} \mathrm{vec}(I_{2})$ with $\delta\geq 0$ and $\Omega_0\in\{\Omega_1,\Omega_2\}$ as in \eqref{Eqn: overreject, var}. We test \eqref{Eqn: hypothesis, intro} with $r = 1$ based on $\hat{\Pi}_{n} = \frac{1}{1000}\sum_{i=1}^{1000}Z_{i}$, at $\alpha = 5\%$. Figure \ref{Fig: CF vs KP 2} shows our tests indeed control size for both choices of $\Omega_0$, while the KP test under-rejects when $\Omega_0=\Omega_1$ and over-rejects when $\Omega_0=\Omega_2$. Note also that the KP-M test is conservative. Next, for the designs in \eqref{Eqn: Motivation: DGPs} and \eqref{Eqn: Motivation: DGPs1}, Figure \ref{Fig: CF vs KP} depicts the power curves of CF-A. For $d=1$, CF-A and KP-M have virtually the same rejection rates across $\delta$. Whenever $d>1$, our test effectively raises the power curves of the KP-M test so that the null rejection rates equal $5\%$, and  the power becomes nontrivial. But it is more than that. The power improvement increases when $d$ gets larger.

To describe our test, let $\hat\Pi_n$ be an estimator of $\Pi_0\in\mathbf M^{m\times k}$ with $\tau_n\{\hat\Pi_n-\Pi_0\}\convl \mathcal M$. The exact characterization of $\mathcal M$ (e.g., the covariance structure) is not required. Here, $\tau_n$ is typically $\sqrt n$ in cross-sectional and stationary time series settings, and may be non-$\sqrt n$ with non-stationary time series. Then our test statistic for \eqref{Eqn: hypothesis, intro} is $\tau_n^2\phi_r(\hat\Pi_n)\equiv \tau_n^2\sum_{j=r+1}^{k}\sigma_j^2(\hat\Pi_n)$. It turns out that, under $\mathrm{H}_0$, we have: for $r_0\equiv\mathrm{rank}(\Pi_0)$,
\begin{align} \label{Eqn: Nulllimt0}
\tau_{n}^{2}\phi_r(\hat{\Pi}_{n})\overset{L}{\rightarrow} \sum_{j=r-r_{0}+1}^{k-r_{0}}\sigma^{2}_j(P_{0,2}^\transpose \mathcal{M} Q_{0,2})~,
\end{align}
where $P_{0,2}\in\mathbb S^{m\times(m-r_0)}$ and $Q_{0,2}\in\mathbb S^{k\times (k-r_0)}$ whose columns are respectively the left and the right singular vectors of $\Pi_{0}$ associated with its zero singular values. Since the limit in \eqref{Eqn: Nulllimt0} depends on the true rank $r_0$ (crucially), $P_{0,2}$, $Q_{0,2}$ and $\mathcal M$, we estimate its law by first estimating these unknown objects, towards constructing critical values.

The rank $r_0$ may be consistently (under $\mathrm H_0$) estimated by: for $\kappa_n\to 0$ and $\tau_n\kappa_n\to\infty$,
\begin{align}\label{Eqn: bootstrap overview0}
\hat r_n=\max\{j=1,\ldots,r: \sigma_j(\hat\Pi_n)\ge \kappa_n\}
\end{align}
if the set is nonempty and $\hat r_n=0$ otherwise. Heuristically, $\kappa_n$ may be thought of as testing which population singular values are zero. Note that by estimating $r_0$ we take into account the possibility $r_0<r$. Next, for a singular value decomposition $\hat\Pi_n=\hat P_n\hat\Sigma_n\hat Q_n^\transpose$, we may respectively estimate $P_{0,2}$ and $Q_{0,2}$ by $\hat{P}_{2,n}$ and $\hat{Q}_{2,n}$, which are respectively formed by the last $(m-\hat r_n)$ and $(k-\hat r_n)$ columns of $\hat P_n$ and $\hat Q_n$. The law of $\mathcal M$ may be consistently estimated by a bootstrap, say, $\hat{\mathcal M}_n^*$. Often, $\hat{\mathcal M}_n^*=\sqrt n\{\hat\Pi_n^*-\hat\Pi_n\}$ with $\hat\Pi_n^*$ computed in the same way as $\hat\Pi_n$ but based on a bootstrap sample. Finally, the law of the limit in \eqref{Eqn: Nulllimt0} is estimated by the conditional distribution (given the data) of
\begin{align}\label{Eqn: bootstrap overview}
\sum_{j=r-\hat r_n+1}^{k-\hat{r}_{n}}\sigma^{2}_j(\hat{P}_{2,n}^\transpose \hat{\mathcal M}_n^* \hat{Q}_{2,n})~.
\end{align}
Given a significance level $\alpha$, the CF-A test rejects $\mathrm H_0$ whenever $\tau_n^2\phi_r(\hat\Pi_n)>\hat c_{n,1-\alpha}$, where $\hat c_{n,1-\alpha}$ is the $1-\alpha$ conditional quantile of \eqref{Eqn: bootstrap overview} given the data.

While we are unable to provide an optimal choice of $\kappa_n$, a two-step test, CF-T, is proposed to mitigate potential concerns. In the first step, we obtain an estimator $\hat r_n$ satisfying $\liminf_{n\to\infty}P(\hat r_n=r_0)\ge 1-\beta$ for some $\beta<\alpha$, and then reject $\mathrm{H}_0$ if $\hat r_n>r$ and move on to the next step if $\hat r_n\le r$. In the second step, we plug $\hat r_n$ into \eqref{Eqn: bootstrap overview} and reject $\mathrm{H}_0$ if $\tau_n^2\phi_r(\hat\Pi_n)>\hat c_{n,1-\alpha+\beta}$, where the significance level is adjusted to be $\alpha-\beta$. The estimator $\hat r_n$ in \eqref{Eqn: bootstrap overview0} now may not be appropriate as it appears challenging to control $P(\hat r_n=r_0)$. Instead, a desired estimator $\hat r_n$ may be obtained by a sequential testing procedure as actually employed in the literature and formalized in Supplemental Appendix \ref{Sec: 4-3}. In this regard, we stress that the KP test may be utilized and is recommended as it is tuning parameter free and does not require additional simulations.

Below we provide an implementation guide for testing \eqref{Eqn: hypothesis, intro} at significance level $\alpha$.

{\centering
\begin{tcolorbox}[enhanced,breakable,boxrule=0.7pt,width=0.95\textwidth]
\underline{\sc Step 1:} Compute a singular value decomposition $\hat\Pi_n=\hat P_n\hat\Sigma_n\hat Q_n^\transpose$.

\vspace{3pt}

\underline{\sc Step 2:} Obtain $\hat r_n$ as in \eqref{Eqn: bootstrap overview0} for a chosen $\kappa_n$ (e.g.\ $\kappa_n=n^{-1/4}$).

\vspace{3pt}

\underline{\sc Step 3:} Bootstrap $B$ times and compute copies of $\hat{\mathcal M}_n^*$, denoted $\{\hat{\mathcal M}_{n,b}^*\}_{b=1}^B$.

\vspace{3pt}

\underline{\sc Step 4:} For $\hat P_{2,n}$ and $\hat Q_{2,n}$ formed by the last $(m-\hat r_n)$ and $(k-\hat r_n)$ columns of $\hat P_n$ and $\hat Q_n$ respectively, set $\hat c_{n,1-\alpha}$ to be the $\lfloor B(1-\alpha)\rfloor$-th largest value in
\begin{align}\label{Eqn: bootstrap overview2}
\sum_{j=r-\hat r_n+1}^{k-\hat{r}_{n}}\sigma^{2}_j(\hat{P}_{2,n}^\transpose \hat{\mathcal M}_{n,1}^* \hat{Q}_{2,n})~,\,\ldots~,\sum_{j=r-\hat r_n+1}^{k-\hat{r}_{n}}\sigma^{2}_j(\hat{P}_{2,n}^\transpose \hat{\mathcal M}_{n,B}^* \hat{Q}_{2,n})~.
\end{align}

\vspace{3pt}

\underline{\sc Step 5:} Reject $\mathrm{H}_0$ if $\tau_n^2\sum_{j=r+1}^{k}\sigma_j^2(\hat\Pi_n)>\hat c_{n,1-\alpha}$.
\end{tcolorbox}
}

Compared to CF-N which is based on the numerical differentiation \citep{Hong_Li2017numerical} (see Sections \ref{Sec: 3} and \ref{Sec: 4} for more details), CF-A is somewhat insensitive to the choice of $\kappa_n$ even in small samples. The two-step test CF-T, on the other hand, is overall the least sensitive, but may be over-sized in small samples $(n\le 100)$. Thus, for practical purpose, we recommend the latter when the sample size is reasonably large. To implement it, one replaces {\sc Steps 2} and 5 with

{\centering
\begin{tcolorbox}[enhanced,breakable,boxrule=0.7pt,width=0.95\textwidth]
\underline{\sc Step 2':} Obtain $\hat r_n$ by sequentially testing $\mathrm{rank}(\Pi_0)=0,1,\ldots,k-1$ at level $\beta$ (e.g., $\beta=\alpha/10$) using the KP test (based on $\hat\Pi_n$), i.e., $\hat r_n=j^*$ if accepting $\mathrm{rank}(\Pi_0)=j^*$ is the first acceptance in the procedure, and $\hat r_n=k$ if all nulls are rejected. Reject $\mathrm H_0$ if $\hat r_n>r$ and move on to Step 3 otherwise.

\vspace{3pt}

\underline{\sc Step 5':} Reject $\mathrm{H}_0$ if $\tau_n^2\sum_{j=r+1}^{k}\sigma_j^2(\hat\Pi_n)>\hat c_{n,1-\alpha+\beta}$.
\end{tcolorbox}
}

\iffalse
\underline{\sc Step 1:} (a) Obtain $\hat r_n$ by sequentially testing $\mathrm{rank}(\Pi_0)=0,1,\ldots,k-1$ at level $\beta$ (e.g., $\beta=\alpha/10$) using the KP test (based on $\hat\Pi_n$), i.e., $\hat r_n=j^*$ if accepting $\mathrm{rank}(\Pi_0)=j^*$ is the first acceptance, and $\hat r_n=k$ if all nulls are rejected.

\textcolor{white}{\underline{\sc Step 1:}} (b) Reject $\mathrm H_0$ if $\hat r_n>r$ and otherwise move on to the next step.

\underline{\sc Step 2:} (a) Draw $B$ bootstrap samples to compute copies of $\hat{\mathcal M}_n^*$, denoted $\{\hat{\mathcal M}_{n,b}^*\}_{b=1}^B$.

\textcolor{white}{\underline{\sc Step 2:}} (b) Compute a singular value decomposition $\hat\Pi_n=\hat P_n\hat\Sigma_n\hat Q_n$, and let $\hat P_{2,n}$ and $Q_{2,n}$ be the last $(m-\hat r_n)$ and $(k-\hat r_n)$ columns of $\hat P_n$ and $\hat Q_n$ respectively.

\textcolor{white}{\underline{\sc Step 2:}} (c) Obtain $\hat c_{n,1-\alpha}$ as the $\lfloor B(1-\alpha+\beta)\rfloor$-th largest number in
\begin{align}\label{Eqn: bootstrap overview2}
\sum_{j=r-\hat r_n+1}^{k-\hat{r}_{n}}\sigma^{2}_j(\hat{P}_{2,n}^\transpose \hat{\mathcal M}_{n,1}^* \hat{Q}_{2,n})~,\,\ldots~,\sum_{j=r-\hat r_n+1}^{k-\hat{r}_{n}}\sigma^{2}_j(\hat{P}_{2,n}^\transpose \hat{\mathcal M}_{n,B}^* \hat{Q}_{2,n})~.
\end{align}

\textcolor{white}{\underline{\sc Step 2:}} (d)  Reject $\mathrm{H}_0$ if $\tau_n^2\sum_{j=r+1}^{k}\sigma_j^2(\hat\Pi_n)>\hat c_{n,1-\alpha+\beta}$.

% \footnote{In Stata, this is produced by command \texttt{matrix svd hatPn hatSigman hatQn = hatPin}.}
\fi

\section{The Inferential Framework}\label{Sec: 3}
%We let $\Pi_0\in\mathbf{M}^{m\times k}$ be a generic $m\times k$ matrix over the real field $\mathbf R$ that depends on the underlying distribution. Specifically, we identify $\Pi_{0}$ as a map $\Pi:\mathbf{P}\to\mathbf{M}^{m\times k}$ such that $\Pi_{0} = \Pi(P_{0})$ where $\mathbf{P}$ is the set of distributions under consideration that govern the data generating process and $P_{0}\in\mathbf{P}$ is the truth. For special matrices such as covariance matrix, various restrictions such as symmetry and positive semidefiniteness can be imposed on the parameter space. For example, $\Pi_0$ is positive semidefinite in testing for perfect multicolinearity in Example \ref{Ex: LinearIV} and in Example \ref{Ex: Conintegration_Nonparametric}. It is without loss of generality to assume $m\geq k$.

In this section, we develop our inferential framework in three steps. First, we derive the differential properties of the map $\phi_r$ given in \eqref{Eqn: hypothesis, phi, intro}, which is nontrivial and the key to our theory. Second, given an estimator $\hat\Pi_n$ of $\Pi_0$, we derive the asymptotic distributions for the plug-in estimator $\phi_r(\hat\Pi_n)$ by invoking the Delta method. These limits turn out to be highly nonstandard whenever $\mathrm{rank}(\Pi_0)<r$. Thus, in the third step, we construct valid and powerful rank tests by appealing  to recent advances on bootstrap in irregular problems \citep{Fang_Santos2014HDD,Chen_Fang2015FOD,Hong_Li2017numerical}. A two-step test is proposed to mitigate potential concerns on sensitivity of our tests to the choices of tuning parameters. Local properties of our tests will also be discussed.

\subsection{Differential Properties}\label{Sec: 3-1}

Let $\Pi_0\in\mathbf{M}^{m\times k}$ be an unknown matrix with $m\ge k$ and $\sigma_1(\Pi_0)\ge\cdots\ge\sigma_k(\Pi_0)\ge 0$ be singular values of $\Pi_0$. Then the rank of $\Pi_0$ is equal to the number of nonzero singular values of $\Pi_0$ -- see, for example, \citet[p.5]{Bhatia1997Matrix} and also Supplemental Appendix for a brief review. Hence, the hypotheses in \eqref{Eqn: hypothesis, intro} are equivalent to
\begin{align}\label{Eqn: hypothesis II}
\mathrm{H}_{0}: \phi_r(\Pi_0)=0 \qquad \text{v.s.} \qquad  \mathrm H_1: \phi_r(\Pi_0)>0~,
\end{align}
where $\phi_r: \mathbf M^{m\times k}\to\mathbf R$ is given by
\begin{align}\label{Eqn: phi, defn}
\phi_r(\Pi)\equiv\sum_{j=r+1}^k\sigma^{2}_j(\Pi)~.
\end{align}
Heuristically, $\phi_r(\Pi)$ simply gives us the sum of the $k-r$ smallest squared singular values of $\Pi$. One may also consider other $L_p$-type functionals such as $\sum_{j=r+1}^k\sigma_j(\Pi)$. Our current focus, however, allows us to uncover $\chi^2$-type limiting distributions when $\mathrm{rank}(\Pi_0)=r$ and in this way facilitates comparisons with existing rank tests.

Towards deriving the asymptotic distributions of the plug-in estimator $\phi_r(\hat\Pi_n)$ for a given estimator $\hat\Pi_n$ of $\Pi_0$, we need to first establish suitable differentiability for the map $\phi_r$. The following lemma shall prove useful in this regard.

\begin{lem}\label{Lem: phi, representation}
For the map $\phi_r$ in \eqref{Eqn: phi, defn}, we have:
\begin{align}\label{Eqn: phi, representation}
\phi_r(\Pi)=\min_{U\in\mathbb S^{k\times(k-r)}}\|\Pi U\|^2~.
\end{align}
\end{lem}

Lemma \ref{Lem: phi, representation} shows that $\phi_r(\Pi)$ can be represented as the minimum of a quadratic form over the space of orthonormal matrices in $\mathbf M^{m\times (k-r)}$. The special case when $r=k-1$ (corresponding to the test of $\Pi$ having full rank) is a well known implication of the classical Courant-Fischer theorem, i.e., $\sigma^{2}_k(\Pi)=\min_{\|U\|=1}\|\Pi U\|^2$. Note that the minimum in \eqref{Eqn: phi, representation} is attained and hence well defined. It turns out that $\phi_r$ is not fully differentiable in general but belongs to a class of directionally differentiable maps. For completeness, we next introduce the relevant notions of directional differentiability.

\begin{defn}\label{Defn: diff}
Let $\phi:\mathbf{M}^{m\times k}\to\mathbf R$ be a generic function.
\begin{itemize}
    \item[(i)] The map $\phi$ is said to be {\it Hadamard directionally differentiable} at $\Pi \in\mathbf{M}^{m\times k}$ if there is a map $\phi_\Pi':\mathbf{M}^{m\times k}\to\mathbf R$ such that:
    \begin{equation}\label{Eqn: HDD 1st}
    \lim_{n\rightarrow \infty}\frac{\phi(\Pi +t_n M_n)-\phi(\Pi)}{t_n} = \phi_\Pi'(M) ~,
    \end{equation}
    whenever $M_n\to M$ in $\mathbf{M}^{m\times k}$ and $t_n\downarrow 0$ for $\{t_n\}$ all strictly positive.
\item[(ii)] If $\phi:\mathbf{M}^{m\times k}\to\mathbf R$ is Hadamard directionally differentiable at $\Pi \in\mathbf{M}^{m\times k}$, then we say that $\phi$ is {\it second order Hadamard directionally differentiable} at $\Pi\in\mathbf{M}^{m\times k}$ if there is a map $\phi_\Pi'':\mathbf{M}^{m\times k} \to\mathbf R$ such that:
\begin{align}\label{Eqn: HDD 2nd}
\lim_{n\rightarrow \infty}\frac{\phi(\Pi +t_n M_n)-\phi(\Pi)-t_n\phi_\Pi'(M_n)}{t_n^2} = \phi_\Pi''(M)~,
\end{align}
whenever $M_n\to M$ in $\mathbf{M}^{m\times k}$ and $t_n\downarrow 0$ for $\{t_n\}$ all strictly positive.
\end{itemize}
\end{defn}

For simplicity, we shall drop the qualifier ``Hadamard'' in what follows, with the understanding that both full differentiability and directional differentiability (both first and second order) are meant in the Hadamard sense. Definition \ref{Defn: diff}(i) generalizes  (full) differentiability which additionally requires the derivative $\phi_\Pi'$ to be linear. By Proposition 2.1 in \citet{Fang_Santos2014HDD}, linearity is precisely the gap between these two notions of differentiability -- see also \citet{Shapiro1990} for more discussions. Despite the relaxation, the Delta method remains valid even when $\phi$ is only directionally differentiable \citep{Shapiro1991,Dumbgen1993}. Unfortunately, as shall be proved, the asymptotic distributions of our statistic $\phi(\hat\Pi_n)$ implied by this generalized Delta method are degenerate under the null. In turn, Definition \ref{Defn: diff}(ii) formulates a suitable second order analog of the directional differentiability, which permits us to obtain nondegenerate asymptotic distributions by a (generalized) second order Delta method \citep{Shapiro2000inference,Chen_Fang2015FOD}. The second order  directional differentiability becomes second order  full differentiability precisely when $\phi_\Pi''$ corresponds to a bilinear form.

The following proposition formally establishes the differentiability of $\phi_r$.

\begin{pro}\label{Pro: phi, differentiability}
Let $\phi_r: \mathbf M^{m\times k}\to\mathbf R$ be defined as in \eqref{Eqn: phi, defn}.
\begin{itemize}
\item[(i)] $\phi_r$ is first order directionally differentiable at any $\Pi\in \mathbf M^{m\times k}$ with the derivative $\phi_{r,\Pi}': \mathbf M^{m\times k}\to\mathbf R$ given by
\begin{align}
\phi_{r,\Pi}'(M)=\min_{U\in\Psi(\Pi)} 2\mathrm{tr}\big((\Pi U)^\transpose MU\big)~,
\end{align}
where $\Psi(\Pi)\equiv\amin_{U\in\mathbb S^{k\times (k-r)}}\|\Pi U\|^{2}$.
\item[(ii)] $\phi_r$ is second order  directionally differentiable at any $\Pi\in \mathbf M^{m\times k}$ satisfying $\phi_r(\Pi)=0$ with the derivative
$\phi_{r,\Pi}^{\prime\prime}: \mathbf M^{m\times k}\to\mathbf R$ given by: for $r_0\equiv\mathrm{rank}(\Pi)$,
\begin{align}\label{Eqn: phi: derivative: representation0}
\phi_{r,\Pi}^{\prime\prime}(M) = \sum_{j=r-r_{0}+1}^{k-r_{0}}\sigma^{2}_j(P_2^\transpose MQ_2)~,
\end{align}
where the columns of $P_2\in\mathbb S^{m\times (m-r_0)}$ and $Q_2\in \mathbb S^{k\times (k-r_0)}$ are left and right singular vectors associated with the zero singular values of $\Pi$.
\end{itemize}
\end{pro}

Proposition \ref{Pro: phi, differentiability}(i) shows that $\phi_r$ is not fully differentiable in general but only directionally differentiable. Moreover, the first order derivative is degenerate at zero whenever $\phi_r(\Pi)=0$ as in this case $\Pi U=0$ for any $U\in\Psi(\Pi)$. Proposition \ref{Pro: phi, differentiability}(ii) indicates that $\phi_r$ is second order  directionally differentiable whenever the degeneracy occurs, and, interestingly, the derivative evaluated at $M$ is simply the sum of the $k-r$ smallest squared singular values of the $(m-r_0)\times (k-r_0)$ matrix $P_2^\transpose MQ_2$. In general, $\phi_r$ is not second order  fully differentiable precisely when $\mathrm{rank}(\Pi)<r$, reflecting a critical irregular nature of our setup -- see Lemma \ref{Lem: phi, socHD} for more details. To gain further intuition, suppose that $\Pi_0=\mathrm{diag}(\pi_{0,1},\pi_{0,2})$ and we want to test if $\mathrm{rank}(\Pi_0)\le 1$. Then by definition
\begin{align}
\phi_r(\Pi_0)=\min\{\pi_{0,1}^2,\pi_{0,2}^2\}~.
\end{align}
Note that if $\mathrm{rank}(\Pi_0)\le 1$, then $\pi_{0,1}^2=\pi_{0,2}^2$ if and only if $\mathrm{rank}(\Pi_0)<1$ in which case $\pi_{0,1}=\pi_{0,2}=0$. Hence, $\phi_r$ is not second order differentiable at $\Pi_0$ if and only if $\mathrm{rank}(\Pi_0)<1$ as the map $(\pi_1,\pi_2)\mapsto\min\{\pi_1,\pi_2\}$ is not differentiable precisely when $\pi_1=\pi_2$. In any case, fortunately, $\phi_r$ is second order  directionally differentiable, which is sufficient to invoke the second order Delta method as we elaborate next.

\iffalse

\begin{align}\label{Eqn: phi: derivative: representation0}
\phi_{\Pi}^{\prime\prime}(M)=\min_{U\in\Psi(\Pi)}\min_{V\in\mathbf{M}^{k\times (k-r)}}\Vert MU+\Pi V\Vert^{2}~.
\end{align}

\fi

\subsection{The Asymptotic Distributions}\label{Sec: 3-2}

With the differentiability established in Proposition \ref{Pro: phi, differentiability}, we now derive the asymptotic distributions for the plug-in statistic $\phi_r(\hat\Pi_n)$ where $\hat\Pi_n$ is a generic estimator of $\Pi_0$. This is achieved by appealing to a generalized Delta method for second order  directionally differentiable maps \citep{Shapiro2000inference,Chen_Fang2015FOD}. Towards this end, we impose the following assumption.

\begin{ass}\label{Ass: Weaklimit: Pihat}
There is an estimator $\hat{\Pi}_{n}: \{X_i\}_{i=1}^n\to\mathbf M^{m\times k}$ of $\Pi_0\in\mathbf{M}^{m\times k}$ (with $m\ge k$) satisfying ${\tau_{n}}\{\hat{\Pi}_{n}-\Pi_{0}\}\overset{L}{\rightarrow}\mathcal{M}$ for some $\tau_{n}\uparrow \infty$ and random matrix $\mathcal{M}\in\mathbf{M}^{m\times k}$.
\end{ass}

Assumption \ref{Ass: Weaklimit: Pihat} simply requires an estimator $\hat\Pi_n$ of $\Pi_0$ that admits an asymptotic distribution. Note that the data need not be i.i.d., $\tau_n$ may be non-$\sqrt n$ and $\mathcal M$ can be non-Gaussian, which is important in, for example, nonstationary time series settings. Moreover, as in \citet{Robin_Smith2000rank} but in contrast to \citet{Cragg_Donald1997infer}, the covariance matrix of $\mathrm{vec}(\mathcal M)$ is not required to be nonsingular. Assumption \ref{Ass: Weaklimit: Pihat} can be relaxed to accommodate settings where convergence rates across entries of $\hat\Pi_n$ are not homogeneous, as in cointegratoin settings -- see Supplemental Appendix \ref{Sec: coint appendix}. For ease of exposition, however, we stick to Assumption \ref{Ass: Weaklimit: Pihat} in the main text.

% Finally, one may standardize the data before forming the matrix estimator -- see Remark \ref{Rem: standardization}.

Given Proposition \ref{Pro: phi, differentiability} and Assumption \ref{Ass: Weaklimit: Pihat}, the following theorem delivers the asymptotic distributions of $\phi_r(\hat{\Pi}_{n})$ by the Delta method.

\begin{thm}\label{thm: Weaklimit: Test}
If Assumption \ref{Ass: Weaklimit: Pihat} holds, then we have, for any $\Pi_0\in\mathbf M^{m\times k}$,
\begin{align}
\tau_{n}\{\phi_r(\hat{\Pi}_{n})-\phi_r(\Pi_{0})\}\overset{L}{\rightarrow}\min_{U\in\Psi(\Pi_{0})} 2\mathrm{tr}(U^\transpose\Pi_{0}^\transpose \mathcal{M} U)~.
\end{align}
If in addition $r_0\equiv \mathrm{rank}(\Pi_0)\le r$, then
\begin{align}\label{Eqn: Nulllimt}
\tau_{n}^{2}\phi_r(\hat{\Pi}_{n})\overset{L}{\rightarrow} \sum_{j=r-r_{0}+1}^{k-r_{0}}\sigma^{2}_j(P_{0,2}^\transpose \mathcal{M} Q_{0,2})~,
\end{align}
where the columns of $P_{0,2}\in\mathbb S^{m\times(m-r_0)}$ and $Q_{0,2}\in\mathbb S^{k\times (k-r_0)}$ are respectively the left and the right singular vectors of $\Pi_{0}$ associated with its zero singular values.
\end{thm}

Theorem \ref{thm: Weaklimit: Test} implies that, under $\mathrm{H}_0$ (and so $\tau_{n}\phi_r(\hat{\Pi}_{n})$ is degenerate), the statistic $\tau_{n}^{2}\phi_r(\hat{\Pi}_{n})$ converges in law to a nondegenerate second order limit. Towards constructing critical values, we would then like to estimate the law of the limit. Unfortunately, as shown by \citet{Chen_Fang2015FOD}, bootstrapping a nondegenerate second order limit is nontrivial; in particular, standard bootstrap schemes such as the nonparametric bootstrap of \citet{Efron1979} are necessarily inconsistent even if they are consistent for $\mathcal M$. This predicament is further intensified by the nondifferentiability nature of the map $\phi_r$ \citep{Dumbgen1993,Fang_Santos2014HDD}, which renders the limits in \eqref{Eqn: Nulllimt} highly nonstandard in general. We shall thus present a consistent bootstrap shortly.

We emphasize that the limit of $\tau_n^2\phi_r(\hat\Pi_n)$ in Theorem \ref{thm: Weaklimit: Test} is obtained pointwise in each $\Pi_0$ under the {\it entire} null, regardless of whether the truth rank of $\Pi_0$ is strictly less than $r$ or not. To the best of our knowledge, this is the first distributional result for a rank test statistic that accommodates the possibility $\mathrm{rank}(\Pi_0)<r$, at the generality of our setup. In turn, such a result permits us to develop a test that has asymptotic null rejection rates exactly equal to the significance level, and hence is more powerful.

In relating our work to the literature, we note that, if $\tau_n=\sqrt n$, then the plug-in statistic $\tau_n^2\phi_r(\hat\Pi_n)$ is precisely a Robin-Smith statistic (see \eqref{Eqn: RS statistic}), while the KP statistic is simply a Wald-type standardization of it. Though standardization can help obtain pivotal asymptotic distributions under $r_0=r$, this is generally not hopeful whenever $r_0<r$. Since we shall reply on bootstrap for inference, non-pivotalness creates no problems for us. Perhaps more importantly, one may be better off without standardization because it entails invertibility of the weighting matrix in the limit, which may be hard to justify. One might nonetheless interpret the inverse in the KP statistic as a generalized inverse, but consistency of the inverse does not automatically follow from consistency of the covariance matrix estimator without further conditions \citep{Andrews1987Wald}.

Finally, the limit of $\tau_n^2\phi_r(\hat\Pi_n)$ obtained under $\mathrm{H}_0$ is in fact a weighted sum of independent $\chi^2(1)$ variables if $r_0=r$ and $\mathcal M$ is centered Gaussian, showing consistency of our work with \citet{Robin_Smith2000rank}. To see this, note that
\begin{align}
\sum_{j=r-r_{0}+1}^{k-r_{0}}\sigma^{2}_j(P_{0,2}^\transpose \mathcal{M} Q_{0,2})=\sum_{j=1}^{k-r}\sigma^{2}_j(P_{0,2}^\transpose \mathcal{M} Q_{0,2})~,
\end{align}
which is simply the sum of all squared singular values of the $(m-r)\times(k-r)$ matrix $P_{0,2}^\transpose \mathcal{M} Q_{0,2}$, or equivalently the squared Frobenius norm of $P_{0,2}^\transpose \mathcal{M} Q_{0,2}$ \citep[p.7]{Bhatia1997Matrix}. Consequently, the limit in \eqref{Eqn: Nulllimt} can be rewritten as
\begin{align}\label{Eqn: RobinSmith, consistency}
\text{vec}(P_{0,2}^{\transpose}\mathcal{M}Q_{0,2})^{\transpose}\text{vec}(P_{0,2}^{\transpose}\mathcal{M}Q_{0,2})
=\mathrm{vec}(\mathcal M)^\transpose (Q_{0,2}\otimes P_{0,2})  (Q_{0,2}\otimes P_{0,2})^\transpose \mathrm{vec}(\mathcal M)~,
\end{align}
as claimed, where we exploited a property of the vec operator \citep[Proposition 10.4]{Hamilton1994}. Our general limit in \eqref{Eqn: Nulllimt} characterizes the channels through which the true rank plays its role, and thus highlights the importance of studying the problem \eqref{Eqn: hypothesis, intro}.

\iffalse
\begin{rem}\label{Rem: standardization}
A standardization remark here.
\end{rem}
\fi

\subsection{The Bootstrap Inference}\label{Sec: 3-3}

Since asymptotic distributions of our statistic $\tau_n^2\phi_r(\hat\Pi_n)$ are not pivotal and highly nonstandard in general, in this section we thus aim to develop a consistent bootstrap. This turns out to be quite challenging due to two complications involved.

First, since under $\mathrm H_0$ the first order derivative of $\phi_r$ is degenerate while second order derivative is not (by Proposition \ref{Pro: phi, differentiability}), $\phi_r(\hat\Pi_n^*)$ is necessarily inconsistent even if $\hat\Pi_n^*$ is a consistent bootstrap (in a sense defined below) in estimating the law of $\mathcal M$ \citep{Chen_Fang2015FOD}, and this remains true in the conventional setup where $\mathrm{rank}(\Pi_0)=r$. Second, the possibility $\mathrm{rank}(\Pi_0)<r$ makes the map $\phi_r$ nondifferentiable -- see Lemma \ref{Lem: phi, socHD}, and hence further complicates the inference \citep{Dumbgen1993,Fang_Santos2014HDD}. One may resort to the $m$ out of $n$ resampling \citep{Shao1994bootstrap} or subsampling \citep{Politis_Romano1994subsample}. However, both methods can be viewed as special cases of our general bootstrap procedure, and that, more importantly, such a perspective enables us to improve upon these existing resampling schemes and to analyze the local properties in a unified and transparent way -- see Remark \ref{Rem: subsampling} and Section \ref{Sec: local power}.

The insight our bootstrap builds on is that the limit $\phi_{r,\Pi_0}''(\mathcal M)$ in Theorem \ref{thm: Weaklimit: Test} is a composition of two unknown components, namely, the limit $\mathcal M$ and the derivative $\phi_{r,\Pi_0}''$. Heuristically, one may therefore obtain a consistent estimator for the law of $\phi_{r,\Pi_0}''(\mathcal M)$ by composing a consistent bootstrap $\hat{\mathcal M}_n^*$ for $\mathcal M$ with an estimator $\hat\phi_{r,n}''$ of $\phi_{r,\Pi_0}''$ that is suitably ``consistent.'' This is precisely the bootstrap initially proposed in \citet{Fang_Santos2014HDD} and further developed in \citet{Chen_Fang2015FOD} and \citet{Hong_Li2017numerical}. In what follows, we thus commence by estimating the two components separately.

Starting with $\mathcal M$, we note that the law of $\mathcal M$ may be estimated by standard bootstrap or variants of it that suit particular settings. To formalize the notion of bootstrap consistency, we employ the bounded Lipschitz metric \citep{Vaart1996} and consider estimating the law of a general random element $\mathbb G$ in a normed space $\mathbb D$ with norm $\|\cdot\|_{\mathbb D}$ -- the space $\mathbb D$ is either $\mathbf M^{m\times k}$ or $\mathbf R$ in this paper. Let $\mathbb G_n^*:\{X_i,W_{ni}\}_{i=1}^n\to\mathbb D$ be a generic bootstrap estimator where $\{W_{ni}\}_{i=1}^{n}$ are bootstrap weights independent of the data $\{X_{i}\}_{i=1}^{n}$. Then we say that the conditional law of $\mathbb G_n^*$ given the data is consistent for the law of $\mathbb G$, or simply $\mathbb G_n^*$ is a consistent bootstrap for $\mathbb G$, if
\begin{align}
\sup_{f\in\mathrm{BL}_1(\mathbb D)}\big |E_W[f(\mathbb G_n^*)]-E[f(\mathbb G)]\big|=o_p(1)~,
\end{align}
where $E_W$ denotes expectation with respect to $\{W_{ni}\}_{i=1}^{n}$ holding $\{X_{i}\}_{i=1}^{n}$ fixed, and
\begin{align}
\mathrm{BL}_1(\mathbb D)\equiv\{f:\mathbb D\to\mathbf R: \sup_{x\in\mathbb D}|f(x)|<\infty,|f(x)-f(y)|\le\|x-y\|_{\mathbb D}\,\forall\,x,y\in\mathbb D\}~.
\end{align}
Given the metric, we now proceed by imposing

\begin{ass}\label{Ass: Boostrap: Pihat}
(i) $\hat{\mathcal M}_{n}^{\ast}: \{X_{i},W_{ni}\}_{i=1}^{n}\to\mathbf{M}^{m\times k}$ is a bootstrap estimator with $\{W_{ni}\}_{i=1}^{n}$ independent of $\{X_{i}\}_{i=1}^{n}$; (ii) $\hat{\mathcal M}_{n}^{\ast}$ is a consistent bootstrap for $\mathcal M$.
\end{ass}
% Need a joint measurability

Assumption \ref{Ass: Boostrap: Pihat}(i) introduces the bootstrap estimator $\hat{\mathcal M}_{n}^{\ast}$, which may be constructed from nonparametric bootstrap, multiplier bootstrap, general exchangeable bootstrap, block bootstrap, score bootstrap, the $m$ out of $n$ resampling or subsampling. The presence of $\{W_{ni}\}_{i=1}^{n}$ simply characterizes the bootstrap randomness given the data -- see \citet{Praestgaard_Wellner1993}. For $\hat\Pi_n^*$ a bootstrap analog of $\hat\Pi_n$, it is common to have $\hat{\mathcal M}_{n}^{\ast}=\tau_n\{\hat\Pi_n^*-\hat\Pi_n\}$; if $\hat\Pi_{m_n}^*$ is an analog of $\hat\Pi_n$ constructed based on a subsample of size $m_n$, then one may instead have $\hat{\mathcal M}_{n}^{\ast}=\tau_{m_n}\{\hat\Pi_{m_n}^*-\hat\Pi_n\}$. Assumption \ref{Ass: Boostrap: Pihat}(ii) requires that $\hat{\mathcal M}_{n}^{\ast}$ be consistent in estimating the law of the target limit $\mathcal M$.

Turning to the estimation of $\phi_{r,\Pi_0}''$, we recall by \citet{Chen_Fang2015FOD} that, given Assumption \ref{Ass: Boostrap: Pihat}, the composition $\hat\phi_{r,n}''(\hat{\mathcal M}_n^*)$ is a consistent bootstrap for $\phi_{r,\Pi_0}''(\mathcal M)$ provided $\hat\phi_{r,n}^{\prime\prime}$ is consistent for $\phi_{r,\Pi_0}''$ in the sense that, whenever $M_n\to M$ as $n\to\infty$,
\begin{align}\label{Eqn: derivative consistency}
\hat\phi_{r,n}^{\prime\prime}(M_n)\convp \phi_{r,\Pi_0}''(M)~.
\end{align}
In this regard, there are two general constructions, namely, the numerical estimator and the analytic estimator, as we elaborate next.

The numerical estimator is simply a finite sample analog of \eqref{Eqn: HDD 2nd} in the definition of second order derivative, i.e., we estimate $\phi^{\prime\prime}_{r,\Pi_{0}}$ by: for any $M\in\mathbf M^{m\times k}$,
\begin{align}\label{Eqn: NumericalDevEst}
\hat\phi_{r,n}^{\prime\prime}(M)=\frac{\phi_r(\hat\Pi_{n}+\kappa_{n}M)-\phi_r(\hat\Pi_{n})}{\kappa_{n}^{2}}~,
\end{align}
for a suitable $\kappa_n\downarrow 0$, where we have exploited $\phi_{r,\Pi_0}'=0$ under the null. By \citet{Chen_Fang2015FOD}, \eqref{Eqn: NumericalDevEst} meets the requirement \eqref{Eqn: derivative consistency} if $\kappa_n\downarrow 0$ and $\tau_n\kappa_n\to\infty$. Numerical differentiation in the general context of the Delta method dates back to \citet{Dumbgen1993}, and is recently extended by \citet{Hong_Li2017numerical}. The numerical estimator enjoys marked simplicity and wide applicability, because it merely requires a sequence $\{\kappa_n\}$ of step sizes satisfying certain rate conditions. There is, however, no general theory to date guiding the choice of $\kappa_n$, a problem that appears challenging \citep{Hong_Li2017numerical}. In this regard, it may be sensible to employ the analytic estimator instead.

The analytic estimator heavily exploits the analytic structure of the derivative $\phi_{r,\Pi_0}''$, which, by Proposition \ref{Pro: phi, differentiability}(ii), involves three unknown objects, namely, the true rank $r_0$, $P_{0,2}$ and $Q_{0,2}$ -- note that the columns of $P_{0,2}$ and $Q_{0,2}$ are the left and the right singular vectors associated with the zero singular values of $\Pi_0$. We may thus estimate $\phi_{r,\Pi_0}''$ by replacing these unknowns with their estimated counterparts. The key is consistent estimation of $r_0$: given a consistent estimator $\hat r_n$ of $r_0$, we may then obtain estimators $\hat{P}_{2,n}$ and $\hat{Q}_{2,n}$ of $P_{0,2}$ and $Q_{0,2}$ respectively in a straightforward manner as described in Section \ref{Sec: 2}. One possible construction of $\hat r_n$ is given by \eqref{Eqn: bootstrap overview0}. Alternatively, $\hat r_n$ may also be constructed by sequential testing, and the tuning parameter then becomes an adjusted significance level -- see Supplemental Appendix \ref{Sec: 4-3}. In any case, by Lemma \ref{Lem: Consistency: DerivEst}, we may then obtain a consistent estimator for $\phi_{r,\Pi_0}''$: for any $M\in\mathbf M^{m\times k}$,
\begin{align}\label{Eqn: StructDevEst}
\hat\phi_{r,n}^{\prime\prime}(M)= \sum_{j=r-\hat r_n+1}^{k-\hat{r}_{n}}\sigma^{2}_j(\hat{P}_{2,n}^\transpose M \hat{Q}_{2,n})~.
\end{align}
Similar to the numerical estimator, the analytic estimator \eqref{Eqn: StructDevEst} also depends on a tuning parameter, but now through consistent estimation of the rank. An advantage of the latter over the former is that the choice of the tuning parameter is easier to motivate. For example, if $\hat r_n$ is given by \eqref{Eqn: bootstrap overview0}, then $\kappa_n$ has a meaningful interpretation, namely, it measures the parsimoniousness in selecting the rank.

\iffalse
If $\hat r_n$ is provided by a sequential procedure -- see Theorem \ref{Thm: rank estimator consistency}, then the tuning parameter becomes the significance level $\alpha_n$ that tends to zero suitably. Alternatively, a consistent estimator $\hat r_n$ may also be obtained by sequential testing (see Section \ref{Sec: 4-3}), in which case the tuning parameter becomes the adjusted significance level. because $\alpha_n$ is free of units and its choice affects our tests to a lesser extent.
\fi

Given a significance level $\alpha$, we now formally define our critical value $\hat c_{n,1-\alpha}$ as
\begin{align}\label{Eqn: Quantile}
\hat{c}_{n,1-\alpha}\equiv\inf\{c\in\mathbf{R}: P_{W}(\hat\phi_{r,n}^{\prime\prime}(\hat{\mathcal M}_n^*)\leq c)\geq 1-\alpha\}~,
\end{align}
where $P_{W}$ denotes the probability evaluated with respect to $\{W_{ni}\}_{i=1}^n$ holding the data fixed. In practice, we often approximate $\hat{c}_{n,1-\alpha}$ using the following algorithm:

{\sc Step 1:} Compute the derivative estimator $\hat\phi_{r,n}''$ by either \eqref{Eqn: NumericalDevEst}, or \eqref{Eqn: bootstrap overview0} and \eqref{Eqn: StructDevEst}.

{\sc Step 2:} Generate $B$ realizations $\{\hat{\mathcal M}_{n,b}^*\}_{b=1}^B$ of $\hat{\mathcal M}_{n}^*$ based on $B$ bootstrap samples.

{\sc Step 3:} Approximate $\hat c_{n,1-\alpha}$ by the $\lfloor B(1-\alpha)\rfloor$ largest number in $\{\hat\phi_{r,n}''(\hat{\mathcal M}_{n,b}^*)\}_{b=1}^B$.

\noindent Our simulations suggest that the analytic method tends to enjoy better size control.

The following theorem establishes that our test has pointwise {\it exact} asymptotic size control under the entire null $\mathrm{H}_0$, and is consistent against any fixed alternatives.

\begin{thm}\label{Thm: SizePowerTest}
Let Assumptions \ref{Ass: Weaklimit: Pihat} and \ref{Ass: Boostrap: Pihat} hold, and $\hat{c}_{n,1-\alpha}$ be as in \eqref{Eqn: Quantile} where $\hat\phi_{r,n}^{\prime\prime}$ is given by either \eqref{Eqn: NumericalDevEst} with $\{\kappa_n\}$ satisfying $\kappa_n\downarrow 0$ and $\tau_n\kappa_n\to\infty$, or \eqref{Eqn: StructDevEst} with $\hat r_n\convp r_0$ under $\mathrm H_0$. If the cdf of the limiting distribution in \eqref{Eqn: Nulllimt} is continuous and strictly increasing at its $(1-\alpha)$-quantile for $\alpha\in(0,1)$, then under $\mathrm{H}_{0}$,
\[\lim_{n\to\infty}P(\tau_{n}^{2}\phi_r(\hat{\Pi}_{n})>\hat{c}_{n,1-\alpha})=\alpha~.\]
Furthermore, under $\mathrm{H}_{1}$,
\[\lim_{n\to\infty}P(\tau_{n}^{2}\phi_r(\hat{\Pi}_{n})>\hat{c}_{n,1-\alpha})=1~.\]
\end{thm}

Theorem \ref{Thm: SizePowerTest} shows that our test is not conservative in the pointwise sense while accommodating the possibility $\mathrm{rank}(\Pi_0)<r$. This roots in the simple fact that our critical values are constructed for the pointwise distributions obtained under $\mathrm{H}_0$. By the same token, the power is nontrivial and tends to one against any fixed alternative. We shall further examine the local power properties in Section \ref{Sec: local power} and provide numerical evidences in Section \ref{Sec: 4}. Overall, the theoretical and numerical results manifest superiority of our test in terms of size control and power performance.

In addition to the attractive features mentioned after Assumption \ref{Ass: Weaklimit: Pihat}, we stress that the bootstrap for $\mathcal M$ may be virtually any consistent resampling scheme, and that no side rank conditions whatsoever are directly imposed beyond those entailed by the restriction that the limiting cdf is continuous and strictly increasing at $c_{1-\alpha}$. Such a quantile restriction is standard as consistent estimation of the limiting laws does not guarantee consistency of critical values -- see, for example, Lemma 11.2.1 in \citet{TSH2005}. To appreciate how weak this condition is, consider the conventional setup \eqref{Eqn: hypothesis, literature, intro} when $\mathcal M$ is Gaussian. Then each limit under $\mathrm{H}_0'$ is a weighted sum of independent $\chi^2(1)$ random variables -- see our discussions towards the end of Section \ref{Sec: 3-2}. Consequently, the quantile condition is automatically satisfied provided the covariance matrix of $\text{vec}(P_{0,2}^{\transpose}\mathcal{M}Q_{0,2})$ is nonzero (i.e., nonzero rank), which is precisely Assumption 2.4 in \citet{Robin_Smith2000rank}. In contrast, \citet{Kleibergen_Paap2006rank} require nonsingularity of the same matrix (i.e., full rank).

Despite the irregular natures of the problem, computation of our testing statistic and the critical values are quite simple as both involve only calculations of singular value decompositions, for which there are commands in common computation softwares. In particular, $\hat c_{n,1-\alpha}$ in practice is set to be the $(1-\alpha)$-quantile of
\begin{align}
\hat\phi_{r,n}^{\prime\prime}(\hat{\mathcal M}_{n,1}^*), \hat\phi_{r,n}^{\prime\prime}(\hat{\mathcal M}_{n,2}^*), \ldots, \hat\phi_{r,n}^{\prime\prime}(\hat{\mathcal M}_{n,B}^*)~.
\end{align}
Therefore, in each repetition, the numerical and the analytic approaches simply entail singular value decompositions of $\hat{\Pi}_{n}+\kappa_{n}\hat{\mathcal M}_{n,b}^*$ and $\hat{P}_{2,n}^\transpose \hat{\mathcal M}_{n,b}^*\hat{Q}_{2,n}$ respectively.

A common feature of our previous two tests is their dependence on a tuning parameter -- see \eqref{Eqn: NumericalDevEst} and \eqref{Eqn: StructDevEst}. To mitigate concerns on sensitivity to the choice of tuning parameters, we next develop a two-step test by exploiting the structure in \eqref{Eqn: StructDevEst}. The intuition is as follows. The estimator \eqref{Eqn: bootstrap overview0}, though consistent, may differ from the truth in finite samples. We would thus like to control $P(\hat r_n= r_0)$, for which \eqref{Eqn: bootstrap overview0} may not be appropriate as it appears challenging to bound $P(\hat r_n= r_0)$. Instead, we may obtain a suitable estimator $\hat r_n$ by a sequential testing procedure -- see Theorem \ref{Thm: Rankdetermination}. Specifically, we sequentially test $\mathrm{rank}(\Pi_0)=0,1,\ldots,k-1$ at level $\beta<\alpha$, and set $\hat r_n=j^*$ if accepting $\mathrm{rank}(\Pi_0)=j^*$ is the first acceptance, and $\hat r_n=k$ if no acceptance occurs. In this regard, we recommend the KP test as it is tuning parameter free and does not require additional simulations.\footnote{If estimation of $r_0$ is one's {\it ultimate} goal (rather than an intermediate step for test), then it may be desirable to instead employ our tests in the sequential procedure, as existing tests may lead to estimators that are not as accurate when $\Pi_0$ is ``local to degeneracy'' -- see Section \ref{Sec: 4} for simulation evidences.} Table \ref{Tab: false selection} compares the empirical probabilities of $\{\hat r_n=r_0\}$ for $\hat r_n$ obtained by \eqref{Eqn: bootstrap overview0} and the sequential KP test respectively, based on the same simulation data from Section \ref{Sec: 2} when $d>1$. The empirical probabilities for \eqref{Eqn: bootstrap overview0} are close to one when $\kappa_n=n^{-1/4}$ (as chosen in Section 2) or $\in\{n^{-1/4}, 1.5n^{-1/4}, n^{-1/5}, 1.5n^{-1/5}\}$ (omitted due to space limitation), but may be far away from one or even close to zero for other choices. On the other hand, the sequential approach leads to rank estimators with empirical probabilities approximately $1-\beta$ across our choices of $\beta$.

{
\setlength{\tabcolsep}{3pt}
\begin{table}[!htbp]
\centering  \small
\caption{Estimation of $\mathrm{rank}(\Pi_0)$ Defined by the Model \eqref{Eqn: Motivation: DGPs}-\eqref{Eqn: Motivation: DGPs1}}
\label{Tab: false selection}
\begin{tabular}{ccccccccccccccc}
\hline\hline
\multirow{2}{*}{\makecell{$d$}} & &\multicolumn{5}{c}{Choices of $\kappa_n$ in \eqref{Eqn: bootstrap overview0}}&&\multicolumn{6}{c}{Choices of $\beta$ for the Sequential Method} \\
\cline{3-7} \cline{9-14}
 & & $n^{-1/4}$ & $n^{-1/3}$&$1.5n^{-1/3}$&$n^{-2/5}$&$1.5n^{-2/5}$& & $\alpha/5$&$\alpha/10$&$\alpha/15$&$\alpha/20$&$\alpha/25$&$\alpha/30$ \\
\cmidrule{1-14}
$2$  &&1.0000&0.9975&1.0000&0.6679&0.9618&& 0.9902&0.9947&0.9965&0.9974&0.9975&0.9979\\
$3$  &&1.0000&0.8516&0.9988&0.2246&0.7862&&0.9908&0.9951&0.9958&0.9963&0.9976&0.9980\\
$4$  &&0.9995&0.5550&0.9922&0.0249&0.4474&&0.9877&0.9949&0.9963&0.9972&0.9976&0.9981\\
$5$  &&0.9977&0.2176&0.9581&0.0003&0.1420&&0.9861&0.9933&0.9958&0.9968&0.9976&0.9979\\
$6$  &&0.9899&0.0422&0.8557&0.0000&0.0203&&0.9840&0.9916&0.9946&0.9960&0.9967&0.9967\\
\hline\hline
\end{tabular}
\end{table}%
}

Given an estimator $\hat r_n$ with $P(\hat r_n=r_0)\ge  1-\beta$ (approximately) for some $\beta<\alpha$, the two-step test now goes as follows. In the first step, we reject $\mathrm{H}_0$ if $\hat r_n>r$; otherwise we plug $\hat r_n$ into \eqref{Eqn: StructDevEst} in the second step and reject $\mathrm{H}_0$ if $\tau_n^2\phi_r(\hat\Pi_n)>\hat c_{n,1-\alpha+\beta}$. Note that the significance level in the second step is adjusted to be $\alpha-\beta$ in order to take into account the event of false selection (which has probability $\beta$). Formally, letting
\begin{align}\label{Eqn: two-step}
\psi_n=1\{\hat r_n>r\text{ or }\tau_n^2\phi_r(\hat\Pi_n)>\hat c_{n,1-\alpha+\beta}\}~,
\end{align}
we then reject the null $\mathrm H_0$ if $\psi_n=1$ and fail to reject otherwise. Our next theorem shows that the two-step procedure controls size and is consistent.

\begin{thm}\label{Thm: two-step}
Suppose that Assumptions \ref{Ass: Weaklimit: Pihat} and \ref{Ass: Boostrap: Pihat} hold, and that the cdf of the limit distribution in \eqref{Eqn: Nulllimt} is continuous and strictly increasing at its $(1-\alpha+\beta)$-quantile for $\alpha\in(0,1)$ and $\beta\in(0,\alpha)$. Let $\psi_n$ be the test given by \eqref{Eqn: two-step}. Then, under $\mathrm{H}_{0}$,
\[
\limsup_{n\to\infty}E[\psi_n]\le \alpha
\]
\iffalse
\[
(\alpha-\beta)(1-\beta)\le \limsup_{n\to\infty}E[\psi_n]\le \alpha
\]
\fi
provided $\liminf_{n\to\infty} P(\hat r_n=r_0)\ge 1-\beta$, and, under $\mathrm{H}_{1}$,
\[
\lim_{n\to\infty}E[\psi_n]=1~.
\]
\end{thm}

The idea of the two-step test may be found in \citet{Loh1985New}, \citet{BergerBoos1994Pvalues}, and \citet{Silvapulle1996Nuisance}, and has recently been employed in the context of moment inequality models \citep{Andrews_Barwick2012Inference,RomanoShaikhWolf2014TwoStep}. A common feature that our test shares here is that the size control is not exact, i.e., we cannot show the size is equal to $\alpha$. This raises the concern that the test may be potentially conservative. Nonetheless, it is possible to derive a lower bound of the asymptotic size which is close to $\alpha$ by choosing a small $\beta$ -- see \citet{RomanoShaikhWolf2014TwoStep} for a similar feature. Summarizing, there are two (types of) test procedures: one rejects $\mathrm{H}_0$ if $\tau_n^2\phi_r(\hat\Pi_n)>\hat c_{n,1-\alpha}$ with $\hat c_{n,1-\alpha}$ computed according to \eqref{Eqn: Quantile}, and the other one applies when one has control over $P(\hat r_n=r_0)$: if $\liminf_{n\to\infty} P(\hat r_n=r_0)\ge 1-\beta$, we reject if $\hat r_n>r$ or $\tau_n^2\phi_r(\hat\Pi_n)>\hat c_{n,1-\alpha+\beta}$. Our simulation results in Section \ref{Sec: 4} show that the two-step procedure produces results that are quite insensitive to our choice of $\beta$.

\begin{rem}\label{Rem: subsampling} %  Double check this remark.
The $m$ out of $n$ bootstrap and the subsampling are special cases of our bootstrap procedure. For example, the former amounts to $\hat{\mathcal M}_{n}^{\ast}=\tau_{m_n}\{\hat\Pi_{m_n}^*-\hat\Pi_n\}$ with $\hat\Pi_{m_n}^*$ constructed based on subsamples of size $m_n$ (obtained through resampling with replacement), and the derivative estimator $\hat\phi_{r,n}''$ given by \eqref{Eqn: NumericalDevEst} with $\kappa_n=m_n^{-1}$. Subsampling is precisely the same procedure except that the subsamples are obtained without replacement. In other words, these procedures estimate the derivative through \eqref{Eqn: NumericalDevEst} implicitly and automatically when the subsample size is properly chosen, combining the two-steps into one single step. By disentangling estimation of the two ingredients, however, we may better estimate both the derivative $\phi_{r,\Pi_0}''$ (through exploiting the structure of the derivative and a choice of the tuning parameter) and the law of the limit $\mathcal M$ (using full samples), which may in turn lead to efficiency improvement. Moreover, such a perspective enables us to establish conditions under which tests based on these resampling schemes have local size control and nontrivial power, properties not guaranteed in general and nontrivial to analyze otherwise \citep{AndrewsandGuggen2010ET}. \qed
\end{rem}

\subsubsection{Local Power Properties}\label{Sec: local power}

Having established size control and consistency, we next aim to obtain a more precise characterization of the quality of our tests by studying the local power properties \citep{Neyman1937LocalPower}. Following \citet{Cragg_Donald1997infer}, we thus proceed by imposing

\begin{assprime}{Ass: Weaklimit: Pihat}\label{Ass: Local}
(i) $\mathrm{rank}(\Pi_{0,n})> r$ for all $n$; (ii) $\tau_n\{\Pi_{0,n}-\Pi_0\}\to \Delta$ for some $\Pi_0$ with $\mathrm{rank}(\Pi_0)\le r$ and nonrandom $\Delta$; (iii) $\tau_n\{\hat\Pi_n-\Pi_{0,n}\}\overset{L_n}{\to}\mathcal M$ for some $\tau_n\uparrow \infty$, where $\overset{L_n}{\to}$ denotes convergence in law along distributions of the data associated with $\{\Pi_{0,n}\}$.
\end{assprime}

Assumption \ref{Ass: Local}(i)(ii) formally defines $\{\Pi_{0,n}\}$ as a sequence of local alternatives that approaches some $\Pi_0$ in the null at the convergence rate $\tau_n$, while Assumption \ref{Ass: Local}(iii) formalizes the notion that the {\it asymptotic} distributions of $\hat\Pi_n$ should remain unchanged in response to {\it small} (finite sample) perturbations of the data generating processes, a property that may be verified through, for example, the framework of limits of statistical experiments \citep{Vaart1998,HallinAkkerWerker2016Coint}.

Our next result characterizes the asymptotic behaviors of the testing statistic $\tau_n^2\phi_r(\hat\Pi_n)$ under local alternatives that satisfy Assumption \ref{Ass: Local}.

\begin{pro}\label{Pro: Local}
If Assumption \ref{Ass: Local} holds, then it follows that
\begin{align}
\tau_n^2\phi_r(\hat\Pi_n)\overset{L_n}{\to} \sum_{j=r-r_{0}+1}^{k-r_{0}}\sigma^{2}_j(P_{0,2}^\transpose (\mathcal M+\Delta) Q_{0,2})~.
\end{align}
\end{pro}

Proposition \ref{Pro: Local} includes Theorem \ref{thm: Weaklimit: Test} as a special case with $\Pi_{0,n}=\Pi_0$ for all $n$ so that $\Delta=0$. The main utility of this result is to analyze the asymptotic local power function. In what follows, we focus on the one-step tests for conciseness and transparency, though analogous results hold for the two-step test $\psi_n$. Thus, if the local alternatives $\{\Pi_{0,n}\}$ in Assumption \ref{Ass: Local} approach $\Pi_0$ in the sense of contiguity \citep{Roussas1972contiguity,Rothenberg1984Handbook},\footnote{This means that if (any) $T_n$ is negligible (i.e., of order $o_p(1)$) under $\Pi_0$ then it remains so under $\Pi_{0,n}$. Thus, contiguity simply formalizes the notion that the effect of ``small'' perturbations is negligible.} then we may obtain a lower bound as follows:
\begin{align}\label{Eqn: local power}
\liminf_{n\to\infty}P_{n}(\tau_n^2\phi_r(\hat\Pi_n)>\hat c_{n,1-\alpha}) \ge P(\sum_{j=r-r_{0}+1}^{k-r_{0}}\sigma^{2}_j(P_{0,2}^\transpose (\mathcal M+\Delta) Q_{0,2})>c_{1-\alpha})~,
\end{align}
where $P_n$ denotes probability evaluated under $\Pi_{0,n}$. While it appears challenging to prove that the asymptotic local power is nontrivial under arbitrary local alternatives, there is, nonetheless, an interesting case under which the asymptotic local power can be proven to be nontrivial. This is the conventional setup where $\mathrm{rank}(\Pi_0)$ is exactly equal to the hypothesized value $r$ and $\mathcal M$ is centered Gaussian. Since the derivative $\phi_{r,\Pi_0}''$ then coincides with the squared Frobenius norm -- see Proposition \ref{Pro: phi, differentiability}(ii), we have along contiguous local alternatives that
\begin{align}
\liminf_{n\to\infty}P_n(\tau_n^2\phi_r(\hat\Pi_n)>\hat c_{n,1-\alpha})\ge P(\|P_{0,2}^\transpose(\mathcal M+\Delta)Q_{0,2}\|^2 >c_{1-\alpha})~.
\end{align}
An application of Anderson's lemma -- see, for example, Lemma 3.11.4 in \citet{Vaart1996} -- then yields
\begin{align}\label{Eqn: local power nontrivial}
P(\|P_{0,2}^\transpose(\mathcal M+\Delta)Q_{0,2}\|^2 >c_{1-\alpha})\ge P(\|P_{0,2}^\transpose \mathcal M Q_{0,2}\|^2 >c_{1-\alpha})=\alpha~.
\end{align}
If the localization parameter $\Delta$ is nontrivial (i.e., $\Delta\neq 0$) and belongs to the support of $\mathcal M$ -- which is the case, for example, if the covariance matrix of $\mathrm{vec}(\mathcal M)$ is nonsingular, then by Lemma B.4 in \citet{Chen_Santos2015} (a strengthening of Anderson's lemma), the asymptotic local lower is in fact nontrivial, i.e.,
\begin{align}
P(\|P_{0,2}^\transpose(\mathcal M+\Delta)Q_{0,2}\|^2 >c_{1-\alpha})>\alpha~.
\end{align}

% Our simulations in Section \ref{Sec: 4} show that, in a number of data generating processes, the local power is indeed above $\alpha$ against our (crude) choices of tuning parameters.

In view of the irregularities of the problem \eqref{Eqn: hypothesis, intro}, one may also be interested in the size control of our test. Under Assumption \ref{Ass: Local} but with (i) replaced by $\mathrm{rank}(\Pi_{0,n})\le r$ for all $n\in\mathbf N$ so that the contiguous perturbations occur under the null, we may obtain
\begin{align}
\limsup_{n\to\infty}P_n(\tau_n^2\phi_r(\hat\Pi_n)>\hat c_{n,1-\alpha}) \le P(\sum_{j=r-r_{0}+1}^{k-r_{0}}\sigma^{2}_j(P_{0,2}^\transpose (\mathcal M+\Delta) Q_{0,2})\ge c_{1-\alpha})~.
\end{align}
Now suppose $\mathrm{rank}(\Pi_0)=r$ but without requiring $\mathcal M$ to be centered nor Gaussian. Since $\phi_r(\Pi_{0,n})=\phi_r(\Pi_0)=0$, it follows by Assumption \ref{Ass: Local}(ii) and Proposition \ref{Pro: phi, differentiability} that
\begin{align}
0=\lim_{n\to\infty}\tau_n^2\{\phi_r(\Pi_{0,n})-\phi_r(\Pi_0)\}=\phi_{r,\Pi_0}''(\Delta)=\|P_{0,2}^\transpose \Delta Q_{0,2}\|^2~.
\end{align}
Hence, we have $P_{0,2}^\transpose \Delta Q_{0,2}=0$ and consequently,
\begin{align}
P(\sum_{j=r-r_{0}+1}^{k-r_{0}}\sigma^{2}_j(P_{0,2}^\transpose (\mathcal M+\Delta) Q_{0,2})\ge c_{1-\alpha})= \alpha~,
\end{align}
if the quantile restrictions on $c_{1-\alpha}$ as in Theorem \ref{Thm: SizePowerTest} hold. Size control under arbitrary local perturbations in $\mathrm H_0$, unfortunately, appears (to us) as challenging as establishing nontrivial local power under arbitrary local alternatives. We pose these as open questions, and leave them for future study.

\subsubsection{Illustration: Identification in Linear IV Models}\label{Sec: linear IV}

We now illustrate how to apply our framework by testing identification in linear IV models due to their simplicity and popularity. Let $(Y,Z^\transpose)^\transpose\in\mathbf R^{1+k}$ satisfy:
\begin{align}
Y=Z^\transpose\beta_0+u~,
\end{align}
where $\beta_0\in\mathbf R^k$ and $u$ is an error term. Let $V\in\mathbf R^m$ be an instrument variable with $E[Vu]=0$ and $m\ge k$. Then global identification of $\beta_0$ requires $E[VZ^\transpose]$ to be of full rank. Thus, identification of $\beta_0$ may be tested by examining \eqref{Eqn: hypothesis, intro} with
\begin{align}
\Pi_0=E[VZ^\transpose] \text{ and } r=k-1~.
\end{align}
The hypotheses in \eqref{Eqn: hypothesis, literature, intro} may be restrictive since it is generally unknown if $\text{rank}(\Pi_{0})\geq k-1$. Analogous rank conditions also arise for global identification in simultaneous linear equation models \citep{KoopmansHood1953Simultaneous,Fisher1961ID} and in models with misclassification errors \citep{Hu2008Identification}, and for local identification in nonlinear/nonparametric models \citep{Rothenberg1971identification,Roehrig1988Conditions,Chesher2003ID,Matzkin2008NPSimul,Chen_Chernozhukov_Lee_Newey2014LocalID} and in DSGE models \citep{Canov_Sala2009,Komunjer_Ng2011DSGE}.

To apply our framework, let $\{V_{i},Z_{i}\}_{i=1}^{n}$ be an i.i.d.\ sample. Then the estimator
\begin{align}\label{Eqn: linear IV}
\hat{\Pi}_{n} = \frac{1}{n}\sum_{i=1}^{n}V_{i}Z_{i}^{\transpose}
\end{align}
satisfies Assumption \ref{Ass: Weaklimit: Pihat} for $\tau_n=\sqrt n$ and some centered Gaussian matrix $\mathcal M$ under suitable moment restrictions. In turn, let $\{Z^{\ast}_{i},V^{\ast}_{i}\}_{i=1}^{n}$ be an i.i.d.\ sample drawn with replacement from $\{Z_{i},V_{i}\}_{i=1}^{n}$. Then $\hat{\mathcal M}_n^*\equiv\sqrt n\{\hat\Pi_n^*-\hat\Pi_n\}$ with $\hat{\Pi}_{n}^{\ast}$ given by
\begin{align}\label{Eqn: linear IV1}
\hat{\Pi}_{n}^{\ast} \equiv \frac{1}{n}\sum_{i=1}^{n}V^{\ast}_{i}Z_{i}^{\ast\transpose}= \frac{1}{n}\sum_{i=1}^{n}W_{ni}V_{i}Z_{i}^{\transpose}~,
\end{align}
where $(W_{n1},\ldots,W_{nn})$ is multinomial over $n$ categories with probabilities $(n^{-1},\ldots,n^{-1})$, satisfies Assumption \ref{Ass: Boostrap: Pihat} -- see, for example, Theorem 23.4 in \citet{Vaart1998}. We have thus verified the main assumptions.

Empirical research, however, is often faced with clustered data; e.g., micro-level data often cluster on geographical regions such as cities or states. To illustrate, suppose that there are $G$ clusters where $G$ is large, and the $g$th cluster has observations $\{V_{gi},Z_{gi}\}_{i=1}^{n_g}$. The data are independent across clusters but may otherwise be correlated within each cluster. Let $n\equiv\sum_{g=1}^{G}n_g$. In these settings, $\Pi_0$ is identified as the probability limit of
\begin{align}\label{Eqn: cluster1}
\hat\Pi_n\equiv \frac{1}{n}\sum_{g=1}^{G}V_{g}^\transpose Z_{g}
\end{align}
as $G\to\infty$, where $V_g\equiv[V_{g1},\ldots,V_{gn_g}]^\transpose$ and $Z_g\equiv[Z_{g1},\ldots,Z_{gn_g}]^\transpose$. Assumption \ref{Ass: Weaklimit: Pihat} holds for $\tau_n=\sqrt n$ and some centered Gaussian matrix $\mathcal M$, by the Lindeberg-Feller type central limit theorem. Following \citet{CameronGelbachMiller2008BootCluster}, we may construct
\begin{align}\label{Eqn: cluster2}
\hat{\mathcal M}_n^*\equiv \frac{1}{n}\sum_{g=1}^{G}W_g\{V_{g}^{\transpose} Z_{g}-\hat\Pi_n\}~,
\end{align}
where $(W_1,\ldots,W_G)$ may be a multinomial vector over $G$ categories with probabilities $(1/G,\ldots,/1G)$ (corresponding to the pairs cluster bootstrap) or other weights (such as those leading to the cluster wild bootstrap); see also \citet{DjogbenouMackinnonNielsen2018WCB}.

For the convenience of practitioners, we next provide an implementation guide of our two-step test at significance level $\alpha$.

{\sc Step 1:} (a) Sequentially test $\mathrm{rank}(\Pi_0)=0,1,\ldots,k-1$ at level $\beta$ (e.g., $\beta=\alpha/10$) based on $\hat\Pi_n$ using the KP test and obtain the rank estimator $\hat r_n$; (b) Reject $\mathrm{H}_0$ if $\hat r_n=k$ and move on to the next step otherwise.

{\sc Step 2:} (a) Draw $B$ bootstrap samples by either the empirical bootstrap or the cluster bootstrap depending on if clustering is present, construct $\{\hat{\mathcal M}_{n,b}^*\}_{b=1}^B$ accordingly (i.e., as in \eqref{Eqn: linear IV1} or \eqref{Eqn: cluster2}), and set $\hat c_{1-\alpha+\beta}$ to be the $\lfloor B(1-\alpha+\beta)\rfloor$ largest number in
\begin{align}
\sum_{j=r-\hat r_n+1}^{k-\hat r_n}\sigma_{j}^2(\hat P_{2,n}^\transpose\hat{\mathcal M}_{n,1}^* \hat Q_{2,n})~,\ldots,\sum_{j=r-\hat r_n+1}^{k-\hat r_n}\sigma_{j}^2(\hat P_{2,n}^\transpose\hat{\mathcal M}_{n,B}^* \hat Q_{2,n})~,
\end{align}
where $\hat P_{2,n}$ and $\hat Q_{2,n}$ are from the singular value decomposition of $\hat\Pi_n$ as before; (b) Reject $\mathrm{H}_0$ if $n\sigma_{\min}^2(\hat\Pi_n)>\hat c_{1-\alpha+\beta}$ with $\sigma_{\min}(\hat\Pi_n)$ the smallest singular value of $\hat\Pi_n$.

\noindent For our one-step test based on \eqref{Eqn: bootstrap overview0} and \eqref{Eqn: StructDevEst}, one may directly proceed with Step 2, but with $\hat r_n$ constructed from \eqref{Eqn: bootstrap overview0} and reject if $n\sigma_{\min}^2(\hat\Pi_n)>\hat c_{1-\alpha}$.

\section{Simulation Studies}\label{Sec: 4}

In this section, we examine the finite sample performance of our inferential framework by Monte Carlo simulations. First, we compare our tests with the multiple KP test in more complicated data environments with heteroskedasticity, serial correlation and different sample sizes. We shall pay special attention to the choices of tuning parameters. We refer the reader to Supplemental Appendix \ref{Sec: comparision with KP} where we provide additional comparisons with \citet{Kleibergen_Paap2006rank} based on their simulation designs and a real dataset that they use. Second, we also conduct simulations to assess the performance of our rank estimators, obtained by a sequential testing procedure employed in the literature and formalized in Supplemental Appendix \ref{Sec: 4-3}.

We commence by considering the following linear model
\begin{align}\label{Eqn: linear model HAC}
Z_{t}  =\Pi_0^\transpose V_{t} +V_{1,t}u_{t} ~,
\end{align}
where $Z_t\in\mathbf R^4$ for all $t$, $\{V_{t}\}\overset{\text{i.i.d.}}{\sim}N(0,I_{4})$ and $\{u_t\}$ are generated according to
\begin{align}
u_{t}=\epsilon_{t}-\frac{1}{4}\mathbf{1}_{4}\mathbf{1}_{4}^{\transpose}\epsilon_{t-1}
\end{align}
with $\{\epsilon_{t}\}\overset{\text{i.i.d.}}{\sim}N(0,I_{4})$ independent of $\{V_{t}\}$, and $V_{1,t}$ the first entry of $V_{t}$. Moreover, we configure $\Pi_0$ as: for $\delta\in\{0, 0.1, 0.3, 0.5\}$,
\begin{align}
\Pi_{0}=\textrm{diag}(\mathbf{1}_{2},\mathbf{0}_{2})+\delta I_{4}~.
\end{align}
We test the hypotheses in \eqref{Eqn: hypothesis, intro} for $r\in\{2,3\}$ at level $\alpha=5\%$. Thus, for both cases, $\mathrm{H}_0$ is true if and only if $\delta=0$, and they respectively correspond to $\mathrm{rank}(\Pi_0)=r$ and $\mathrm{rank}(\Pi_0)<r$ under $\mathrm H_0$. We estimate $\Pi_{0}$ by $\hat{\Pi}_{n}=\frac{1}{n}\sum_{t=1}^{n}V_{t}Z_{t}^{\transpose}$ for sample sizes $n\in\{50,100,300,1000\}$, and for each $n$, the number of simulation replications is set to be 5,000 with 500 bootstrap repetitions for each replication. As the data exhibit first order autocorrelation, we adopt the circular block bootstrap \citep{Politis_Romano1992circular} with block size $b=2$. To implement the multiple KP test, labelled KP-M, we estimate the variance of $\mathrm{vec}(\hat\Pi_n)$ by the HACC estimator with one lag \citep{West1997AnotherHAC}. To carry out our tests, we choose $\kappa_n\in\{n^{-2/5},1.5n^{-2/5}, n^{-1/5}, n^{-1/4}, n^{-1/3}, 1.5n^{-1/5}$, $1.5n^{-1/4}, 1.5n^{-1/3}\}$ for both the numerical estimator in \eqref{Eqn: NumericalDevEst} and the analytic estimator in \eqref{Eqn: bootstrap overview0} and \eqref{Eqn: StructDevEst}, and $\beta\in\{\alpha/5,\alpha/10,\alpha/15,\alpha/20$, $\alpha/25,\alpha/30\}$ for the two-step test. As in Section \ref{Sec: 2}, we respectively label these three tests as CF-N, CF-A, and CF-T.

\iffalse
Due to space limitation,  we only report here the results for $\kappa_n\in\{n^{-1/4}, n^{-1/4}, n^{-1/3}\}$ and $\beta\in\{\alpha/10,\alpha/15,\alpha/20\}$, relegating the remaining results to Supplemental Appendix \ref{Sec: additional simulations}.
\fi

\setlength{\tabcolsep}{3pt}
\begin{table}[!htbp]
\centering
\begin{threeparttable}
\caption{Rejection rates of rank tests for the model \eqref{Eqn: linear model HAC} at $\alpha=5\%$\tnote{\dag}}\label{Tab: Simulation1}
\begin{tabular}{cccccccccccccccc}
\hline\hline
&\multirow{2}{*}{\makecell{Sample\\ size}} &\multicolumn{3}{c}{CF-T}&&\multicolumn{3}{c}{CF-A}&&\multicolumn{3}{c}{CF-N}&&\multirow{2}{*}{KP-M}&\\
\cline{3-5}\cline{7-9}\cline{11-13}
&&$\alpha/10$&$\alpha/15$&$\alpha/20$ &&$n^{-1/5}$&$n^{-1/4}$&$n^{-1/3}$ &&$n^{-1/5}$&$n^{-1/4}$&$n^{-1/3}$ &&\\
\cmidrule{2-15}
&&\multicolumn{13}{c}{Rejection rates for $r=2$}&\\
\cline{2-15}
&&\multicolumn{13}{c}{$\delta=0$}&\\
\cline{2-15}
&$50$   &0.17&0.17&0.17&&0.04&0.04&0.04&&0.29&0.28&0.21&&0.08&\\
&$100$  &0.08&0.08&0.08&&0.04&0.04&0.04&&0.23&0.20&0.12&&0.08&\\
&$300$  &0.04&0.04&0.04&&0.05&0.05&0.05&&0.16&0.12&0.05&&0.06&\\
&$1000$ &0.04&0.04&0.04&&0.05&0.05&0.05&&0.11&0.08&0.04&&0.05&\\
\cline{2-15}
&&\multicolumn{13}{c}{$\delta=0.1$}&\\
\cline{2-15}
 &$50$   &0.23&0.23&0.23&&0.08&0.08&0.08&&0.37&0.35&0.27&&0.13&\\
 &$100$  &0.18&0.17&0.17&&0.12&0.12&0.12&&0.38&0.34&0.23&&0.19&\\
 &$300$  &0.34&0.34&0.34&&0.35&0.35&0.35&&0.57&0.51&0.36&&0.44&\\
 &$1000$ &0.89&0.90&0.90&&0.90&0.90&0.90&&0.95&0.92&0.88&&0.92&\\
%\hline
%&\multicolumn{10}{c}{$\delta=0.2$}\\
%\hline
%$50$   &0.4310&0.4302&&0.2248&0.2248&&0.5714&0.4880&&0.1904\\
%$100$  &0.5066&0.4994&&0.4254&0.4254&&0.6950&0.5750&&0.4410\\
% $300$  &0.9362&0.9340&&0.9350&0.9350&&0.9694&0.9366&&0.9526\\
% $1000$ &1.0000&1.0000&&1.0000&1.0000&&1.0000&1.0000&&1.0000\\
\cline{2-15}
&&\multicolumn{13}{c}{$\delta=0.3$}&\\
\cline{2-15}
&$50$  &0.67&0.67&0.66&&0.48&0.48&0.48&&0.80&0.79&0.72& &0.40&\\
&$100$  &0.85&0.85&0.85&&0.80&0.80&0.80&&0.95&0.93&0.89&&0.77&\\
&$300$  &1.00&1.00&1.00&&1.00&1.00&1.00&&1.00&1.00&1.00&&1.00&\\
&$1000$ &1.00&1.00&1.00&&1.00&1.00&1.00&&1.00&1.00&1.00&&1.00&\\
%\hline
%&\multicolumn{10}{c}{$\delta=0.4$}\\
%\hline
%$50$   &0.8546&0.8536&&0.7220&0.7220&&0.9236&0.8896&&0.5380\\
%$100$  &0.9722&0.9714&&0.9618&0.9618&&0.9954&0.9832&&0.9456\\
%$300$  &1.0000&1.0000&&1.0000&1.0000&&1.0000&1.0000&&1.0000\\
%$1000$ &1.0000&1.0000&&1.0000&1.0000&&1.0000&1.0000&&1.0000\\
\cline{2-15}
&&\multicolumn{13}{c}{$\delta=0.5$}&\\
\cline{2-15}
&$50$   &0.95&0.95&0.94&&0.89&0.89&0.89&&0.98&0.98&0.97&&0.55&\\
&$100$  &1.00&1.00&1.00&&1.00&1.00&1.00&&1.00&1.00&1.00&&0.87&\\
&$300$  &1.00&1.00&1.00&&1.00&1.00&1.00&&1.00&1.00&1.00&&1.00&\\
&$1000$ &1.00&1.00&1.00&&1.00&1.00&1.00&&1.00&1.00&1.00&&1.00&\\
\cmidrule{1-16}
&&\multicolumn{13}{c}{Rejection rates for $r=3$}&\\
\cline{2-15}
&&\multicolumn{13}{c}{$\delta=0$}&\\
\cline{2-15}
&$50$   &0.09&0.09&0.10&&0.07&0.06&0.04&&0.14&0.14&0.12&&0.01&\\
&$100$  &0.06&0.07&0.07&&0.06&0.06&0.03&&0.12&0.12&0.09&&0.01&\\
&$300$  &0.04&0.05&0.05&&0.05&0.05&0.03&&0.09&0.08&0.06&&0.01&\\
&$1000$ &0.05&0.05&0.05&&0.06&0.06&0.05&&0.08&0.07&0.05&&0.00&\\
\cline{2-15}
&&\multicolumn{13}{c}{$\delta=0.1$}&\\
\cline{2-15}
&$50$   &0.12&0.12&0.12&&0.10&0.09&0.05&&0.18&0.18&0.16&&0.01&\\
&$100$  &0.12&0.13&0.13&&0.13&0.11&0.06&&0.21&0.19&0.16&&0.02&\\
&$300$  &0.25&0.26&0.27&&0.32&0.29&0.16&&0.38&0.36&0.31&&0.09&\\
&$1000$ &0.63&0.65&0.67&&0.82&0.81&0.59&&0.84&0.82&0.77&&0.54&\\
\cline{2-15}
&&\multicolumn{13}{c}{$\delta=0.3$}&\\
\cline{2-15}
&$50$   &0.43&0.44&0.45&&0.39&0.33&0.25&&0.57&0.56&0.52&&0.12&\\
&$100$  &0.61&0.63&0.64&&0.66&0.57&0.50&&0.80&0.79&0.74&&0.43&\\
&$300$  &0.96&0.96&0.96&&0.98&0.96&0.96&&1.00&0.99&0.99&&0.96&\\
&$1000$ &1.00&1.00&1.00&&1.00&1.00&1.00&&1.00&1.00&1.00&&1.00&\\
\cline{2-15}
&&\multicolumn{13}{c}{$\delta=0.5$}&\\
\cline{2-15}
&$50$   &0.76&0.77&0.78&&0.68&0.64&0.63&&0.88&0.88&0.84&&0.37&\\
&$100$  &0.92&0.93&0.93&&0.92&0.91&0.91&&0.99&0.99&0.98&&0.79&\\
&$300$  &1.00&1.00&1.00&&1.00&1.00&1.00&&1.00&1.00&1.00&&1.00&\\
&$1000$ &1.00&1.00&1.00&&1.00&1.00&1.00&&1.00&1.00&1.00&&1.00&\\
\hline\hline
\end{tabular}
\begin{tablenotes}
      \small
      \item[\dag] The three values under CF-T are the choices of $\beta$, and those under CF-A and CF-N are the choices of $\kappa_n$ as in \ref{Eqn: bootstrap overview0} and \eqref{Eqn: NumericalDevEst} respectively.
    \end{tablenotes}
\end{threeparttable}
\end{table}%

\setlength{\tabcolsep}{4pt}
\begin{sidewaystable}[!htbp]
\centering
\begin{threeparttable}
\caption{Additional results on rejection rates of rank tests for the model \eqref{Eqn: linear model HAC} with $r=2$, at $\alpha=5\%$\tnote{\dag}}
\label{Tab: r2}
\begin{tabular}{cccccccccccccccccccc}
\hline\hline
&\multirow{2}{*}{\makecell{Sample\\ size}} &\multicolumn{3}{c}{CF-T}&&\multicolumn{5}{c}{CF-A}&&\multicolumn{5}{c}{CF-N}&&\multirow{2}{*}{KP-M}&\\
\cline{3-5}\cline{7-11}\cline{13-17}
&&$\alpha/5$&$\alpha/25$&$\alpha/30$ &&$1.5n^{-1/5}$&$1.5n^{-1/4}$&$1.5n^{-1/3}$&$n^{-2/5}$&$1.5n^{-2/5}$ &&$1.5n^{-1/5}$&$1.5n^{-1/4}$&$1.5n^{-1/3}$&$n^{-2/5}$&$1.5n^{-2/5}$ & & & \\
\cline{2-19}
&&\multicolumn{17}{c}{$\delta=0$}&\\
\cline{2-19}
&$50$   &0.16&0.17&0.17&&0.08&0.05&0.04&0.04&0.04&&0.31&0.30&0.29&0.13&0.25&&0.08&\\
&$100$  &0.08&0.08&0.08&&0.04&0.04&0.04&0.04&0.04&&0.26&0.24&0.20&0.06&0.15&&0.08&\\
&$300$  &0.03&0.04&0.04&&0.05&0.05&0.05&0.05&0.05&&0.20&0.18&0.11&0.02&0.06&&0.06&\\
&$1000$ &0.04&0.05&0.05&&0.05&0.05&0.05&0.05&0.05&&0.15&0.12&0.06&0.02&0.03&&0.05&\\
\cline{2-19}
&&\multicolumn{17}{c}{$\delta=0.1$}&\\
\cline{2-19}
&$50$   &0.23&0.23&0.24&&0.11&0.09&0.08&0.08&0.08&&0.39&0.39&0.36&0.17&0.31&&0.13&\\
&$100$  &0.17&0.17&0.17&&0.12&0.12&0.12&0.12&0.12&&0.41&0.40&0.34&0.12&0.26&&0.19&\\
&$300$  &0.34&0.34&0.34&&0.35&0.35&0.35&0.35&0.35&&0.63&0.60&0.50&0.21&0.37&&0.44&\\
&$1000$ &0.88&0.90&0.90&&0.90&0.90&0.90&0.90&0.90&&0.97&0.96&0.91&0.80&0.87&&0.92&\\
%\hline
%&\multicolumn{10}{c}{$\delta=0.2$}\\
%\hline
%$50$   &0.4310&0.4302&&0.2248&0.2248&&0.5714&0.4880&&0.1904\\
%$100$  &0.5066&0.4994&&0.4254&0.4254&&0.6950&0.5750&&0.4410\\
% $300$  &0.9362&0.9340&&0.9350&0.9350&&0.9694&0.9366&&0.9526\\
% $1000$ &1.0000&1.0000&&1.0000&1.0000&&1.0000&1.0000&&1.0000\\
\cline{2-19}
&&\multicolumn{17}{c}{$\delta=0.3$}&\\
\cline{2-19}
&$50$   &0.67&0.66&0.66&&0.49&0.48&0.48&0.48&0.48&&0.81&0.81&0.79&0.61&0.76&&0.40&\\
&$100$  &0.86&0.84&0.84&&0.80&0.80&0.80&0.80&0.80&&0.95&0.95&0.94&0.79&0.90&&0.77&\\
&$300$  &1.00&1.00&1.00&&1.00&1.00&1.00&1.00&1.00&&1.00&1.00&1.00&1.00&1.00&&1.00&\\
&$1000$ &1.00&1.00&1.00&&1.00&1.00&1.00&1.00&1.00&&1.00&1.00&1.00&1.00&1.00&&1.00&\\
%\hline
%&\multicolumn{10}{c}{$\delta=0.4$}\\
%\hline
%$50$   &0.8546&0.8536&&0.7220&0.7220&&0.9236&0.8896&&0.5380\\
%$100$  &0.9722&0.9714&&0.9618&0.9618&&0.9954&0.9832&&0.9456\\
%$300$  &1.0000&1.0000&&1.0000&1.0000&&1.0000&1.0000&&1.0000\\
%$1000$ &1.0000&1.0000&&1.0000&1.0000&&1.0000&1.0000&&1.0000\\
\cline{2-19}
&&\multicolumn{17}{c}{$\delta=0.5$}&\\
\cline{2-19}
&$50$   &0.95&0.94&0.94&&0.89&0.89&0.89&0.89&0.89&&0.98&0.99&0.98&0.93&0.97 &&0.55&\\
&$100$  &1.00&1.00&1.00&&1.00&1.00&1.00&1.00&1.00&&1.00&1.00&1.00&1.00&1.00 &&0.87&\\
&$300$  &1.00&1.00&1.00&&1.00&1.00&1.00&1.00&1.00&&1.00&1.00&1.00&1.00&1.00 &&1.00&\\
&$1000$ &1.00&1.00&1.00&&1.00&1.00&1.00&1.00&1.00&&1.00&1.00&1.00&1.00&1.00 &&1.00&\\ \hline\hline
  \end{tabular}
  \begin{tablenotes}
      \small
      \item[\dag] The three values under CF-T are the choices of $\beta$, and those under CF-A and CF-N are the choices of $\kappa_n$ as in \ref{Eqn: bootstrap overview0} and \eqref{Eqn: NumericalDevEst} respectively.
    \end{tablenotes}
\end{threeparttable}
\end{sidewaystable}%

\setlength{\tabcolsep}{4pt}
\begin{sidewaystable}[!htbp]
\centering
\begin{threeparttable}
\caption{Additional results on rejection rates of rank tests for the model \eqref{Eqn: linear model HAC} with $r=3$, at $\alpha=5\%$\tnote{\dag}}
\label{Tab: r3}
\begin{tabular}{cccccccccccccccccccc}
\hline\hline
&\multirow{2}{*}{\makecell{Sample\\ size}} &\multicolumn{3}{c}{CF-T}&&\multicolumn{5}{c}{CF-A}&&\multicolumn{5}{c}{CF-N}&&\multirow{2}{*}{KP-M}&\\
\cline{3-5}\cline{7-11}\cline{13-17}
&&$\alpha/5$&$\alpha/25$&$\alpha/30$ &&$1.5n^{-1/5}$&$1.5n^{-1/4}$&$1.5n^{-1/3}$&$n^{-2/5}$&$1.5n^{-2/5}$ &&$1.5n^{-1/5}$&$1.5n^{-1/4}$&$1.5n^{-1/3}$&$n^{-2/5}$&$1.5n^{-2/5}$ & & &\\
\cline{2-19}
&&\multicolumn{17}{c}{$\delta=0$}&\\
\cline{2-19}
&$50$   &0.08&0.10&0.10&&0.08&0.08&0.06&0.02&0.05&&0.14&0.14&0.14&0.10&0.14&&0.01&\\
&$100$  &0.05&0.07&0.07&&0.06&0.06&0.06&0.01&0.04&&0.13&0.13&0.12&0.07&0.10&&0.01&\\
&$300$  &0.04&0.05&0.05&&0.05&0.05&0.05&0.01&0.03&&0.10&0.09&0.07&0.04&0.06&&0.01&\\
&$1000$ &0.04&0.05&0.05&&0.06&0.06&0.05&0.01&0.05&&0.10&0.08&0.06&0.04&0.05&&0.00&\\
\cline{2-19}
&&\multicolumn{17}{c}{$\delta=0.1$}&\\
\cline{2-19}
&$50$   &0.10&0.13&0.13&&0.11&0.17&0.09&0.03&0.07&&0.18&0.18&0.18&0.13&0.17 &&0.01&\\
&$100$  &0.10&0.13&0.14&&0.13&0.13&0.11&0.04&0.07&&0.21&0.21&0.20&0.11&0.17 &&0.02&\\
&$300$  &0.21&0.28&0.28&&0.32&0.32&0.28&0.10&0.16&&0.41&0.39&0.35&0.23&0.31 &&0.09&\\
&$1000$ &0.58&0.68&0.68&&0.88&0.82&0.76&0.54&0.58&&0.86&0.84&0.81&0.68&0.77 &&0.54&\\
%\hline
%&\multicolumn{10}{c}{$\delta=0.2$}\\
%\hline
%$50$   &0.4310&0.4302&&0.2248&0.2248&&0.5714&0.4880&&0.1904\\
%$100$  &0.5066&0.4994&&0.4254&0.4254&&0.6950&0.5750&&0.4410\\
% $300$  &0.9362&0.9340&&0.9350&0.9350&&0.9694&0.9366&&0.9526\\
% $1000$ &1.0000&1.0000&&1.0000&1.0000&&1.0000&1.0000&&1.0000\\
\cline{2-19}
&&\multicolumn{17}{c}{$\delta=0.3$}&\\
\cline{2-19}
&$50$   &0.40&0.45&0.45&&0.47&0.44&0.35&0.23&0.28&&0.57&0.57&0.56&0.44&0.54&&0.12&\\
&$100$  &0.56&0.65&0.65&&0.75&0.72&0.58&0.49&0.51&&0.81&0.80&0.79&0.65&0.76&&0.43&\\
&$300$  &0.95&0.96&0.96&&0.99&0.99&0.96&0.96&0.96&&1.00&1.00&0.99&0.97&0.99&&0.96&\\
&$1000$ &1.00&1.00&1.00&&1.00&1.00&1.00&1.00&1.00&&1.00&1.00&1.00&1.00&1.00&&1.00&\\
%\hline
%&\multicolumn{10}{c}{$\delta=0.4$}\\
%\hline
%$50$   &0.8546&0.8536&&0.7220&0.7220&&0.9236&0.8896&&0.5380\\
%$100$  &0.9722&0.9714&&0.9618&0.9618&&0.9954&0.9832&&0.9456\\
%$300$  &1.0000&1.0000&&1.0000&1.0000&&1.0000&1.0000&&1.0000\\
%$1000$ &1.0000&1.0000&&1.0000&1.0000&&1.0000&1.0000&&1.0000\\
\cline{2-19}
&&\multicolumn{17}{c}{$\delta=0.5$}&\\
\cline{2-19}
&$50$    &0.74&0.78&0.79&&0.80&0.74&0.65&0.63&0.63&&0.89&0.89&0.88&0.77&0.87&&0.37&\\
&$100$   &0.91&0.93&0.93&&0.96&0.94&0.92&0.91&0.91&&0.99&0.99&0.99&0.95&0.98&&0.79&\\
&$300$   &1.00&1.00&1.00&&1.00&1.00&1.00&1.00&1.00&&1.00&1.00&1.00&1.00&1.00&&1.00&\\
&$1000$  &1.00&1.00&1.00&&1.00&1.00&1.00&1.00&1.00&&1.00&1.00&1.00&1.00&1.00&&1.00&\\
\hline\hline
\end{tabular}
\begin{tablenotes}
      \small
      \item[\dag] The three values under CF-T are the choices of $\beta$, and those under CF-A and CF-N are the choices of $\kappa_n$ as in \ref{Eqn: bootstrap overview0} and \eqref{Eqn: NumericalDevEst} respectively.
    \end{tablenotes}
\end{threeparttable}
\end{sidewaystable}%

Table \ref{Tab: Simulation1} summarizes the simulation results for tuning parameters in the middle range of the choices, while Tables \ref{Tab: r2} and \ref{Tab: r3} collect results for the remaining choices. For the case of $r=2$ (so that $\mathrm{rank}(\Pi_0)=r$ under $\mathrm H_0$), the performance of CF-A and CF-T is comparable with that of KP-M especially when the sample size is large, though CF-T exhibits more size distortion than KP-M for $n=50$ and CF-N appears to be somewhat sensitive to the choice of $\kappa_n$. For the case of $r=3$ (so that $\mathrm{rank}(\Pi_0)<r$ under $\mathrm H_0$), KP-M is markedly under-sized even in large samples, while its local power is uniformly dominated by our three tests, across all the choices of the tuning parameters, sample sizes, and the local parameter $\delta$. With regards to comparisons among our three tests, there are also some persistent patterns. First, CF-N overall tends to be the most over-sized especially in small samples, and the most sensitive to the choice of the tuning parameters. Second, between CF-A and CF-T, one does not seem to dominate the other. The former appears to perform better overall in terms of size control and local power in small samples, though the differences become smaller as the sample size increases. The latter, on the other hand, seems to be the least sensitive to the choice of the tuning parameters especially in the irregular case when $r=3$, as desired. Thus, it seems sensible to employ CF-A in small samples and CF-T instead in large samples.

{\setlength\intextsep{0pt}

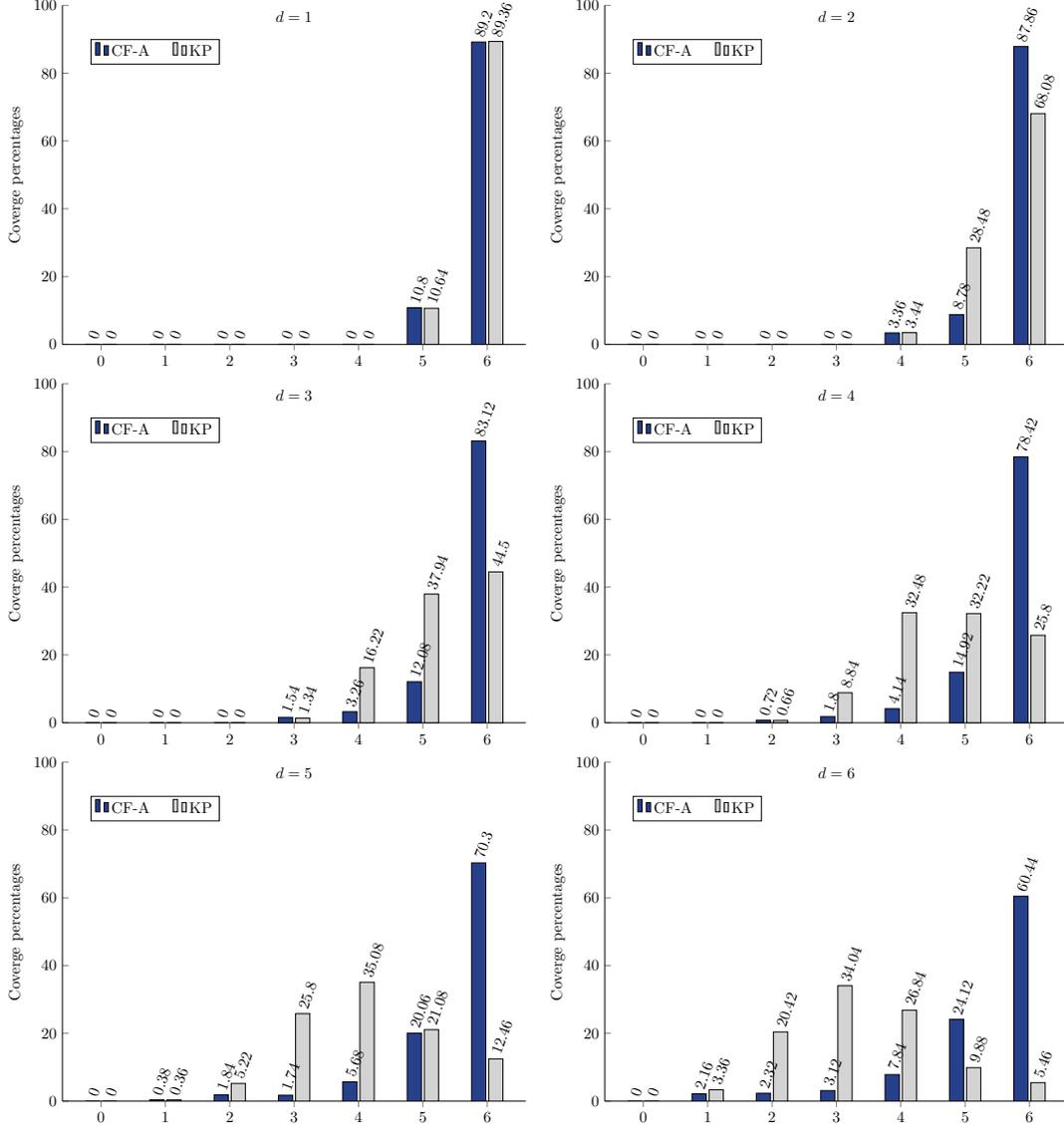
\begin{figure}[!h]
\begin{subfigure}[b]{0.48\textwidth}
\centering
\resizebox{\linewidth}{!}{
\pgfplotsset{title style={at={(0.5,0.91)}}}
\begin{tikzpicture}
\begin{axis}[
        ybar,
    height=10cm,
    width=13cm,
    enlarge y limits=false,
    axis lines*=left,
    ymin=0,
    ymax=100,
        title={$d=1$},
     legend style={at={(0.2,0.9)},
        anchor=north,legend columns=-1,
        /tikz/every even column/.append style={column sep=0.5cm}
        },
        ylabel={Coverge percentages},
        symbolic x coords={0,1,2,3,4,5,6},
     xtick=data,
        nodes near coords,
    every node near coord/.append style={
        anchor=mid west,
        rotate=70
    }
    ]

    \addplot [fill=RoyalBlue4] coordinates {(0,0) (1,0) (2,0)
      (3,0) (4,0) (5,10.80)(6,89.20)};
    \addplot [fill=LightGrey] coordinates {(0,0) (1,0) (2,0)
      (3,0) (4,0) (5,10.64) (6,89.36)};

       \legend{CF-A,KP}
  \end{axis}
\end{tikzpicture}}
\end{subfigure}
\begin{subfigure}[b]{0.48\textwidth}
\centering
\resizebox{\linewidth}{!}{
\pgfplotsset{title style={at={(0.5,0.91)}}}
\begin{tikzpicture}
\begin{axis}[
        ybar,
    height=10cm,
    width=13cm,
    enlarge y limits=false,
    axis lines*=left,
    ymin=0,
    ymax=100,
    title={$d=2$},
     legend style={at={(0.2,0.9)},
        anchor=north,legend columns=-1,
        /tikz/every even column/.append style={column sep=0.5cm}
        },
        ylabel={Coverge percentages},
        symbolic x coords={0,1,2,3,4,5,6},
     xtick=data,
        nodes near coords,
    every node near coord/.append style={
        anchor=mid west,
        rotate=70
    }
    ]
    \addplot [fill=RoyalBlue4] coordinates {(0,0) (1,0) (2,0)
      (3,0) (4,3.36) (5,8.78)(6,87.86)};
    \addplot [fill=LightGrey] coordinates {(0,0) (1,0) (2,0)
      (3,0) (4,3.44) (5,28.48) (6,68.08)};
    \legend{CF-A,KP}
  \end{axis}
\end{tikzpicture}}
\end{subfigure}

\begin{subfigure}[b]{0.48\textwidth}
\centering
\resizebox{\linewidth}{!}{
\pgfplotsset{title style={at={(0.5,0.91)}}}
\begin{tikzpicture}
\begin{axis}[
        ybar,
    height=10cm,
    width=13cm,
    enlarge y limits=false,
    axis lines*=left,
    ymin=0,
    ymax=100,
    title={$d=3$},
     legend style={at={(0.2,0.9)},
        anchor=north,legend columns=-1,
        /tikz/every even column/.append style={column sep=0.5cm}
        },
        ylabel={Coverge percentages},
        symbolic x coords={0,1,2,3,4,5,6},
     xtick=data,
        nodes near coords,
    every node near coord/.append style={
        anchor=mid west,
        rotate=70
    }
    ]
    \addplot [fill=RoyalBlue4] coordinates {(0,0) (1,0) (2,0)
      (3,1.54) (4,3.26) (5,12.08)(6,83.12)};
    \addplot [fill=LightGrey] coordinates {(0,0) (1,0) (2,0)
      (3,1.34) (4,16.22) (5,37.94) (6,44.50)};
    \legend{CF-A,KP}
  \end{axis}
\end{tikzpicture}}
\end{subfigure}
\begin{subfigure}[b]{0.48\textwidth}
\centering
\resizebox{\linewidth}{!}{
\pgfplotsset{title style={at={(0.5,0.91)}}}
\begin{tikzpicture}
\begin{axis}[
        ybar,
    height=10cm,
    width=13cm,
    enlarge y limits=false,
    axis lines*=left,
    ymin=0,
    ymax=100,
    title={$d=4$},
     legend style={at={(0.2,0.9)},
        anchor=north,legend columns=-1,
        /tikz/every even column/.append style={column sep=0.5cm}
        },
        ylabel={Coverge percentages},
        symbolic x coords={0,1,2,3,4,5,6},
     xtick=data,
        nodes near coords,
    every node near coord/.append style={
        anchor=mid west,
        rotate=70
    }
    ]
    \addplot [fill=RoyalBlue4] coordinates {(0,0) (1,0) (2,0.72)
      (3,1.80) (4,4.14) (5,14.92)(6,78.42)};
    \addplot [fill=LightGrey] coordinates {(0,0) (1,0) (2,0.66)
      (3,8.84) (4,32.48) (5,32.22) (6,25.80)};
    \legend{CF-A,KP}
  \end{axis}
\end{tikzpicture}}
\end{subfigure}

\begin{subfigure}[b]{0.48\textwidth}
\centering
\resizebox{\linewidth}{!}{
\pgfplotsset{title style={at={(0.5,0.91)}}}
\begin{tikzpicture}
\begin{axis}[
        ybar,
    height=10cm,
    width=13cm,
    enlarge y limits=false,
    axis lines*=left,
    ymin=0,
    ymax=100,
    title={$d=5$},
     legend style={at={(0.2,0.9)},
        anchor=north,legend columns=-1,
        /tikz/every even column/.append style={column sep=0.5cm}
        },
        ylabel={Coverge percentages},
        symbolic x coords={0,1,2,3,4,5,6},
     xtick=data,
        nodes near coords,
    every node near coord/.append style={
        anchor=mid west,
        rotate=70
    }
    ]
    \addplot [fill=RoyalBlue4] coordinates {(0,0) (1,0.38) (2,1.84)
      (3,1.74) (4,5.68) (5,20.06)(6,70.30)};
    \addplot [fill=LightGrey] coordinates {(0,0) (1,0.36) (2,5.22)
      (3,25.80) (4,35.08) (5,21.08) (6,12.46)};
    \legend{CF-A,KP}
  \end{axis}
\end{tikzpicture}}
\end{subfigure}
\begin{subfigure}[b]{0.48\textwidth}
\centering
\resizebox{\linewidth}{!}{
\pgfplotsset{title style={at={(0.5,0.91)}}}
\begin{tikzpicture}
\begin{axis}[
        ybar,
    height=10cm,
    width=13cm,
    enlarge y limits=false,
    axis lines*=left,
    ymin=0,
    ymax=100,
    title={$d=6$},
     legend style={at={(0.2,0.9)},
        anchor=north,legend columns=-1,
        /tikz/every even column/.append style={column sep=0.5cm}
        },
        ylabel={Coverge percentages},
        symbolic x coords={0,1,2,3,4,5,6},
     xtick=data,
        nodes near coords,
    every node near coord/.append style={
        anchor=mid west,
        rotate=70
    }
    ]
    \addplot [fill=RoyalBlue4] coordinates {(0,0) (1,2.16) (2,2.32)
      (3,3.12) (4,7.84) (5,24.12)(6,60.44)};
    \addplot [fill=LightGrey] coordinates {(0,0) (1,3.36) (2,20.42)
      (3,34.04) (4,26.84) (5,9.88) (6,5.46)};
    \legend{CF-A,KP}
  \end{axis}
\end{tikzpicture}}
\end{subfigure}
\caption{The rank estimation: $\mathrm{rank}(\Pi_0)=6$, $\alpha = 5\%$ and $\delta=0.1$} \label{Fig: rank estimation1}
\end{figure}
}

{\setlength\intextsep{0pt}
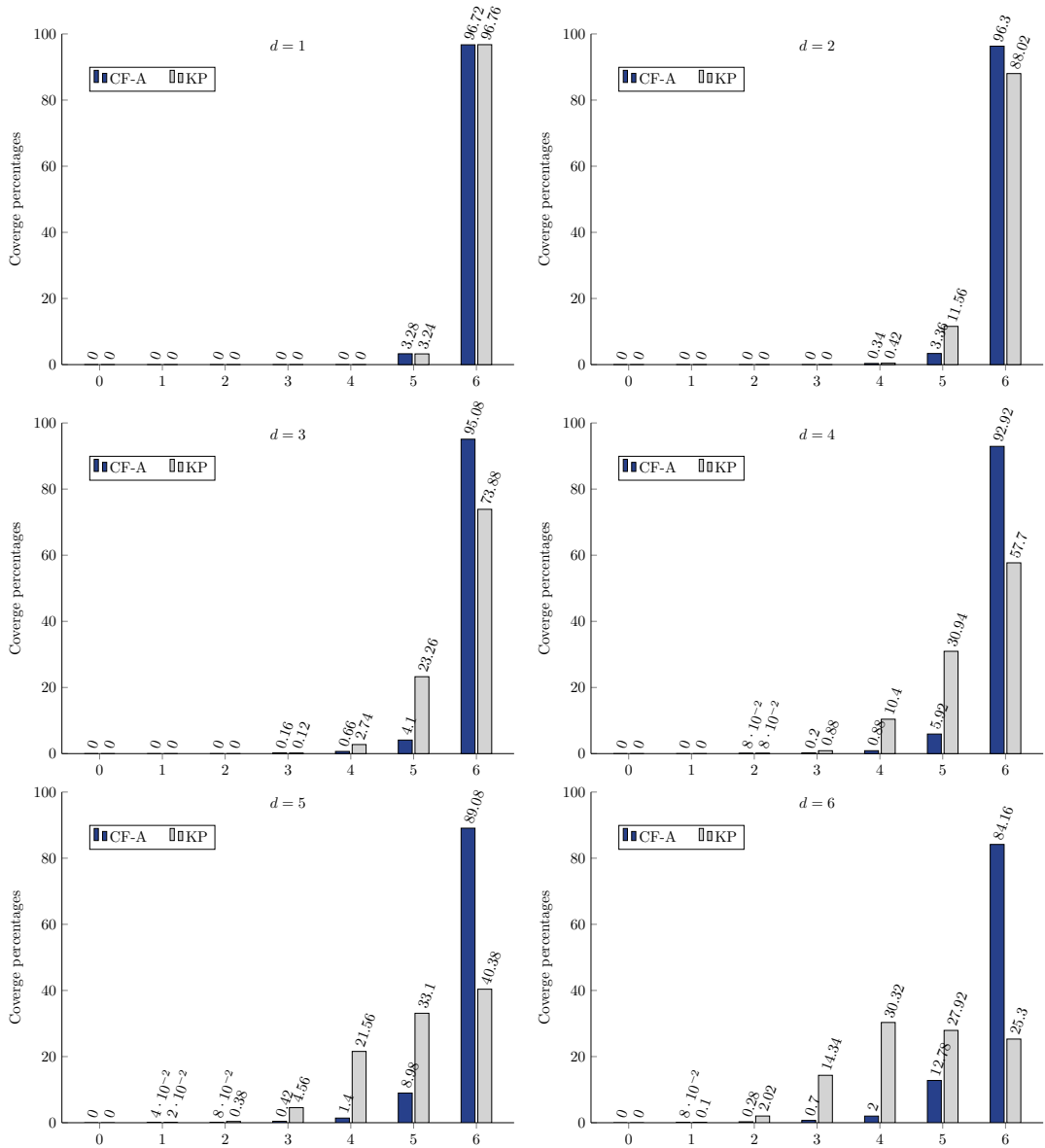
\begin{figure}[!h]
\begin{subfigure}[b]{0.48\textwidth}
\centering
\resizebox{\linewidth}{!}{
\pgfplotsset{title style={at={(0.5,0.91)}}}
\begin{tikzpicture}
\begin{axis}[
        ybar,
    height=10cm,
    width=13cm,
    enlarge y limits=false,
    axis lines*=left,
    ymin=0,
    ymax=100,
    title={$d=1$},
     legend style={at={(0.2,0.9)},
        anchor=north,legend columns=-1,
        /tikz/every even column/.append style={column sep=0.5cm}
        },
        ylabel={Coverge percentages},
        symbolic x coords={0,1,2,3,4,5,6},
     xtick=data,
        nodes near coords,
    every node near coord/.append style={
        anchor=mid west,
        rotate=70
    }
    ]
    \addplot [fill=RoyalBlue4] coordinates {(0,0) (1,0) (2,0)
      (3,0) (4,0) (5,3.28)(6,96.72)};
    \addplot [fill=LightGrey] coordinates {(0,0) (1,0) (2,0)
      (3,0) (4,0) (5,3.24) (6,96.76)};
    \legend{CF-A,KP}
  \end{axis}
\end{tikzpicture}}
\end{subfigure}
\begin{subfigure}[b]{0.48\textwidth}
\centering
\resizebox{\linewidth}{!}{
\pgfplotsset{title style={at={(0.5,0.91)}}}
\begin{tikzpicture}
\begin{axis}[
        ybar,
    height=10cm,
    width=13cm,
    enlarge y limits=false,
    axis lines*=left,
    ymin=0,
    ymax=100,
    title={$d=2$},
     legend style={at={(0.2,0.9)},
        anchor=north,legend columns=-1,
        /tikz/every even column/.append style={column sep=0.5cm}
        },
        ylabel={Coverge percentages},
        symbolic x coords={0,1,2,3,4,5,6},
     xtick=data,
        nodes near coords,
    every node near coord/.append style={
        anchor=mid west,
        rotate=70
    }
    ]
    \addplot [fill=RoyalBlue4] coordinates {(0,0) (1,0) (2,0)
      (3,0) (4,0.34) (5,3.36)(6,96.30)};
    \addplot [fill=LightGrey] coordinates {(0,0) (1,0) (2,0)
      (3,0) (4,0.42) (5,11.56) (6,88.02)};
    \legend{CF-A,KP}
  \end{axis}
\end{tikzpicture}}
\end{subfigure}

\begin{subfigure}[b]{0.48\textwidth}
\centering
\resizebox{\linewidth}{!}{
\pgfplotsset{title style={at={(0.5,0.91)}}}
\begin{tikzpicture}
\begin{axis}[
        ybar,
    height=10cm,
    width=13cm,
    enlarge y limits=false,
    axis lines*=left,
    ymin=0,
    ymax=100,
    title={$d=3$},
     legend style={at={(0.2,0.9)},
        anchor=north,legend columns=-1,
        /tikz/every even column/.append style={column sep=0.5cm}
        },
        ylabel={Coverge percentages},
        symbolic x coords={0,1,2,3,4,5,6},
     xtick=data,
        nodes near coords,
    every node near coord/.append style={
        anchor=mid west,
        rotate=70
    }
    ]
    \addplot [fill=RoyalBlue4] coordinates {(0,0) (1,0) (2,0)
      (3,0.16) (4,0.66) (5,4.10)(6,95.08)};
    \addplot [fill=LightGrey] coordinates {(0,0) (1,0) (2,0)
      (3,0.12) (4,2.74) (5,23.26) (6,73.88)};
    \legend{CF-A,KP}
  \end{axis}
\end{tikzpicture}}
\end{subfigure}
\begin{subfigure}[b]{0.48\textwidth}
\centering
\resizebox{\linewidth}{!}{
\pgfplotsset{title style={at={(0.5,0.91)}}}
\begin{tikzpicture}
\begin{axis}[
        ybar,
    height=10cm,
    width=13cm,
    enlarge y limits=false,
    axis lines*=left,
    ymin=0,
    ymax=100,
    title={$d=4$},
     legend style={at={(0.2,0.9)},
        anchor=north,legend columns=-1,
        /tikz/every even column/.append style={column sep=0.5cm}
        },
        ylabel={Coverge percentages},
        symbolic x coords={0,1,2,3,4,5,6},
     xtick=data,
        nodes near coords,
    every node near coord/.append style={
        anchor=mid west,
        rotate=70
    }
    ]
    \addplot [fill=RoyalBlue4] coordinates {(0,0) (1,0) (2,0.08)
      (3,0.20) (4,0.88) (5,5.92)(6,92.92)};
    \addplot [fill=LightGrey] coordinates {(0,0) (1,0) (2,0.08)
      (3,0.88) (4,10.40) (5,30.94) (6,57.70)};
    \legend{CF-A,KP}
  \end{axis}
\end{tikzpicture}}
\end{subfigure}

\begin{subfigure}[b]{0.48\textwidth}
\centering
\resizebox{\linewidth}{!}{
\pgfplotsset{title style={at={(0.5,0.91)}}}
\begin{tikzpicture}
\begin{axis}[
        ybar,
    height=10cm,
    width=13cm,
    enlarge y limits=false,
    axis lines*=left,
    ymin=0,
    ymax=100,
    title={$d=5$},
     legend style={at={(0.2,0.9)},
        anchor=north,legend columns=-1,
        /tikz/every even column/.append style={column sep=0.5cm}
        },
        ylabel={Coverge percentages},
        symbolic x coords={0,1,2,3,4,5,6},
     xtick=data,
        nodes near coords,
    every node near coord/.append style={
        anchor=mid west,
        rotate=70
    }
    ]
    \addplot [fill=RoyalBlue4] coordinates {(0,0) (1,0.04) (2,0.08)
      (3,0.42) (4,1.40) (5,8.98)(6,89.08)};
    \addplot [fill=LightGrey] coordinates {(0,0) (1,0.02) (2,0.38)
      (3,4.56) (4,21.56) (5,33.10) (6,40.38)};
    \legend{CF-A,KP}
  \end{axis}
\end{tikzpicture}}
\end{subfigure}
\begin{subfigure}[b]{0.48\textwidth}
\centering
\resizebox{\linewidth}{!}{
\pgfplotsset{title style={at={(0.5,0.91)}}}
\begin{tikzpicture}
\begin{axis}[
        ybar,
    height=10cm,
    width=13cm,
    enlarge y limits=false,
    axis lines*=left,
    ymin=0,
    ymax=100,
    title={$d=6$},
     legend style={at={(0.2,0.9)},
        anchor=north,legend columns=-1,
        /tikz/every even column/.append style={column sep=0.5cm}
        },
        ylabel={Coverge percentages},
        symbolic x coords={0,1,2,3,4,5,6},
     xtick=data,
        nodes near coords,
    every node near coord/.append style={
        anchor=mid west,
        rotate=70
    }
    ]
    \addplot [fill=RoyalBlue4] coordinates {(0,0) (1,0.08) (2,0.28)
      (3,0.70) (4,2.00) (5,12.78)(6,84.16)};
    \addplot [fill=LightGrey] coordinates {(0,0) (1,0.10) (2,2.02)
      (3,14.34) (4,30.32) (5,27.92) (6,25.30 )};
    \legend{CF-A,KP}
  \end{axis}
\end{tikzpicture}}
\end{subfigure}
\caption{The rank estimation: $\mathrm{rank}(\Pi_0)=6$, $\alpha = 5\%$ and $\delta=0.12$} \label{Fig: rank estimation2}
\end{figure}
}

We now compare with \citet{Kleibergen_Paap2006rank} in terms of estimation by making use of the same data generating process as specified by \eqref{Eqn: Motivation: DGPs} and \eqref{Eqn: Motivation: DGPs1} with $\delta=0.1$ and $0.12$ so that $\mathrm{rank}(\Pi_0)=6$ (i.e., full rank) in both cases for all $d=1,\ldots, 6$. Our estimation is based on the analytic derivative estimator \eqref{Eqn: StructDevEst} with $\hat r_n$ given by \eqref{Eqn: bootstrap overview0} and $\kappa_n=n^{-1/4}$ -- the results for $\kappa_n=n^{-1/3}$ are similar and available upon request. In each configuration, we depict the empirical distributions of the estimators based on 5,000 simulations, 500 bootstrap repetitions for each simulation, and $\alpha=5\%$. As shown by Figures \ref{Fig: rank estimation1} and \ref{Fig: rank estimation2}, our rank estimators, labelled CF-A, pick up the truth with probabilities higher than the KP estimators, uniformly over $d\in\{2,\ldots, 6\}$ and $\delta\in\{0.1,0.12\}$; when $d=1$, the two sets of estimators are very similar. Note that the empirical probabilities of $\hat r_n=r_0$ are lower in Figure \ref{Fig: rank estimation1} than in Figure \ref{Fig: rank estimation2} because $\Pi_0$ is closer to a lower rank matrix (due to a smaller value of $\delta$), and in each figure, the probabilities for both sets of estimators decrease as $\Pi_0$ becomes more degenerate (as $d$ increases). There are two additional interesting persistent patterns. First, the distributions of the KP estimators are more spread out and tend to underestimate the true rank, especially when $d$ is large, i.e., when $\Pi_0$ is local to a matrix whose rank is small. This is in accord with the trivial power of the KP test in this scenario -- see Figure \ref{Fig: CF vs KP}. Second, the probability of our rank estimators equal to the truth can exceed that of the KP rank estimator by as high as 57.84\%, and in 5 out of the 12 data generating processes considered, the probabilities of our rank estimator covering the truth are at least 48.70\% higher. Once again, this happens especially when $\Pi_0$ is local to a matrix whose rank is small. These observations suggest that our estimators are more robust to local-to-degeneracy.

\section{Saliency Analysis in Matching Models}\label{Sec: mathching}

In this section, we study a one-to-one, bipartite matching model with transferable utility, where a central question is how many attributes are statistically relevant for the sorting of agents \citep{DupuyGalichon2014Personality,CiscatoGalichonGousse2015Like}. As shall be seen shortly, this question can be answered by appealing to our framework developed previously. Following the literature, we shall call the two sets of agents men and women, though the theory obviously extends under the general setup.

\subsection{The Model Setup and Saliency Analysis}

Let $X\in\mathcal X\subset\mathbf R^m$ and $Y\in\mathcal Y\subset\mathbf R^k$ be vectors of attributes of men and women respectively, with $P_0$ and $Q_0$ the probability distributions of $X$ and $Y$ respectively. A matching is then characterized by a probability distribution $\pi$ on $\mathcal X\times\mathcal Y$ such that its density $f_\pi(x,y)$ describes the probability of occurrence of a couple with attributes $(x,y)$. Since we only consider matched couples and matching is one-to-one, $\pi$ must have marginals $P_0$ and $Q_0$. A defining feature of the transferrable utility framework is that matched couples behave unitarily, i.e., there is a single surplus function $s: \mathcal X\times\mathcal Y\to\mathbf R$ generated by the matching, and how the surplus is shared between the spouses is endogenous. A final ingredient crucial to the matching game is the equilibrium concept. As standard in the literature, we employ the notion of stability \citep{GaleShapley1962College}, and call a matching stable if (i) no matched individual would rather be single and (ii) no pair of individuals would {\it both} like being matched together better than their current situation. It is well known that stability (a game theoretical concept) and surplus maximization (a social planner's problem) are equivalent \citep{ShapleyShubik1971Assignment,ChiapporiMcCannNesheim2010Hedonic}. Consequently, the matching $\pi_0$ in equilibrium can be characterized by the centralized problem:
\begin{align}\label{Eqn: saliency, transport}
\max_{\pi\in\bm\Pi(P_0,Q_0)} E_\pi[s(X,Y)]~,
\end{align}
where $\bm\Pi(P_0,Q_0)$ is the family of distributions on $\mathcal X\times\mathcal Y$ with marginals $P_0$ and $Q_0$.

Without further appropriate modelling, the optimal transport problem \eqref{Eqn: saliency, transport} implies pure matching under regularity conditions \citep{Becker1973Marriage,ChiapporiMcCannNesheim2010Hedonic}, i.e., a certain type of men is for sure going to be matched with a certain type of women. One empirical strategy to reconcile such unrealistic predictions with data is to incorporate unobserved heterogeneity into the surplus function. Following \citet{ChooSiow2006Marries} and \citet{ChiapporiSalanieWeiss2017Partner}, we assume that
\begin{align}\label{Eqn: saliency, transport, surplus}
s(x,y)=\Phi(x,y)+\epsilon_m(y)+\epsilon_w(x)~,
\end{align} % See p.846 in the survey paper.
where $\Phi(x,y)$ is the systematic part of the surplus, and $\epsilon_m(y)$ and $\epsilon_w(x)$ are unobserved random shocks. Note that $\epsilon_m(y)$ and $\epsilon_w(x)$ enter the surplus function additively and separably, which is by no means a haphazard restriction: it makes an otherwise extremely difficult problem more tractable \citep{ChiapporiSalanie2016Matching,Chiappori2017Matching}. Nonparametric identification of both $\Phi$ and the error distributions, however, remains a challenging task. Following \citet{Dagsvik2000Aggregation} and \citet{ChooSiow2006Marries}, we thus further assume that the errors follow the type-I extreme value distribution, though we note that such distributional assumption can be completely dispensed with \citep{GalichonSalanie2015Cupid}. The matching distribution $\pi_0$ can in turn be characterized by
\begin{align}\label{Eqn: saliency transport with entropy}
\max_{\pi\in\bm\Pi(P_0,Q_0)} E_\pi[\Phi(X,Y)]-E_\pi[\log f_\pi(X,Y)]~,
\end{align}
and $\Phi$ is nonparametrically identified \citep{GalichonSalanie2015Cupid}. For the purpose of estimation, we further assume that, for some $A_0\in\mathbf M^{m\times k}$ and any $(x,y)\in\mathcal X\times\mathcal Y$,
\begin{align}\label{Eqn: saliency, bilinear form}
\Phi(x,y)\equiv\Phi_{A_0}(x,y)=x^\transpose A_0 y~,
\end{align}
where $A_0$ is called the affinity matrix. Such a parametric specification has also been employed by \citet{GalichonSalanie2010Tradeoff,GalichonSalanie2015Cupid} and \citet{DupuyGalichon2014Personality}.

Heuristically, the $(i,j)$th entry $a_{ij}$ of $A_0$ measures the strength of mutual attractiveness between attributes $x_i$ and $y_j$. The rank of $A_0$ provides valuable information on the number of dimensions on which sorting occurs, and helps construct indices of mutual attractiveness \citep{DupuyGalichon2014Personality,DupuyGalichon2015Canonical}. Following \citet{DupuyGalichon2014Personality} and \citet{GalichonSalanie2015Cupid}, we estimate $A_0$ by matching moments:
\begin{align}\label{Eqn: saliency, mathcing moments}
E_{\pi(A_0,P_0,Q_0)}[XY^\transpose]= E[XY^\transpose]~,
\end{align}
where $\pi_0\equiv \pi(A_0,P_0,Q_0)$ is the matching distribution in equilibrium. By Lemma \ref{Lem: saliency, affinity diff}, if $X$ and $Y$ are finitely discrete-valued with probability mass functions $p_0$ and $q_0$, then equation \eqref{Eqn: saliency, mathcing moments} defines not only a unique $A_0$, but also an implicit map $(p_0,q_0,E[XY^\transpose])\mapsto  A(p_0,q_0,E[XY^\transpose])\equiv A_0$ which is  differentiable. This has two immediate implications. First, the estimator $\hat A_n$ defined by the sample analog of \eqref{Eqn: saliency, mathcing moments}, i.e.,
\begin{align}\label{Eqn: saliency, mathcing moments, sample}
E_{\pi(\hat A_n,\hat p_n,\hat q_n)}[XY^\transpose]= \frac{1}{n}\sum_{i=1}^{n}X_iY_i^\transpose~,
\end{align}
where $\hat p_n$ and $\hat q_n$ are sample analogs of $p_0$ and $q_0$ respectively, is asymptotically normal. Second, the bootstrap estimator $\hat A_n^*$ defined by the bootstrap analog of \eqref{Eqn: saliency, mathcing moments, sample}, i.e.,
\begin{align}\label{Eqn: saliency, mathcing moments, bootstrap}
E_{\pi(\hat A_n^*,\hat p_n^*,\hat q_n^*)}[XY^\transpose]= \frac{1}{n}\sum_{i=1}^{n}X_i^*Y_i^{*\transpose}~,
\end{align} % Not sure if the footnote is really needed.
where $\hat p_n^*$ and $\hat q_n^*$ are bootstrap analogs of $\hat p_n$ and $\hat q_n$ respectively, is consistent in estimating the asymptotic distribution of $\hat A_n$. We have thus verified the main assumptions in order to apply our framework. We note in passing that it appears challenging to verify Assumption \ref{Ass: Boostrap: Pihat} when $X$ and $Y$ are continuous, and we believe it should be based on arguments different from those above.

% see also \citet{ArconesGine1992Bootstrap} and \citet{Hahn1996note} for bootstrap results on general M and GMM estimators.

Alternatively, \citet{DupuyGalichon2014Personality} estimate the rank of $A_0$ by employing the test of \citet{Kleibergen_Paap2006rank}, which they call the saliency analysis. There are two motivations of using our inferential procedure. First, as argued previously, the KP test is designed for the more restrictive setup \eqref{Eqn: hypothesis, literature, intro} and can be invalid and/or conservative for the hypotheses in \eqref{Eqn: hypothesis, intro}. Consequently, estimation of $\mathrm{rank}(A_0)$ by sequentially conducting the KP tests may be less accurate. Second, the KP test relies on an estimator of the asymptotic variance of $\hat A_n$ which appears to be somewhat complicated -- see the formula (B18) in \citet{DupuyGalichon2014Personality}, while one generic merit of bootstrap inference is to avoid analytic complications by repetitive resampling \citep{HorowitzBoot}.

\subsection{Data and Empirical Results}

We use the same data source as \citet{DupuyGalichon2014Personality}, i.e., the 1993-2002 waves of the DNB Household Survey, to estimate preferences in the marriage market in Dutch. The panel contains rich information about individual attributes such as demographic variables (e.g., education), anthropometry parameters (e.g., height and weight), personality traits (e.g., emotional stability, extraversion, conscientiousness, agreeableness, autonomy) and risk attitude -- see \citet{Nyhus1996VSB} for more detailed descriptions of the data. In order to apply our framework, we have discretized the variables in the following way: (i) BMI\footnote{The body mass index (BMI) is defined as the body mass divided by the square of the body height, which provides a simple numeric measure of a person's thinness.} is converted into a trinary variable according to the international BMI classification, i.e., BMI is set to be 1 if BMI $<$18.50, 2 if $18.50\le$ BMI $< 24.99$, and 3 if BMI $\ge 24.99$; (ii) Five personal traits variables and risk aversion are also converted into trinary data by taking the value 1 if they are below the corresponding 25\% quantiles, 2 if they are between the 25\% and the 75\% quantiles, and 3 if they are strictly larger than the 75\% quantiles; (iii) Education remains unchanged since it is discrete as it is. We make use of the same sample as \citet{DupuyGalichon2014Personality} which has 1158 couples, but only include subsets of the 10 attribute variables that they considered to reduce the computational burden -- see Table \ref{Tab: mathching, models}. Following \citet{DupuyGalichon2014Personality} still, we demean and standardize the data beforehand, and then compute the optimal matching distribution by the iterative projection fitting procedure \citep{Ruschendorf1995Iterative}.

% Look up the difference between (8) and (9); might want to remove (8).

\begin{table}[!h]
\caption{Model specifications}\label{Tab: mathching, models}
\begin{center}
\begin{tabular}{ccl}
\hline\hline
Model &&  Attributes included\\
\hline
(1) && Education, BMI, Risk aversion\\
(2) && Education, BMI, Risk aversion, Conscientiousness\\
(3) && Education, BMI, Risk aversion, Extraversion\\
(4) && Education, BMI, Risk aversion, Agreeableness\\
(5) && Education, BMI, Risk aversion, Emotional stability\\
(6) && Education, BMI, Risk aversion, Autonomy\\
(7) && Education, BMI, Risk aversion, Conscientiousness, Extraversion\\
(8) && Education, BMI, Risk aversion, Conscientiousness, Autonomy\\
(9) && Education, BMI, Risk aversion, Extraversion, Autonomy\\
\hline\hline
\end{tabular}
\end{center}
\end{table}

\setlength{\tabcolsep}{2pt}
\begin{table}[!h]
\caption{Empirical results}\label{Tab2: emp, matching}
\begin{center}
\begin{threeparttable}
\begin{tabular}{ccccccccccccccccc}
\hline\hline
& &  & \multicolumn{13}{c}{The $p$-values for full rank tests\tnote{\dag}}\\
\cmidrule{2-16}
&\multirow{2}{*}{Model} &\multirow{2}{*}{\makecell{Maximum\\ rank}} & \multicolumn{3}{c}{CF-T}& & \multicolumn{3}{c}{CF-A}& &\multicolumn{3}{c}{CF-N} & &\multirow{2}{*}{KP-M\tnote{\ddag}}&\\
\cmidrule{4-6}\cmidrule{8-10} \cmidrule{12-14}
&&&$\alpha/10$&$\alpha/15$&$\alpha/20$ & &$n^{-1/5}$&$n^{-1/4}$&$n^{-1/3}$ & &$n^{-1/5}$& $n^{-1/4}$ & $n^{-1/3}$&&&\\
\cmidrule{2-16}
&$(1)$ & 3 &0.00 &0.00& 0.00 &&0.00& 0.00 & 0.00 &&0.00&0.00 &0.00  &&0.00 &\\
&$(2)$ & 4 &0.01 &0.01& 0.01 &&0.00& 0.00 &0.01  &&0.00&0.00 &0.00  &&0.03 &\\
&$(3)$ & 4 &0.04 &0.04& 0.04 &&0.01& 0.04 &0.18  &&0.01&0.02 &0.04  &&0.25 &\\
&$(4)$ & 4 &0.88 &0.88& 0.88 &&0.86& 0.88 &0.92  &&0.85&0.86 &0.87  &&0.94 &\\
&$(5)$ & 4 &0.23 &0.08& 0.08 &&0.03& 0.08 &0.23  &&0.02&0.03 &0.06  &&0.35 &\\
&$(6)$ & 4 &0.01 &0.01& 0.01 &&0.00& 0.00 &0.01  &&0.00&0.00 &0.00  &&0.03 &\\
&$(7)$ & 5 &0.02 &0.02& 0.02 &&0.00& 0.02 &0.14  &&0.00&0.00 &0.01  &&0.19 &\\
&$(8)$ & 5 &0.00 &0.00& 0.00 &&0.00& 0.00 &0.01  &&0.00&0.00 &0.00  &&0.03 &\\
&$(9)$ & 5 &0.00 &0.00& 0.00 &&0.00& 0.00 &0.03  &&0.00&0.00 &0.00  &&0.22 &\\
\cmidrule{2-16}
&&  & \multicolumn{13}{c}{Estimates of the true rank ($\alpha=5\%$)}&\\
\cmidrule{2-16}
% &&&& \multicolumn{2}{c}{CF-T}& & \multicolumn{2}{c}{CF-A}& &\multicolumn{2}{c}{CF-N} & & \multicolumn{2}{c}{KP}\\
% \cmidrule{5-6}\cmidrule{8-9} \cmidrule{11-12}
% &&&& $\beta=\alpha/10$ & $\beta=\alpha/20$ & & $\kappa_n=n^{-1/4}$ & $\kappa_n=n^{-1/3}$& & $\kappa_n=n^{-1/4}$ & $\kappa_n=n^{-1/3}$ & &  \\
% \cmidrule{3-14}
&$(1)$ & 3 &3 &3 & 3 && 3& 3&3 && 3& 3&3 &&3&\\
&$(2)$ & 4 &4 &4 & 4 && 4& 4&4 && 4& 4&4 &&4&\\
&$(3)$ & 4 &4 &4 & 4 && 4& 4&3 && 4& 4&4 &&3&\\
&$(4)$ & 4 &3 &3 & 3 && 3& 3&3 && 3& 3&3 &&3&\\
&$(5)$ & 4 &3 &3 & 3 && 4& 3&3 && 4& 4&3 &&3&\\
&$(6)$ & 4 &4 &4 & 4 && 4& 4&4 && 4& 4&4 &&4&\\
&$(7)$ & 5 &5 &5 & 5 && 5& 5&4 && 5& 5&5 &&3&\\
&$(8)$ & 5 &5 &5 & 5 && 5& 5&5 && 5& 5&5 &&5&\\
&$(9)$ & 5 &5 &5 & 5 && 5& 5&5 && 5& 5&5 &&3&\\
\hline\hline
\end{tabular}
\begin{tablenotes}
      \small
      \item[\dag] The three values under CF-T are the choices of $\beta$, and those under CF-A and CF-N are the choices of $\kappa_n$ as in \ref{Eqn: bootstrap overview0} and \eqref{Eqn: NumericalDevEst} respectively.
      \item[\ddag] The $p$-value for KP-M is given by the smallest significance level such that the null hypothesis is rejected, which is equal to the maximum $p$-value of all \citet{Kleibergen_Paap2006rank}'s tests implemented by the multiple testing method.
    \end{tablenotes}
\end{threeparttable}
\end{center}
\end{table}

For each model specification, we study two problems: testing singularity of the corresponding affinity matrix and estimating its true rank. In carrying out our inferential procedures, we estimate the derivative through either \eqref{Eqn: StructDevEst} or \eqref{Eqn: NumericalDevEst}, for which we choose the tuning parameter $\kappa_n\in\{n^{-1/5},n^{-1/4},n^{-1/3}\}$. The corresponding results are labelled as CF-A and CF-N respectively. We also implement the two-step procedure with $\beta\in\{\alpha/10,\alpha/15,\alpha/20\}$, labelled as CF-T. The significance level is $\alpha=5\%$. As shown by Table \ref{Tab2: emp, matching}, our three inferential procedures yield overall consistent results, with the exception of models (3), (5) and (7). For example, for model (3), all our procedures estimate the rank to be 4, except CF-A with $\kappa_n=n^{-1/3}$ which estimates the rank to be $3$. Such discrepancies may be due to the choices of tuning parameters or finite sample variations. Nonetheless, what is comforting to us is that, in the three models, the majority of the 9 estimates point to the same rank. We also note that the $p$-values and estimates of the rank based on CF-T are the same across all three choices of $\beta$, for all model specifications except for model (5).

There are, however, noticeable differences between our results and those obtained by the KP test. First, there are sizable discrepancies between the $p$-values of our tests and those for the KP-M tests, especially for model specifications (3), (5), (7) and (9). Second, in terms of estimation, there are also marked differences. For example, for model (9), our tests unanimously estimate the rank to be 5, while the KP test estimates the rank to be 3. Similar patterns occur for models (3) and (7) for which the KP test provides a smaller rank estimator. Inspecting these differences, it seems that Extraversion is not important for matching in the Dutch marriage market according to the KP results, while our results show that it is important. Overall, we obtain estimates different from those based on \citet{Kleibergen_Paap2006rank} in 3 out of the 9 model specifications.

\section{Conclusion}\label{Sec: 5}

In this paper, we have developed a general framework for conducting inference on the rank of a matrix $\Pi_0$. The problem is of first order importance because we have shown, through an analytic example and simulation evidences, that existing tests may be invalid due to over-rejections when in truth $\mathrm{rank}(\Pi_0)$ is strictly less than the hypothesized value $r$, while their multiple testing versions, though valid, can be substantially conservative. We have then developed a testing procedure that has asymptotic exact size control, is consistent, and meanwhile accommodates the possibility $\mathrm{rank}(\Pi_0)<r$. A two-step test is proposed to mitigate the concern on tuning parameters. We also characterized classes of local perturbations under which our tests have local size control and nontrivial local power. These attractive testing properties in turn lead to more accurate rank estimators. We illustrated the empirical relevance of our results by conducting inference on the rank of an affinity matrix in a two-sided matching model.

We stress that our framework is limited to matrices of fixed dimensions and inapplicable to examples where the dimensions diverge as sample size increases. This is because Assumption \ref{Ass: Weaklimit: Pihat} is being violated in these settings, as $\Pi_0$ typically does not admit weakly convergent estimators. While we find extensions allowing varying dimensions important in, for example, many IV problems and high dimensional factor models, a thorough treatment is beyond the scope of this paper and hence left for future study.

%\phantomsection
\addcontentsline{toc}{section}{References}
\putbib
\end{bibunit}

\newpage

\begin{bibunit}

\setcounter{page}{1}
\begin{appendices}
\titleformat{\section}{\Large\center}{{\sc Appendix} \thesection}{1em}{}
\setcounter{equation}{0}
\renewcommand{\theequation}{A.\arabic{equation}}
\numberwithin{table}{section}
\setcounter{table}{0}
\renewcommand{\thetable}{\thesection.\arabic{table}}
\setcounter{figure}{0}
\renewcommand\thefigure{\thesection.\arabic{figure}}

\emptythanks

\phantomsection
\pdfbookmark[1]{Appendix Title}{title1}
\title{Supplemental Appendix to ``Improved Inference on the Rank of a Matrix''}
\author{
Qihui Chen\\ School of Management and Economics\\The \hspace{-0.03cm}Chinese \hspace{-0.03cm}University \hspace{-0.03cm}of \hspace{-0.03cm}Hong \hspace{-0.03cm}Kong, \hspace{-0.03cm}Shenzhen\\ qihuichen@cuhk.edu.cn
\and
Zheng Fang \\ Department of Economics \\ Texas A\&M University\\ zfang@tamu.edu}
\date{\today}
\maketitle

For convenience of the reader, we commence by gathering some notation that appear in the paper, most of which are standard in the literature.

\renewcommand{\arraystretch}{1}
\begin{table}[h]
\begin{center}
\begin{tabular}{ccl}
$\mathbf M^{m\times k}$ & & The space of $m\times k$ real matrices for $m,k\in\mathbf N$.\\
$I_k$ & & The identity matrix of size $k$.\\
$\mathbf 0_k, \mathbf 1_k$ &&  The $k\times 1$ vectors of zeros and ones.\\
$A^\transpose$ && The transpose of a matrix $A\in\mathbf M^{m\times k}$.\\
$\text{tr}(A)$ && The trace of a square matrix $A\in\mathbf M^{k\times k}$.\\
$\text{vec}(A)$&& The column vectorization of $A\in\mathbf M^{m\times k}$.\\
$\|A\|$ && The Frobenius norm of a matrix $A\in\mathbf M^{m\times k}$.\\
$\sigma_j(A)$ && The $j$th largest singular value of a matrix $A\in\mathbf M^{m\times k}$.\\
$\mathbb S^{m\times k}$ &&  A subset of $\mathbf M^{m\times k}$: $\mathbb S^{m\times k}\equiv\{U\in\mathbf M^{m\times k}: U^\transpose U=I_k\}$.\\
$C(T)$ && The space of continuous functions on a (topological) space $T$. \\
$\varphi: \mathbb D\twoheadrightarrow\mathbb E$ && A correspondence from a set $\mathbb D$ to another set $\mathbb E$.
\end{tabular}
\end{center}
\end{table}

Due to the fundamental role played by the singular value decomposition in the paper, we next provide a brief review and emphasize facts that are relevant to our development. Conceptually, the singular value decomposition generalizes the spectral decomposition to arbitrary (possibly rectangular) matrices. Let $\Pi\in\mathbf M^{m\times k}$ with $m\ge k$. Then the singular value decomposition of $\Pi$ is
\begin{align}
\Pi=P \Sigma Q^\transpose~,
\end{align}
where $P\in\mathbb S^{m\times m}$ and $Q\in\mathbb S^{k\times k}$ are orthornormal, and $\Sigma\in\mathbf M^{m\times k}$ is a diagonal matrix with its diagonal entries in descending order -- throughout the paper such a decomposition format is silently understood. The columns of $P$, called the left singular vectors of $\Pi$, are eigenvectors of $\Pi\Pi^\transpose$ (which is symmetric), the columns of $Q$, called the right singular vectors of $\Pi$, are eigenvectors of $\Pi^\transpose\Pi$ (which is also symmetric), and the diagonal entries of $\Sigma$, called the singular values, are the corresponding square roots of the eigenvalues of $\Pi\Pi^\transpose$ and also of $\Pi^\transpose\Pi$. Such a decomposition allows us to conclude that $\mathrm{rank}(\Pi)$ is precisely equal to the number of nonzero singular values.

The matrix $\Sigma$ is uniquely determined, though not the matrices $P$ and $Q$. If $\mathrm{rank}(\Pi)=r_0$, then we may partition $P$ as $P=[P_1,P_2]$ such that $P_1$ consists of precisely the first $r_0$ columns of $P$ that are associated with the nonzero singular values of $\Pi$; similarly we may partition $Q$ as $Q=[Q_1,Q_2]$. Then the null space of $\Pi$ is precisely the column space of $Q_2$, and the null space of $\Pi^\transpose$ is precisely the column space of $P_2$. Moreover, $P_2$ and $Q_2$ are uniquely determined respectively up to postmultiplication by $(m-r_0)\times (m-r_0)$ and $(k-r_0)\times(k-r_0)$ orthonormal matrices. Fortunately, the singular values $\sigma_j(P_2^\transpose MQ_2)$ (as in \eqref{Eqn: Nulllimt}) for any $M\in\mathbf M^{m\times k}$ are invariant to such transformations.

For convenience of applied researchers who work with Stata, we have developed a command \texttt{bootranktest} that may be used to test whether a matrix of the form $E[VZ^\transpose]$ has full rank based on our two-step test. In the first step, we use the KP test to obtain the rank estimator by choosing $\beta=0.05/15$. Its syntax is as follows:
\begin{verbatim}
bootranktest (varlist1) (varlist2) [if] [in]
\end{verbatim}
where \texttt{varlist1} should have more variables than \texttt{varlist2}. As of now, this command is designed for i.i.d.\ data and employs \citet{Efron1979}'s empirical bootstrap with 500 bootstrap repetitions. We plan to refine it by adding more features in future.

The remainder of the supplement is organized as follows. Appendix \ref{Sec: Main results} presents the proofs of our main results. Appendix \ref{Sec: comparision with KP} provides additional details and discussions regarding comparisons with \citet{Kleibergen_Paap2006rank}, while Appendix \ref{Sec: 4-3} derives some estimation results based on a sequential testing procedure. Appendix \ref{Sec: aux} contains some supporting lemmas. Additional examples are presented in Appendix \ref{Sec: B} where special attention is paid to inference on cointegration rank.

\section{Proofs of Main Results}\label{Sec: Main results}

\noindent{\sc Proof of Lemma \ref{Lem: phi, representation}:} The proof is based on a simple application of the representation of extremal partial trace. Recall that $\sigma^{2}_1(\Pi),\ldots,\sigma^{2}_k(\Pi)$ are eigenvalues of $\Pi^\transpose\Pi$ in descending order. Let $d\equiv k-r$. It follows by Proposition 1.3.4 in \citet{Tao2012Matrix} that
\begin{align}\label{Eqn: extremal partial trace}
\phi_r(\Pi)=\sum_{j=r+1}^k\sigma^{2}_j(\Pi)=\inf_{u_1,\ldots,u_{d}}\sum_{j=1}^{d}u_j^\transpose\Pi^\transpose\Pi u_j~,
\end{align}
where the infimum is taken over all $u_1,\ldots,u_{d}\in\mathbf R^k$ that are orthonormal. Noting $U\equiv [u_1,\ldots,u_{d}]\in\mathbb S^{k\times {d}}$, we obtain by \eqref{Eqn: extremal partial trace} and the definition of Frobenius norm that
\begin{align}\label{Eqn: extremal partial trace2}
\phi_r(\Pi)=\inf_{U\in\mathbb S^{k\times {d}}}\text{tr}(U^\transpose\Pi^\transpose\Pi U)=\inf_{U\in\mathbb S^{k\times {d}}}\|\Pi U\|^2~.
\end{align}
The infimum in \eqref{Eqn: extremal partial trace2} is achieved on $\mathbb S^{k\times {d}}$ because $U\mapsto\|\Pi U\|^2$ is continuous, and $\mathbb S^{k\times {d}}$ is compact since it is closed and bounded. This completes the proof of the lemma. \qed

\noindent{\sc Proof of Proposition \ref{Pro: phi, differentiability}:} Let $d\equiv k-r$, and define $\psi_1:\mathbf M^{m\times k}\to C(\mathbb S^{k\times d})$ by $\psi_1(\Pi)(U)=\|\Pi U\|^2$, and $\psi_2: C(\mathbb S^{k\times d})\to \mathbf R$ by $\psi_2(f)=\min\{f(U): U\in\mathbb S^{k\times d}\}$, so that $\phi_r=\psi_2\circ\psi_1$ by Lemma \ref{Lem: phi, representation}. For part (i), we proceed by verifying first order Hadamard directional differentiability of $\psi_1$ and $\psi_2$, and then conclude by the chain rule.

Let $\{M_n\}\subset\mathbf M^{m\times k}$ be a sequence satisfying $M_n\to M\in \mathbf M^{m\times k}$, and $t_n\downarrow 0$ as $n\to\infty$. For each $n\in\mathbf N$, define $g_n:\mathbb S^{k\times d}\to\mathbf R$ by
\[
g_n(U)=\frac{\|(\Pi+t_n M_n)U\|^2-\|\Pi U\|^2}{t_n}=\frac{\|\Pi U+t_n M_nU\|^2-\|\Pi U\|^2}{t_n}~,
\]
and $g:\mathbb S^{k\times d}\to\mathbf R$ by $g(U)=2\text{tr}((\Pi U)^\transpose MU)$. Then by simple algebra we have
\begin{multline}\label{Eqn: Rankcon HDD, aux1}
\sup_{U\in\mathbb S^{k\times d}}|g_n(U) - g(U)|=\sup_{U\in\mathbb S^{k\times d}}| 2\text{tr}((\Pi U)^\transpose (M_{n} - M)U) + t_n\|M_n U\|^2 | \\
\le \sup_{U\in\mathbb S^{k\times d}}\{2\|\Pi U\|\|(M_{n} - M)U\|+t_n\|M_nU\|^2\}~,
\end{multline}
where the inequality follows by the triangle inequality and the Cauchy-Schwarz inequality for the trace operator. For the right hand side of \eqref{Eqn: Rankcon HDD, aux1}, we further have
\begin{multline}\label{Eqn: Rankcon HDD, aux1-1}
 \sup_{U\in\mathbb S^{k\times d}}\{2\|\Pi U\|\|(M_{n} - M)U\|+t_n\|M_nU\|^2\}\\
\le \sup_{U\in\mathbb S^{k\times d}}\{2\|\Pi\|\| U\|\|M_{n} - M\|\|U\|+t_n\|M_n\|^2\|U\|^2\}=o(1)~,
\end{multline}
where we exploited the sub-multiplicativity of Frobenius norm and the facts that $\|U\|=\sqrt{d}$, $M_n\to M$ and $t_n\downarrow 0$ as $n\to\infty$. We thus conclude from \eqref{Eqn: Rankcon HDD, aux1} and \eqref{Eqn: Rankcon HDD, aux1-1} that $g_n\to g$ uniformly in $C(\mathbb S^{k\times d})$, or equivalently $\psi_1$ is first order Hadamard directionally differentiable at $\Pi$ with derivative $\psi_{1,\Pi}': \mathbf M^{m\times k}\to C(\mathbb S^{k\times d})$ given by
\begin{align}\label{Eqn: Rankcon HDD, phi1}
\psi_{1,\Pi}'(M)(U)=2\text{tr}((\Pi U)^\transpose MU)~.
\end{align}
On the other hand, Theorem 3.1 in \citet{Shapiro1991} implies that $\psi_2: C(\mathbb S^{k\times d})\to \mathbf R$ is first order Hadamard directionally differentiable at any $f\in C(\mathbb S^{k\times d})$ with derivative $\psi_{2,f}': C(\mathbb S^{k\times d})\to \mathbf R$ given by: for $\Psi(f)\equiv\arg\min_{U\in\mathbb S^{k\times d}}f(U)$,
\begin{align}\label{Eqn: Rankcon HDD, phi2}
\psi_{2,f}'(h)=\min_{U\in\Psi(f)}h(U)~.
\end{align}
Combining \eqref{Eqn: Rankcon HDD, phi1}, \eqref{Eqn: Rankcon HDD, phi2} and the chain rule \citep[Proposition 3.6]{Shapiro1990}, we may now conclude that $\phi_r: \mathbf M^{m\times k} \to\mathbf R$ is first order Hadamard directionally differentiable at any $\Pi\in\mathbf M^{m\times k}$ with the derivative $\phi_{r,\Pi}': \mathbf M^{m\times k}\to\mathbf R$ given by
\begin{align*}
\phi_{r,\Pi}'(M)=\psi_{2,\psi_1(\Pi)}'\circ\psi_{1,\Pi}'(M)=\min_{U\in\Psi(\Pi)} 2\text{tr}((\Pi U)^\transpose M U)~.
\end{align*}
This completes the proof of part (i) of the proposition.

For part (ii), note that $\phi_r(\Pi)=0$ implies that $\Pi U=0$ for all $U\in\Psi(\Pi)$ and hence $\phi_{r,\Pi}'(M)=0$ for all $M\in \mathbf M^{m\times k}$. Recall that $\{M_n\}\subset\mathbf M^{m\times k}$ with $M_n\to M\in \mathbf M^{m\times k}$ and $t_n\downarrow 0$ as $n\to\infty$. By Lemma \ref{Lem: phi, representation} we have
\begin{multline}\label{Eqn: Rankcon HDD, aux3}
\vert \phi_r(\Pi+t_{n}M_{n})-\phi_r( \Pi+t_{n}M)\vert
=\big\vert \min_{U\in \mathbb S^{k\times d}}\Vert(\Pi +t_{n}M_{n})U\Vert-\min_{U\in \mathbb S^{k\times d}}\Vert ( \Pi +t_{n}M)U\Vert \big\vert \\
\times \big(\min_{U\in \mathbb S^{k\times d}}\Vert(\Pi +t_{n}M_{n})U\Vert+\min_{U\in \mathbb S^{k\times d}}\Vert ( \Pi +t_{n}M)U\Vert\big)~,
\end{multline}
where the equality also exploited the elementary formula $a^{2}-b^{2} = (a+b)(a-b)$. For the first term on the right hand side of \eqref{Eqn: Rankcon HDD, aux3}, we have
\begin{align}\label{Eqn: Rankcon HDD, aux3-1}
\big\vert \min_{U\in \mathbb S^{k\times d}}\Vert(\Pi +t_{n}M_{n})U\Vert-\min_{U\in \mathbb S^{k\times d}}\Vert ( \Pi +t_{n}M)U\Vert \big\vert\leq t_{n}\sqrt{d}\Vert M_{n}-M\Vert=o(t_{n})~,
\end{align}
where the inequality follows by the Lipschitz continuity of the infimum operator, the triangle inequality, $\|\cdot\|$ being sub-multiplicative, and $\|U\|=\sqrt{d}$ for $U\in\mathbb S^{k\times d}$. For the second term on the right hand side of \eqref{Eqn: Rankcon HDD, aux3}, we have: for any fixed $U^{\ast}\in\Psi(\Pi)$,
\begin{multline}\label{Eqn: Rankcon HDD, aux3-2}
\min_{U\in \mathbb S^{k\times d}}\Vert(\Pi +t_{n}M_{n})U\Vert+\min_{U\in \mathbb S^{k\times d}}\Vert ( \Pi +t_{n}M)U\Vert\leq \Vert(\Pi +t_{n}M_{n})U^{\ast}\Vert\\
+\Vert ( \Pi +t_{n}M)U^{\ast}\Vert\leq t_{n}\Vert M_{n}\Vert \Vert U^{\ast}\Vert + t_{n}\Vert M\Vert \Vert U^{\ast}\Vert=O(t_n)~,
\end{multline}
where we exploited $\Pi U^{\ast}=0$, the sub-multiplicativity of Frobenius norm, $\|U^{\ast}\|=\sqrt{d}$ and $M_n\to M$ as $n\to\infty$. Combining \eqref{Eqn: Rankcon HDD, aux3}-\eqref{Eqn: Rankcon HDD, aux3-2}, we thus obtain
\begin{align}\label{Eqn: Rankcon HDD, aux3-3}
\vert \phi_r(\Pi+t_{n}M_{n})-\phi_r( \Pi+t_{n}M)\vert = o(t_{n}^{2})~.
\end{align}

Next, for $\epsilon>0$, let $\Psi(\Pi)^{\epsilon}\equiv \{U\in \mathbb S^{k\times d}: \min_{U'\in\Psi(\Pi)}\Vert U'-U\Vert\leq\epsilon\}$ and $\Psi(\Pi)_{1}^{\epsilon}\equiv \{U\in \mathbb S^{k\times d}: \min_{U'\in\Psi(\Pi)}\Vert U'-U\Vert\geq\epsilon\}$. In what follows we consider the nontrivial case when $\Pi\neq 0$ and $M\neq 0$. Then we must have $\Psi(\Pi)\subsetneqq \mathbb S^{k\times d}$ and hence $\Psi(\Pi)_1^\epsilon\neq \emptyset$ for $\epsilon$ sufficiently small. Let $\sigma_{\min}^{+}(\Pi)$ denote the smallest positive singular value of $\Pi$ which exists since $\Pi\neq 0$, and $\Delta\equiv3\sqrt{2}[\sigma_{\min}^{+}(\Pi)]^{-1}\max_{U\in \mathbb S^{k\times d}}$ $\Vert MU\Vert>0$ since $M\neq 0$. Then it follows that for all $n$ sufficiently large
\begin{align}\label{Eqn: Rankcon HDD, aux4}
\min_{U\in \Psi(\Pi)_{1}^{t_{n}\Delta}}\Vert ( \Pi +t_{n} M)U\Vert & \ge \min_{U\in\Psi(\Pi)_{1}^{t_{n}\Delta}}\Vert\Pi U\Vert-t_{n}\max_{U\in \mathbb S^{k\times d}}\Vert M U\Vert \notag\\
  & \ge\frac{\sqrt 2}{2} t_{n}\sigma_{\min}^{+}(\Pi)\Delta -t_{n}\max_{U\in \mathbb{S}^{k\times d}}\Vert MU\Vert > t_{n}\max_{U\in \mathbb S^{k\times d}}\Vert MU\Vert \notag\\
  & \ge \min_{U\in \Psi(\Pi)}\|( \Pi +t_{n} M)U\|\ge\sqrt{\phi_r(\Pi+t_nM)}~,
\end{align}
where the first inequality follows by the triangle inequality and the fact that $\Psi(\Pi)_{1}^{t_{n}\Delta}\subset S^{k\times d}$, the second inequality follows by Lemma \ref{Lem: phi, diff lemma, ID}, the third inequality is due to the definition of $\Delta$, and the fourth inequality holds by the fact that $\Pi U=0$ for $U\in \Psi(\Pi)$. By \eqref{Eqn: Rankcon HDD, aux4}, we thus obtain that, for all $n$ sufficiently large
\begin{eqnarray}\label{Eqn: Rankcon HDD, aux5}
\phi_r( \Pi+t_{n} M) = \min_{U\in \Psi(\Pi)^{t_{n}\Delta}}\Vert ( \Pi +t_{n} M)U\Vert^{2}~.
\end{eqnarray}

Now, for fixed $U\in\Psi(\Pi)$, $\Delta>0$ and $t\in\mathbf R$, let $\Gamma^\Delta\equiv\{V\in\mathbf M^{k\times d}: \|V\|\le\Delta\}$ and $\Gamma_{U}^{\Delta}(t)\equiv \{V\in \Gamma^\Delta: U+tV\in \mathbb{S}^{k\times d}\}=\{V\in \Gamma^\Delta: V^\transpose U+U^\transpose V=-tV^\transpose V\}$. Define a correspondence $\varphi: \mathbf R\twoheadrightarrow\mathbb S^{k\times d}\times\Gamma^\Delta$ by $\varphi(t)=\{(U,V): U\in\Psi(\Pi), V\in\Gamma_U^\Delta(t)\}$. Then the right hand side of \eqref{Eqn: Rankcon HDD, aux5} can be written as
\begin{multline}\label{Eqn: Rankcon HDD, aux6}
 \min_{U\in \Psi(\Pi)^{t_{n}\Delta}}\Vert ( \Pi +t_{n} M)U\Vert^{2} = \min_{(U,V)\in\varphi(t_n)}\Vert( \Pi +t_{n}M)(U+t_{n}V)\Vert^{2}\\
=t_{n}^{2}\min_{(U,V)\in\varphi(t_n)}\Vert\Pi V+MU\Vert^{2}+o(t_{n}^{2})~,
\end{multline}
where we exploited $\Pi U = 0$ for all $U\in\Psi(\Pi)$ and $\|M V\|\leq \|M\|\Delta$ for all $V\in\Gamma^\Delta$. By Lemma \ref{Lem: phi, auxlem, correspondence}, $\varphi(t)$ is continuous at $t=0$. Since $\varphi$ is obviously compact-valued, we may then obtain by Theorem 17.31 in \citet{AliprantisandBorder2006} that
\begin{multline}\label{Eqn: Rankcon HDD, aux8}
\min_{(U,V)\in\varphi(t_n)}\Vert\Pi V+MU\Vert^{2} =\min_{(U,V)\in\varphi(0)}\Vert\Pi V+MU\Vert^{2}+o(1)\\
 =\min_{U\in \Psi(\Pi)}\min_{V\in\mathbf{M}^{k\times d}}\Vert\Pi V+MU\Vert^{2}+o(1)~,
\end{multline}
where the second equality holds by letting $\Delta$ sufficiently large in view of Lemma \ref{Lem: Rankcon, projection}. Combining \eqref{Eqn: Rankcon HDD, aux5}, \eqref{Eqn: Rankcon HDD, aux6} and \eqref{Eqn: Rankcon HDD, aux8} then yields
\begin{align}\label{Eqn: Rankcon HDD, aux9}
\phi_r( \Pi+t_{n} M)=t_n^2 \min_{U\in \Psi(\Pi)}\min_{V\in\mathbf{M}^{k\times d}}\Vert\Pi V+MU\Vert^{2}+o(t_n^2)~.
\end{align}
The proposition now follows from result \eqref{Eqn: Rankcon HDD, aux9} and Lemma \ref{Lem: phi, derivative, reprentation}. \qed

\noindent{\sc Proof of Theorem \ref{thm: Weaklimit: Test}:} The first and second results are respectively straightforward implications of Theorems 2.1 in \citet{Fang_Santos2014HDD} and \citet{Chen_Fang2015FOD} by noting that $\phi^{\prime}_{r,\Pi_{0}}=0$ under $\mathrm{H}_{0}$. In particular, their Assumptions 2.1 are satisfied in view of Proposition \ref{Pro: phi, differentiability} and their Assumptions 2.2 are satisfied by Assumption \ref{Ass: Weaklimit: Pihat}. \qed

\noindent{\sc Proof of Theorem \ref{Thm: SizePowerTest}:} By the rate conditions on $\{\kappa_n\}$ and Assumption \ref{Ass: Weaklimit: Pihat}, the numerical estimator \eqref{Eqn: NumericalDevEst} satisfies the condition \eqref{Eqn: derivative consistency} by Proposition 3.1 in \citet{Chen_Fang2015FOD}, while the analytic estimator in \eqref{Eqn: StructDevEst} and \eqref{Eqn: bootstrap overview0} does so by Lemma \ref{Lem: Consistency: DerivEst}. In turn, following exactly the same proof of Corollary 3.2 in \citet{Fang_Santos2014HDD}, we obtain that $\hat{c}_{n,1-\alpha}\convp c_{1-\alpha}$ by Assumption \ref{Ass: Boostrap: Pihat} and the quantile restrictions on $c_{1-\alpha}$. Thus, under $\mathrm{H}_{0}$, the first claim follows from combining Theorem \ref{thm: Weaklimit: Test}, Slutsky's lemma, $c_{1-\alpha}$ being a continuity point of the limiting law and the portmanteau theorem.

For the second claim, Consider first the numerical estimator \eqref{Eqn: NumericalDevEst}. Note that by Assumption \ref{Ass: Boostrap: Pihat}, $\hat{\mathcal M}_n^*=O_{P_W}(1)$ in $P_X$-probability. Together with Assumption \ref{Ass: Weaklimit: Pihat}, $\kappa_n=o(1)$ as $n\to\infty$ and continuity of $\phi_r$, we in turn see that, in $P_X$-probability,
\begin{align}\label{Eqn: test consistency, aux44}
\phi_r(\hat\Pi_n+\kappa_n\hat{\mathcal M}_n^*)=O_{P_W}(1)~.
\end{align}
By the definition of $\hat c_{n,1-\alpha}$, it follows from \eqref{Eqn: test consistency, aux44} and $\phi_r(\hat\Pi_n)\ge 0$ that
\begin{align}\label{Eqn: test consistency, aux444}
\kappa_n^2 \hat c_{n,1-\alpha}\le O_{P_W}(1)
\end{align}
in $P_X$-probability. By Assumption \ref{Ass: Weaklimit: Pihat} and continuity of $\phi_r$ at $\Pi_0$, we have: under $\mathrm H_1$,
\begin{align}\label{Eqn: test consistency, aux4}
\phi_r(\hat\Pi_n)\convp \phi_r(\Pi_0)>0~.
\end{align}
Combining results \eqref{Eqn: test consistency, aux444} and \eqref{Eqn: test consistency, aux4}, together with $\tau_n\kappa_n\to\infty$, we thus conclude that
\begin{align}\label{Eqn: test consistency, aux2}
P(\tau_{n}^2\phi_r(\hat\Pi_n)> \hat c_{n,1-\alpha})=P((\tau_{n}\kappa_n)^2\phi_r(\hat\Pi_n)> \kappa_n^2\hat c_{n,1-\alpha})=1~.
\end{align}

For the analytic estimator, let $\hat{d}_{n}\equiv k -\hat{r}_{n}$ and $d\equiv k-r$. By Lemma \ref{Lem: phi, representation}, we have
\begin{align}\label{Eqn: test consistency, aux22}
\hat\phi_{r,n}''(\hat{\mathcal M}_n^*)=\min_{U\in \mathbb S^{\hat{d}_{n}\times d}}\|\hat P_{2,n}^\transpose \hat{\mathcal M}_n^*\hat Q_{2,n} U\|^{2}\le \|\hat{\mathcal M}_n^*\|^{2} m k d~,
\end{align}
where the second inequality exploited $\|\hat{P}_{2,n}^{\transpose}\|^{2}\|\hat{Q}_{2,n}\|^{2}\leq mk$ and $\|U\|^2= d$. Since $\hat{\mathcal M}_n^*=O_{P_W}(1)$ in $P_X$-probability by Assumption \ref{Ass: Boostrap: Pihat}, it follows from \eqref{Eqn: test consistency, aux22} that
\begin{align}\label{Eqn: test consistency, aux3}
\hat c_{n,1-\alpha}\le O_{P_W}(1)
\end{align}
in $P_X$-probability. Combining \eqref{Eqn: test consistency, aux4} and \eqref{Eqn: test consistency, aux3}, together with $\tau_n\to\infty$, we thus obtain
\begin{align}
P(\tau_{n}^2\phi_r(\hat\Pi_n)> \hat c_{n,1-\alpha})=1~.
\end{align}
This completes the proof of the second claim. \qed

\noindent{\sc Proof of Theorem \ref{Thm: two-step}:} For notational simplicity, define
\begin{align}
A_n=\{\hat r_n>r\}~, B_n=\{\tau_n^2\phi_r(\hat\Pi_n)>\hat c_{n,1-\alpha+\beta}\}~, C_n=\{\hat r_n=r_0\}~.
\end{align}
It follows that, under $\mathrm H_0$,
\begin{align}
\limsup_{n\to\infty} E[\psi_n] & \le \limsup_{n\to\infty} P((A_n\cup B_n)\cap C_n) + \limsup_{n\to\infty}  P((A_n\cup B_n)\cap C_n^c)\notag\\
&\le \limsup_{n\to\infty}  P(A_n\cap C_n) + \limsup_{n\to\infty}  P(B_n\cap C_n) + \limsup_{n\to\infty}  P(C_n^c)\notag\\
&\le 0+\alpha-\beta+\beta=\alpha~,
\end{align}
where we exploited $A_n\cap C_n=\emptyset$ under $\mathrm{H}_0$, $\limsup_{n\to\infty} P(B_n\cap C_n)\le \alpha-\beta$ by Theorem \ref{Thm: SizePowerTest}, and $\limsup_{n\to\infty}  P(C_n^c)\le\beta$. This completes the proof of the first claim. For the second claim of the theorem, note that
\begin{align}
\liminf_{n\to\infty} E[\psi_n]\ge \liminf_{n\to\infty} P(\tau_n^2\phi_r(\hat\Pi_n)>\hat c_{n,1-\alpha+\beta})=1~,
\end{align}
where the equality follows by the proof of Theorem \ref{Thm: SizePowerTest}. \qed

\noindent{\sc Proof of Proposition \ref{Pro: Local}:} By Assumption \ref{Ass: Local}(ii)(iii), we have
\begin{align}
\tau\{\hat\Pi_n-\Pi_0\}=\tau_n\{\hat\Pi_n-\Pi_{0,n}\}+\tau_n\{\Pi_{0,n}-\Pi_0\}\convl \mathcal M+\Delta~.
\end{align}
This in turn allows us to conclude by Proposition \ref{Pro: phi, differentiability} and $\phi_r(\Pi_0)=0$. \qed

\section{Comparisons with the KP Test}\label{Sec: comparision with KP}
\renewcommand{\theequation}{B.\arabic{equation}}
\setcounter{equation}{0}

In this section, we first review the KP test for the reader's convenience, and then provide additional results regarding comparisons with \citet{Kleibergen_Paap2006rank}.

\iffalse
Suppose that the singular value decomposition of $\hat{\Pi}_{n}$ is given by
\begin{align}\footnotesize
\hat\Pi_n=\underbracket[0.4pt][1pt]{\hat{P}_{n}}_{m\times m}\underbracket[0.4pt][1pt]{\hat{\Sigma}_{n}}_{m\times k}\underbracket[0.4pt][1pt]{\hat{Q}_{n}^{\transpose}}_{k\times k}=\begin{bmatrix}
\underbracket[0.4pt][1pt]{\hat{P}_{1,n}}_{m\times r} & \underbracket[0.4pt][1pt]{\hat{P}_{2,n}}_{m\times (m-r)}
\end{bmatrix}\begin{bmatrix}
\underbracket[0.4pt][1pt]{\hat\Sigma_{1n}}_{r\times r} & 0\\
 0 & \underbracket[0.4pt][1pt]{\hat\Sigma_{2n}}_{(m-r)\times (k-r)}
\end{bmatrix}\begin{bmatrix}
\underbracket[0.4pt][1pt]{\hat{Q}_{1,n}}_{k\times r} & \underbracket[0.4pt][1pt]{\hat{Q}_{2,n}}_{k\times (k-r)}
\end{bmatrix}^\transpose
\end{align}
\fi

To describe the KP test, let $\hat{\Pi}_{n}$ be an estimator for $\Pi_{0}\in\mathbf{M}^{m\times k}$ such that
\begin{align}\label{Eqn: KP statistic, vec limit}
\sqrt{n}\{\mathrm{vec}(\hat\Pi_n)-\mathrm{vec}(\Pi_0)\}\convl N(0, \Omega_0)~,
\end{align}
where the covariance matrix $\Omega_0$ admits a consistent estimator $\hat\Omega_n$. Let $\hat{\Pi}_{n}=\hat{P}_{n}\hat{\Sigma}_{n}\hat{Q}_{n}^{\transpose}$ be a singular value decomposition of $\hat{\Pi}_{n}$, where $\hat{P}_{n}\in\mathbb{S}^{m\times m}$, $\hat{Q}_{n}\in\mathbb{S}^{k\times k}$, and $\hat{\Sigma}_{n}\in\mathbf{M}^{m\times k}$ is diagonal with diagonal entries in descending order. For $r$ the hypothesized value in \eqref{Eqn: hypothesis, literature, intro}, rewrite $\hat{P}_{n}=[\hat{P}_{1,n},\hat{P}_{2,n}]$ and $\hat{Q}_{n}=[\hat{Q}_{1,n},\hat{Q}_{2,n}]$ with $\hat{P}_{1,n}\in\mathbf{M}^{m\times r}$ and $\hat{Q}_{1,n}\in\mathbf{M}^{k\times r}$, and let $\hat{\Sigma}_{2,n}$ be the right bottom $(m-r)\times(k-r)$ submatrix of $\hat{\Sigma}_{n}$. Then the testing statistic proposed by \citet{Kleibergen_Paap2006rank} for the hypotheses \eqref{Eqn: hypothesis, literature, intro} is
\begin{align}\label{Eqn: KP statistic}
T_{n,\mathrm{kp}}=n \cdot\mathrm{vec}(\hat{\Sigma}_{2,n})^{\transpose}[(\hat{Q}_{2,n}\otimes\hat{P}_{2,n})^{\transpose}\hat{\Omega}_{n}(\hat{Q}_{2,n}\otimes\hat{P}_{2,n})]^{-1}\mathrm{vec}(\hat{\Sigma}_{2,n})~,
\end{align}
where $\otimes$ signifies the Kronecker product, and the inverse is assumed to exist asymptotically. A special case of the testing statistic designed by \citet{Robin_Smith2000rank} shares exactly the same form but without the weighting matrix,\footnote{\citet{Robin_Smith2000rank} propose a class of testing statistics (indexed by functions $h$ in their paper) which are asymptotically equivalent.} i.e.,
\begin{align}\label{Eqn: RS statistic}
T_{n,\mathrm{rs}}=n \cdot\mathrm{vec}(\hat{\Sigma}_{2,n})^{\transpose}\mathrm{vec}(\hat{\Sigma}_{2,n})~.
\end{align}
\citet{Kleibergen_Paap2006rank} show that if $\mathrm{rank}(\Pi_0)=r$, then
\begin{align}\label{Eqn: KPTestStat: WeakLimit0}
T_{n,\mathrm{kp}}\convl\chi^2((m-r)(k-r))~.
\end{align}
Thus, the KP test rejects the null $\mathrm{H}_0'$ in \eqref{Eqn: hypothesis, literature, intro} at the significance level $\alpha$ if $T_{n,\mathrm{kp}}$ is larger than the $(1-\alpha)$-quantile of $\chi^2((m-r)(k-r))$.

In Section \ref{Sec: 2}, we have shown that the KP test may be invalid since the $\chi^2$-limit of the KP statistic is derived under $\mathrm{H}_0'$, ignoring the possibility $\mathrm{rank}(\Pi_0)<r$. As an alternative, one may construct a valid test for \eqref{Eqn: hypothesis, intro} by a multiple test on $\mathrm{rank}(\Pi_0)=0,1,\ldots,r$. Indeed, to show the validity of a multiple test, let $\psi_{n,r}$ be a nonrandomized test for hypotheses of the form \eqref{Eqn: hypothesis, literature, intro} that rejects the null if $\psi_{n,r}=1$ and fails to reject if $\psi_{n,r}=0$. Moreover, suppose that $\psi_{n,r}$ is a consistent test that has asymptotic null rejection rates exactly equal to $\alpha$. Then one may design a valid multiple test $\psi_n$ for \eqref{Eqn: hypothesis, intro} by setting $\psi_n=\prod_{j=0}^{r}\psi_{n,j}$, i.e., $\psi_n$ rejects $\mathrm H_0$ if and only if all $\psi_{n,j}$'s reject. It follows that $\psi_n$ has size control because, under $\mathrm H_0$ and for $r_0\equiv\mathrm{rank}(\Pi_0)$,
\begin{align}\label{Eqn: multiple test, valid}
\limsup_{n\to\infty} E[\psi_n]=\limsup_{n\to\infty} P(\psi_{n,0}=1,\ldots,\psi_{n,r}=1)\le \limsup_{n\to\infty} P(\psi_{n,r_0}=1)=\alpha~,
\end{align}
and that $\psi_n$ is also consistent because, under $\mathrm H_1$,
\begin{multline}
\liminf_{n\to\infty} E[\psi_n]=\liminf_{n\to\infty} P(\psi_{n,0}=1,\ldots,\psi_{n,r}=1)\\ \ge 1-\sum_{j=1}^{r}[1-\liminf_{n\to\infty} P(\psi_{n,j}=1)]=1~,
\end{multline}
where the inequality holds by the Boole's inequality and consistency of each $\psi_{n,j}$. This shows that $\psi_n$ is valid, in fact consistent but may be conservative. The source of conservativeness of $\psi_n$ is inherent in the inequality of \eqref{Eqn: multiple test, valid} which is generically strict. Moreover, $\psi_n$ is conservative whenever $\psi_{n,r}$ is, because
\begin{align}
\limsup_{n\to\infty} E[\psi_n]=\limsup_{n\to\infty} P(\psi_{n,0}=1,\ldots,\psi_{n,r}=1)\le \limsup_{n\to\infty} P(\psi_{n,r}=1)<\alpha~.
\end{align}

The remainder of this section is devoted to additional comparisons of our tests with the KP test based on the simulation designs and empirical application in \citet{Kleibergen_Paap2006rank}. First, following those authors, we consider, for $R_t\in\mathbf R^{10}, F_t\in\mathbf R^{4}$,
\begin{align}\label{Eqn: DGP2}
R_{t} = \Pi_{0}F_{t}+\varepsilon_{t}~,
\end{align}
where $\{F_{t}\}\overset{i.i.d.}{\sim}N(0,\Sigma_{F})$ and $\{\epsilon_t\}$ are independently generated according to
\begin{align}
\varepsilon_{t} =v_{t}+\Gamma v_{t-1}
\end{align}
with $\{v_{t}\}\overset{i.i.d.}{\sim}N(0,\Sigma_{v})$. We are interested in $\Pi_0$ which is specified as
\begin{align}
\Pi_{0}=\beta\alpha^{\transpose}+\delta\Pi_{1}~,
\end{align}
where $\delta\in\mathbf R$, $\alpha\in\mathbf{R}^{4}$, $\beta\in\mathbf{R}^{10}$ and $\Pi_{1}\in\mathbf{M}^{10\times 4}$. \citet{Kleibergen_Paap2006rank} try a wide range of values for $\delta$; we shall focus on $\delta=0,0.01,\ldots,0.1$ since we are concerned with local power. Other unknown parameters involved are configured to be exactly the same as those in \citet{Kleibergen_Paap2006rank}:
\begin{itemize}
\item $\Sigma_{F}$ is specified as the sample correlation matrix of $\{F_{t}\}_{t=1}^{n}$, where $\{F_{t}\}_{t=1}^{n}$ is the real data to be studied for the empirical application;

\item $\alpha =(0.0813, -0.0271, -0.6203, -0.0460)^{\transpose}$;

\item $\beta =  (-0.3411,  -0.1277,  -0.3838,  -0.5312,  -0.2728,  -0.3527, -0.2188, -0.293,$\\ $ -0.2035,-0.3427)^{\transpose}$;

\item $\Pi_{1}=\bar{\Pi}_{n}-\beta\alpha^{\transpose}$, where $\bar{\Pi}_{n}=\sum_{t=1}^{n}R_{t}F_{t}^{\transpose}(\sum_{t=1}^{n}F_{t}F_{t}^{\transpose})^{-1}$ with $\{R_{t},F_{t}\}_{t=1}^{n}$ being the real data in the empirical application;

\item $\Gamma$ is specified as {\setlength{\arraycolsep}{0.5pt}\small
\begin{align*}
\Gamma & = \left[
                   \begin{array}{rrrrrrrrrr}
0.0312  &0.0255 &-0.0185&0.0591 &0.0389 &0.0953	&-0.1515&0.2286	&-0.0806&-0.1659\\
0.0346	&-0.0166&-0.0608&0.0743	&0.0794 &-0.0043&-0.2194&0.2959	&-0.0043&0.0016\\
-0.0304	&0.0624	&-0.1347&0.1054	&-0.0369&-0.0187&-0.0989&0.3571	&0.0133	&-0.1731\\
-0.0414	&0.0951	&0.0029	&-0.0497&-0.0586&0.0910	&-0.0903&0.1850	&0.0616	&-0.0865\\
-0.0570	&-0.0845&0.0606	&-0.0143&-0.1971&0.0528	&0.0403	&0.1935	&-0.0114&0.1141\\
-0.0649	&-0.0738&0.0030	&0.0335	&0.0346 &-0.0432&-0.0787&0.2199	&-0.0266&-0.0013\\
-0.0334	&-0.1163&-0.0139&-0.0218&-0.0390&0.0128	&-0.0645&0.1299	&0.1105	&0.0097\\
-0.1029	&0.0368	&0.0737	&-0.0005&-0.1686&0.0254	&0.0184	&0.0966	&-0.0176&0.0596\\
-0.1153	&0.0008	&0.0373	&0.0185	&-0.0927&0.1029	&0.0546	&0.0529	&-0.1792&0.0798\\
-0.0737	&-0.0669&0.0500	&0.1466	&-0.1359&0.0617	&0.1090	&0.0402	&-0.0659&-0.0440\\
                   \end{array}
                 \right]~;
\end{align*}}

\item $\Sigma_{v}$ is specified as {\small
\begin{align*}
\Sigma_{v} & = \frac{1}{100}\left[
             \begin{array}{rrrrrrrrrr}
0.19&	0.09&	0.07&	0.05&	0.04&	0.03&	0.02&	-0.01&	0.00&	-0.01\\
0.09&	0.11	&0.06&	0.05&	0.04&	0.04&	0.03&	0.01&	0.02&	0.01\\
0.07&	0.06&	0.10&	0.05&	0.04&	0.04&	0.03&	0.03&	0.02&	0.01\\
0.05&	0.05&	0.05&	0.08&	0.04&	0.04&	0.04&	0.03&	0.02&	0.01\\
0.04&	0.04	&0.04	&0.04	&0.08	&0.05&	0.05&	0.05&	0.04&	0.03\\
0.03&	0.04&	0.04&	0.04&	0.05&	0.08&	0.06&	0.05&	0.05&	0.03\\
0.02&	0.03&	0.03&	0.04&	0.05&	0.06&	0.08&	0.06&	0.05&	0.03\\
-0.01&	0.01&	0.03&	0.03&	0.05&	0.05&	0.06&	0.10&	0.07&	0.05\\
0.00&	0.02	&0.02&	0.02	&0.04	&0.05&	0.05&	0.07&	0.09&	0.04\\
-0.01&	0.01&	0.01&	0.01&	0.03&	0.03&	0.03&	0.05&	0.04&	0.07\\
\end{array}
           \right]~.
\end{align*}}
\end{itemize}

Given the above configurations, we test the hypotheses $\mathrm H_0: \mathrm{rank}(\Pi_0)\le r$ v.s.\ $\mathrm H_1: \mathrm{rank}(\Pi_0)> r$ for $r=3$ at $\alpha=5\%$. Thus, $\mathrm{H}_0$ holds if and only if $\delta=0$, in which case $\mathrm{rank}(\Pi_0)<r$. Note that \citet{Kleibergen_Paap2006rank} instead consider $\mathrm H_0': \mathrm{rank}(\Pi_0)= 1$ v.s.\ $\mathrm H_1': \mathrm{rank}(\Pi_0)> 1$ so that the possibility $\mathrm{rank}(\Pi_0)< 1$ is excluded. We estimate $\Pi_0$ based a sample $\{R_t,F_t\}_{t=1}^n$ of size $n=330$ (as in \citet{Kleibergen_Paap2006rank}) that is generated according to the process \eqref{Eqn: DGP2}. The number of simulation replications is set to be 5,000, while the number of block bootstrap repetitions (with block size 2) is 500 for each simulation replication. We implement the three of our tests in same manner as we did in Section \ref{Sec: 4}, and compare with the multiple KP test (based on the HACC estimator for the long run variance), although the results for the direct application of the KP test are similar and available upon request.

Table \ref{Tab: Simulation3} summarizes the simulation results. We find patterns similar to those exhibited in Table \ref{Tab: Simulation1}. In particular, the multiple KP test is severely undersized, and its local power is overall dominated by our tests, though again the test based on numerical derivative estimators (CF-N) is somewhat sensitive to the choices of the step size. The two-step test (CF-T) and the test based on numerical derivative estimators (CF-A), on the other hand, show strong insensitivity to the choices of the tuning parameters.

\setlength{\tabcolsep}{5pt}
\begin{sidewaystable}[htbp]
\centering
\begin{threeparttable}
\renewcommand{\arraystretch}{1.2}
\caption{Rejection rates of rank tests for the model \eqref{Eqn: DGP2} with $r=3$, at $\alpha=5\%$\tnote{\dag}}\label{Tab: Simulation3}
\begin{tabular}{ccccccccccccccccccc}
\hline\hline
&\multirow{2}{*}{$\delta$}&&\multicolumn{6}{c}{CF-T}&&\multicolumn{8}{c}{CF-A}&\\
\cline{4-9}\cline{11-18}
& &&$\alpha/5$&$\alpha/10$ &$\alpha/15$&$\alpha/20$ &$\alpha/25$&$\alpha/30$ &&$n^{-1/5}$&$ 1.5n^{-1/5}$ &$n^{-1/4}$&$1.5n^{-1/4}$ &$n^{-1/3}$&$1.5n^{-1/3}$&$ n^{-2/5}$&$1.5n^{-2/5}$ &\\
\cline{2-18}
&$0.00$&&0.03&0.03&0.04&0.04&0.04&0.04&&0.05&0.05&0.05&0.05&0.05&0.05&0.05&0.05&\\
&$0.01$&&0.06&0.07&0.07&0.07&0.08&0.08&&0.28&0.28&0.28&0.28&0.28&0.28&0.28&0.28&\\
&$0.02$&&0.10&0.10&0.10&0.10&0.10&0.10&&0.42&0.42&0.42&0.42&0.42&0.42&0.42&0.42&\\
&$0.03$&&0.20&0.20&0.20&0.20&0.20&0.20&&0.59&0.59&0.59&0.59&0.59&0.59&0.59&0.59&\\
&$0.04$&&0.35&0.35&0.35&0.35&0.35&0.35&&0.75&0.75&0.75&0.75&0.75&0.75&0.75&0.75&\\
&$0.05$&&0.54&0.54&0.53&0.53&0.53&0.53&&0.87&0.87&0.87&0.87&0.87&0.87&0.79&0.87&\\
&$0.06$&&0.71&0.71&0.71&0.71&0.71&0.71&&0.95&0.95&0.95&0.95&0.95&0.95&0.89&0.95&\\
&$0.07$&&0.85&0.85&0.85&0.84&0.84&0.84&&0.98&0.98&0.98&0.98&0.96&0.98&0.96&0.96&\\
&$0.08$&&0.93&0.93&0.93&0.93&0.92&0.92&&0.99&0.99&0.99&0.99&0.98&0.99&0.99&0.99&\\
&$0.09$&&0.97&0.97&0.97&0.97&0.97&0.97&&1.00&1.00&1.00&1.00&0.99&1.00&0.99&0.99&\\
&$0.10$&&0.99&0.99&0.99&0.99&0.99&0.99&&1.00&1.00&1.00&1.00&1.00&1.00&1.00&1.00&\\
\cline{2-18}
&\multirow{2}{*}{$\delta$}&&\multicolumn{6}{c}{\multirow{2}{*}{KP-M}}&&\multicolumn{8}{c}{CF-N}&\\
\cline{11-18}
& &&&&&&&&&$n^{-1/5}$&$ 1.5n^{-1/5}$ &$n^{-1/4}$&$1.5n^{-1/4}$ &$n^{-1/3}$&$1.5n^{-1/3}$&$ n^{-2/5}$&$1.5n^{-2/5}$ &\\
\cline{2-18}
&$0.00$&&\multicolumn{6}{c}{0.00}&&0.05&0.05&0.05&0.05&0.04&0.05&0.03&0.04&\\
&$0.01$&&\multicolumn{6}{c}{0.05}&&0.21&0.25&0.18&0.22&0.11&0.17&0.06&0.11&\\
&$0.02$&&\multicolumn{6}{c}{0.10}&&0.22&0.29&0.16&0.24&0.09&0.15&0.03&0.09&\\
&$0.03$&&\multicolumn{6}{c}{0.20}&&0.28&0.37&0.22&0.31&0.12&0.20&0.07&0.13&\\
&$0.04$&&\multicolumn{6}{c}{0.36}&&0.39&0.49&0.33&0.42&0.24&0.32&0.14&0.24&\\
&$0.05$&&\multicolumn{6}{c}{0.53}&&0.54&0.62&0.49&0.56&0.40&0.48&0.28&0.41&\\
&$0.06$&&\multicolumn{6}{c}{0.69}&&0.69&0.75&0.66&0.71&0.58&0.65&0.45&0.58&\\
&$0.07$&&\multicolumn{6}{c}{0.80}&&0.83&0.86&0.81&0.84&0.74&0.80&0.63&0.74&\\
&$0.08$&&\multicolumn{6}{c}{0.87}&&0.92&0.93&0.90&0.92&0.86&0.90&0.77&0.87&\\
&$0.09$&&\multicolumn{6}{c}{0.91}&&0.96&0.97&0.96&0.97&0.94&0.96&0.88&0.94&\\
&$0.10$&&\multicolumn{6}{c}{0.93}&&0.99&0.99&0.98&0.99&0.98&0.98&0.95&0.98&\\
\hline\hline
\end{tabular}
  \begin{tablenotes}
      \small
      \item[\dag] The six values under CF-T are the choices of $\beta$, and those under CF-A and CF-N are the choices of $\kappa_n$ as in \ref{Eqn: bootstrap overview0} and \eqref{Eqn: NumericalDevEst} respectively.
    \end{tablenotes}
\end{threeparttable}
\end{sidewaystable}%

\setlength{\tabcolsep}{5pt}
\begin{table}[!htbp]
\centering
\begin{threeparttable}
\renewcommand{\arraystretch}{1.1}
\caption{The $p$-values for different tests}\label{Tab: pvalues}
\begin{tabular}{ccccccccccccccc}
\hline\hline
&\multicolumn{13}{c}{Panel A: our tests\tnote{\dag}}&\\
\cline{2-14}
&\multirow{2}{*}{\makecell{Block\\ size}}&&\multicolumn{3}{c}{CF-T}&&\multicolumn{3}{c}{CF-A}&&\multicolumn{3}{c}{CF-N}&\\
\cline{4-6}\cline{8-10}\cline{12-14}
&&&$\alpha/10$ & $\alpha/15$ &$\alpha/20$&&$n^{-1/5}$&$n^{-1/4}$&$n^{-1/3}$ &&$n^{-1/5}$&$n^{-1/4}$&$n^{-1/3}$ \\
\cline{2-14}
&$b=1$&&0.15 &0.15& 0.15 &&0.08& 0.08&0.08&&0.11&0.12&0.12&\\
&$b=2$&&0.15 &0.15& 0.15 &&0.10& 0.09&0.09&&0.11&0.11&0.12&\\
&$b=3$&&0.18 &0.18& 0.18 &&0.10& 0.10&0.10&&0.13&0.13&0.14&\\
&$b=4$&&0.16 &0.16& 0.16 &&0.08& 0.08&0.08&&0.13&0.13&0.14&\\
\cline{2-14}
& \multicolumn{13}{c}{Panel B: the KP-M test\tnote{\ddag}}\\
\cline{2-14}
&\multicolumn{13}{c}{0.91}\\
\hline\hline
\end{tabular}
\begin{tablenotes}
      \small
      \item[\dag] The three values under CF-T are the choices of $\beta$, and those under CF-A and CF-N are the choices of $\kappa_n$ as in \ref{Eqn: bootstrap overview0} and \eqref{Eqn: NumericalDevEst} respectively.
      \item[\ddag] The $p$-value for KP-M is given by the smallest significance level such that the null hypothesis is rejected, which is equal to the maximum $p$-value of all \citet{Kleibergen_Paap2006rank}'s tests implemented by the multiple testing method.
    \end{tablenotes}
\end{threeparttable}
\end{table}%

Finally, following \citet{Kleibergen_Paap2006rank}, we study a stochastic discount factor model based on the conditional capital asset pricing model proposed in the influential work of \citet{JagannathanWang1996CCAPM}. Suppose that $R_t\in\mathbf R^m$ is a vector of returns on $m$ assets at time $t$ and $F_{t}\in\mathbf{R}^{k}$ is a vector of $k$ common factors at time $t$. According to the stochastic discount factor model, $R_t$ and $F_t$ are related through
\begin{align}
E[R_{t+1}F_{t+1}^{\transpose}\gamma_{0}|\mathcal{I}_{t}]=\mathbf{1}_{m}~,
\end{align}
where $\mathcal{I}_{t}$ represents information at time $t$, and $\gamma_{0}\in\mathbf{R}^{k}$ is a vector of risk premia. If $\{R_t,F_t\}$ is governed by a stationary linear process:
\begin{align}
R_{t}=\Pi_{0}F_{t}+\varepsilon_{t}
\end{align}
where $E[\epsilon_{t+1}F_{t+1}|\mathcal I_t]=0$ and $E[F_{t+1}F_{t+1}^\transpose]$ is nonsingular, then $\gamma_0$ is identified if and only if the coefficient matrix $\Pi_0$ is of full rank. For this, we may test $\mathrm H_0: \mathrm{rank}(\Pi_0)\le r$ v.s.\ $\mathrm H_1: \mathrm{rank}(\Pi_0)> r$ with $r=k-1$.

We use the same data set as in \citet{Kleibergen_Paap2006rank}. There are returns $R_{t}$ on $10$ portfolios and $4$ factors in $F_{t}$ with monthly observations from July $1963$ to December $1990$, so $m=10$, $k=4$ and $n=330$. The factors in $F_{t}$ consist of constant, the return on a value-weighted portfolio, a corporate bond yield spread and a measure of per capita labor income growth. We estimate $\Pi_{0}$ by
\begin{align}
\hat{\Pi}_{n}=\sum_{t=1}^{n}R_{t}F_{t}^{\transpose}(\sum_{t=1}^{n}F_{t}F_{t}^{\transpose})^{-1}~.
\end{align}
Since the return sequence $\{R_{t}\}$ exhibits first order autocorrelation, we thus follow \citet{Kleibergen_Paap2006rank} and compute the KP statistic by employing the HACC estimator with one lag \citep{West1997AnotherHAC} for the long run covariance matrix. We implement our CF-T, CF-A and CF-N tests by adopting the block bootstrap \citep{lahiri2003Resampling} with block size $b=1,2,3,4$, employing the same choices of tuning parameters as before, and setting the number of bootstrap repetitions to be 1,000.

Table \ref{Tab: pvalues} reports the $p$-values of CF-T, CF-A and CF-N, as well as that of the KP-M test. The differences between our $p$-values and those of the KP-M tests are substantial: ours are uniformly less than 20\% while the latter are over 90\%. Thus, while the KP-M test strongly support the null, our tests are inconclusive depending on the significance levels and of course also the choices of the tuning parameters. It is worth noting that our three tests are quite insensitive across all choices of tuning parameters and the block size; in particular, the $p$-values of CF-T and CF-A are invariant to these choices.

\section{Estimation of the Rank}\label{Sec: 4-3}
\renewcommand{\theequation}{C.\arabic{equation}}
\setcounter{equation}{0}

There are settings as evident in Examples \ref{Ex: game} and \ref{Ex: ConsumerDemand}-\ref{Ex: ModelSelection} in Appendix \ref{Sec: examples more} where one would like to construct an estimate of the rank. The need of rank estimation is further reinforced should one deem our test based on \eqref{Eqn: StructDevEst} desirable. Following  \citet{Cragg_Donald1997infer} and \citet{Robin_Smith2000rank}, we adopt a sequential testing procedure that has been previously employed in the literature of model selection \citep{Potscher1983ARMAorder,BauerPotscherHackl1988MultipleTest,Hosoya1989Hierarchical}.\footnote{One may alternatively employ information criteria as in \citet{Cragg_Donald1997infer}. We do not pursue this possibility here in order to coherently present what is essential to our paper.}

Specifically, one may progressively test if the true rank is equal to 0, 1,\ldots, $k-1$ and set the estimator $\hat r_n$ to be the smallest $r\in\{0,1,\ldots,k-1\}$ that cannot be rejected if such a $r$ exists and to be $k$ if it does not. The conventional setup \eqref{Eqn: hypothesis, literature, intro} then suits well to this end because the possibility of $\mathrm{rank(\Pi_0)}$ strictly smaller than the hypothesized value is ``ruled out'' in each step by previous test(s). However, we argue that accommodating the possibility $\mathrm{rank}(\Pi_0)<r$, as we do in what follows, may once again lead to more reliable results. Heuristically, there are two possible errors involved in the procedure, namely, falsely rejecting a true null (i.e., type I error) and not rejecting a false null (i.e., type II error). Sequentially testing nulls of the form \eqref{Eqn: hypothesis, literature, intro} ignores type I errors potentially made in previous steps, and may have trivial or poor power when $\Pi_0$ is local to a matrix whose rank is ``small'', i.e., the capability of controlling type II error is limited. These are the two channels through which our rank estimator improves upon existing ones. Given a confidence level $1-\alpha$, we formally define the rank estimator $\hat r_n$ as
\begin{align}\label{Eqn: rank CI}
\hat r_n= \min\{r=0,\ldots,k-1: \tau_n^2\phi_r(\hat\Pi_n)\le \hat c_{n,1-\alpha}(r)\}
\end{align}
if the set is nonempty, and $\hat r_n= k$ if the set is empty, where $\hat c_{n,1-\alpha}(r)$ is defined by \eqref{Eqn: Quantile} for which we also make its dependence on $r$ explicit.

The following theorem shows that the estimator $\hat r_n$ in \eqref{Eqn: rank CI} picks up the true rank with probability at least $1-\alpha$ (asymptotically).

\iffalse
\begin{thm}\label{Thm: Rankdetermination}
For $\alpha\in(0,1)$, let $\psi_{n}^{(r)}$ be a test for the hypotheses \eqref{Eqn: hypothesis, intro} or \eqref{Eqn: hypothesis, literature, intro} such that $\lim_{n\to\infty}P(\psi_{n}^{(r)}=1)=\alpha$ when $r_{0}=r$, and $\lim_{n\to\infty}P(\psi_{n}^{(r)}=1)= 1$ when $r_{0}>r$. Then $\lim_{n\to\infty}P(\hat{r}^{\ast}_{n}<r_{0})= 0$,
\[\lim_{n\to\infty}P(\hat{r}^{\ast}_{n}=r_{0})= 1-\alpha \text{ if } r_{0}<k \text{ and } 1 \text{ if } r_{0}=k~,\]
and
\[\lim_{n\to\infty}P(\hat{r}^{\ast}_{n}>r_{0})=\alpha \text{ if } r_{0}<k \text{ and } 0 \text{ if } r_{0}=k~.\]
\end{thm}
\fi

\begin{thm}\label{Thm: Rankdetermination}
Let Assumptions \ref{Ass: Weaklimit: Pihat} and \ref{Ass: Boostrap: Pihat} hold, and the cdf of the limiting law in \eqref{Eqn: Nulllimt} when $r=r_0$ be continuous and strictly increasing at its $(1-\alpha)$-quantile for $\alpha\in(0,1)$. Then the rank estimator $\hat r_n$ defined by \eqref{Eqn: rank CI} satisfies
\begin{align}\label{Eqn: rank estimator, CI}
\lim_{n\to\infty}P(\hat r_n=r_0)=\begin{cases}
1-\alpha &\text{ if }r_0<k\\
1 & \text{ if } r_0=k
\end{cases}~,
\end{align}
$\lim_{n\to\infty} P(\hat r_n<r_0)=0$, and $\lim_{n\to\infty} P(\hat r_n>r_0)=\alpha$ (for $r_0<k$).
\end{thm}

Theorem \ref{Thm: Rankdetermination} implies that the procedure will select an estimator that is no smaller than the truth (asymptotically), and the probability of choosing a larger value (i.e., false selection) is controlled by the significance level $\alpha$ -- see \citet{Johansen1995likelihood} for related results in cointegration settings. These properties are intrinsically connected to the size control and consistency of our test. Moreover, by Theorem \ref{Thm: Rankdetermination}, the sequential procedure can be utilized in our two-step test to provide a preliminary rank estimator, although we stress that existing tests can also be employed in this regard -- the downside of these tests is that they may yield less accurate rank estimators as argued previously.

While the construction of a ``confidence set/singleton'' is of interest in its own right, one may also be interested in obtaining a consistent estimator, for which the probability of false selection should be negligible. One such an estimator is given by \eqref{Eqn: bootstrap overview0} or Lemma \ref{Lem: Consistency: RankEst}, where a tuning parameter is involved. This estimator is somewhat crude in that the probability of false selection is unclear and appears challenging to control. Employing the sequential procedure, we may achieve consistency while controlling the estimation error. As suggested by \eqref{Eqn: rank estimator, CI} and noted in the literature \citep{Potscher1983ARMAorder}, we must adjust the significance level $\alpha=\alpha_n$ according to the sample size so that $\alpha_n\to 0$ at a suitable rate, in order to obtain a consistent estimator. This turns out to be nontrivial in the current setup (where $\mathrm{rank}(\Pi_0)\le r$ is tested in each step) as we elaborate next.

If one sequentially tests $\mathrm{rank}(\Pi_0)= r$ for $r=0,\ldots,k-1$ based on, for example, \citet{Cragg_Donald1997infer} or \citet{Kleibergen_Paap2006rank}, the critical values are then obtained from chi-squared distributions. The rate at which $\alpha_n$ should tend to zero in order to deliver consistency has been well understood in this case by exploiting the analytic expansions of the cdfs of chi-squares -- see Theorem 5.8 in \citet{Potscher1983ARMAorder} for this result, \citet{Cragg_Donald1997infer} for an application of it in rank estimation, and \citet{Andrews1999GMS} in moment selection. There are, unfortunately, two challenges for us. First, the limiting distributions whose critical values we aim to approximate is highly nonstandard in general, and as a result, deriving rate conditions on $\alpha_n$ through analytic expansions appears challenging to us. Second, our critical values are obtained through bootstrap, and we believe that it is nontrivial to control the sample uncertainty embodied in these critical values as $\alpha_n\downarrow 0$. Nonetheless, we show that the our rank estimator is consistent under the same rate conditions on $\alpha_n$ as in \citet{Cragg_Donald1997infer} and \citet{Robin_Smith2000rank}. To formalize our discussions below, we thus impose:

\begin{ass}\label{Ass: alpha adjusted}
$\{\alpha_n\}_{n=1}^\infty$ satisfy (i) $\alpha_n\downarrow 0$, and (ii) $\tau_n^{-2}\log\alpha_n\to 0$.
\end{ass}

Assumption \ref{Ass: alpha adjusted} is quite mild in that it merely requires that, loosely speaking, $\alpha_n$ approach zero slower than exponentially decaying rates (not too fast). In this way, it encompasses a wide range of choices for $\alpha_n$. Given the adjusted significance level $\alpha_n$, we may now formally define the rank estimator to be
\begin{align}\label{Eqn: rank estimator}
\tilde r_n= \min\{r=0,\ldots,k-1: \tau_n^2\phi_r(\hat\Pi_n)\le \hat c_{n,1-\alpha_n}(r)\}
\end{align}
if the set is nonempty, and $\tilde r_n= k$ if the set is empty.

The next theorem establishes if Assumption \ref{Ass: alpha adjusted} holds and $\mathcal M$ is Gaussian (in addition to previous assumptions), then the estimator $\tilde r_n$ is indeed consistent.

\begin{thm}\label{Thm: rank estimator consistency}
Suppose that Assumptions \ref{Ass: Weaklimit: Pihat}, \ref{Ass: Boostrap: Pihat} and \ref{Ass: alpha adjusted} hold. Let $\tilde r_n$ be given by \eqref{Eqn: rank estimator}. If $\mathcal M$ is Gaussian but not constant (in $\mathbf M^{m\times k}$), then $\lim_{n\to\infty}P(\tilde r_n=r_0)= 1$.
\end{thm}

% It can be treated as a very special case of stepdown procedures in Chapter 9 of TSH. The element in the family of null hypotheses can be those in \eqref{Eqn: hypothesis, intro} or \eqref{Eqn: hypothesis, literature, intro}. The distinct feature is as follows. For the former, at most one is true. For the latter, the null hypotheses are increasing and more than one of them can be true unless only the last one is true. Yes, stepdown procedures are more accurate than stepup ones. For the latter (i.e. our tests), the procedure can be improved by controlling FWER and using improved method in Chapter 9 of TSH.

%In particular, one trivial modification for improvement is first using our test to determine whether $r_{0}=k$. Specifically, implement our test for the hypotheses \eqref{Eqn: hypothesis, intro} with $r=k-1$, and let $\bar{r}_{n} = k$ if the null hypotheses is rejected and otherwise implement the sequential testing procedure. Furthermore, our test can be used in the sequential testing procedure. Thus, we have the following procedure. % tend to choose a larger rank.
%
%\begin{defn}
%The sequential testing procedure based on our test is defined as: $\bar{r}_{n} = k$ if our test for the hypotheses \eqref{Eqn: hypothesis, intro} with $r=k-1$ is rejected and otherwise $\bar{r}_{n}$ is given by \eqref{Eqn: rank: determination} with $\psi_{n}^{(q)}$ being our test.
%\end{defn}

We reiterate that Theorem \ref{Thm: rank estimator consistency} may be of use not only in estimation problems but also in conducting our rank test based on the analytic derivative estimator -- see \eqref{Eqn: StructDevEst} and Lemma \ref{Lem: Consistency: DerivEst}. On a technical note, the Gaussianity condition plays an instrumental but not essential role. Concretely, it allows us to relate the significance levels to the corresponding critical values through a concentration inequality for Gaussian random vectors/matrices -- see Lemmas \ref{Lem: concentration inequality}-\ref{Lem: alpha adjustment2}. Thus, this condition can be relaxed whenever a suitable concentration inequality for $\mathcal M$ is available \citep{Ledoux2001concentration}.

\noindent{\sc Proof of Theorem \ref{Thm: Rankdetermination}:} For notational simplicity, define: for $r=0,\ldots,k-1$,
\begin{align}\label{Aux: Eqn: Thm: Rankdetermination: 0}
A_{n,r}=\{\tau_n^2\phi_r(\hat\Pi_n)> \hat c_{n,1-\alpha}(r)\}~,
\end{align}
i.e., $A_{n,r}$ are the events of rejecting the nulls. Consider the first the case when $r_0=k$. Then we must have $\{\hat r_n=r_0\}= A_{n,0}\cap A_{n,1}\cap\cdots\cap A_{n,k-1}$ and hence
\begin{multline}\label{Aux: Eqn: Thm: Rankdetermination: 1}
\liminf_{n\to\infty} P(\hat r_n=r_0) = \liminf_{n\to\infty} P(A_{n,0}\cap A_{n,1}\cap\cdots\cap A_{n,k-1})\\
\ge 1-\sum_{r=0}^{k-1}[1- \liminf_{n\to\infty} P(A_{n,r})]=1~,
\end{multline}
where the inequality follows from the Boole's inequality, and the last step is because of the consistency result of Theorem \ref{Thm: SizePowerTest}.

Next, suppose $r_0<k$. Then $\{\hat r_n=r_0\}= A_{n,0}\cap \cdots\cap A_{n,r_0-1}\cap A_{n,r_0}^c$ and hence
\begin{multline}\label{Aux: Eqn: Thm: Rankdetermination: 2}
\limsup_{n\to\infty} P(\hat r_n=r_0) = \limsup_{n\to\infty} P(A_{n,0}\cap \cdots\cap A_{n,r_0-1}\cap A_{n,r_0}^c)\\
\le \limsup_{n\to\infty} P(A_{n,r_0}^c)=1-\liminf_{n\to\infty} P(A_{n,r_0})=1-\alpha~,
\end{multline}
where the last step follows from the first claim of Theorem \ref{Thm: SizePowerTest}. Moreover,
\begin{multline}\label{Aux: Eqn: Thm: Rankdetermination: 3}
\liminf_{n\to\infty} P(\hat r_n=r_0) = \liminf_{n\to\infty} P(A_{n,0}\cap \cdots\cap A_{n,r_0-1}\cap A_{n,r_0}^c)\\
\ge 1-\sum_{r=0}^{r_0-1}[1- \liminf_{n\to\infty} P(A_{n,r})]-\limsup_{n\to\infty} P(A_{n,r_0}) =1-\alpha~,
\end{multline}
where we exploited the size control and the consistency results in Theorem \ref{Thm: SizePowerTest}.

Turning to the second claim, note that if $\hat r_n<r_0$, then $r_0>0$ and $\{\hat r_n<r_0\}\subset A_{n,0}^c\cup \cdots\cup A_{n,r_0-1}^c$. It follows that
\begin{multline}\label{Aux: Eqn: Thm: Rankdetermination: 4}
\limsup_{n\to\infty} P(\hat r_n<r_0) \le  \limsup_{n\to\infty} P(A_{n,0}^c\cup \cdots\cup A_{n,r_0-1}^c)\\
\le \sum_{r=0}^{r_0-1} \limsup_{n\to\infty} P(A_{n,r}^c)= \sum_{r=0}^{r_0-1} [1- \liminf_{n\to\infty} P(A_{n,r})] =0~,
\end{multline}
where the last step is because of the consistency result of Theorem \ref{Thm: SizePowerTest}. The last claim is a simple implication of the first two claims. We are thus done. \qed

\noindent{\sc Proof of Theorem \ref{Thm: rank estimator consistency}:} For notational simplicity, define: for $r=0,\ldots,k-1$,
\begin{align}\label{Aux: Eqn: Thm: Rankdetermination: 0}
A_{n,r}=\{\tau_n^2\phi_r(\hat\Pi_n)> \hat c_{n,1-\alpha_n}(r)\}~.
\end{align}
First, note that $\tilde r_n<r_0$ if and only if $r_0\ge 1$ and
\begin{align}\label{Eqn: rank estimator consistency, aux1}
\{\tilde r_n=r\}=A_{n,0}\cap \cdots\cap A_{n,r-1}\cap A_{n,r}^c
\end{align}
for some $r=0,\ldots, r_0-1$. Fix $r\in\{0,1,\ldots,r_0-1\}$. It follows from \eqref{Eqn: rank estimator consistency, aux1} that
\begin{align}\label{Eqn: rank estimator consistency, aux2}
P(\tilde r_n=r) \le P(A_{n,r}^c)=1-P(\phi_r(\hat\Pi_n)>\frac{\hat c_{n,1-\alpha_n}}{\tau_n^2})\to 0~,
\end{align}
where we exploited $\hat c_{n,1-\alpha_n}/\tau_n^2=o_p(1)$ by Assumption \ref{Ass: alpha adjusted}(ii), Lemma \ref{Lem: alpha adjustment2} and $\phi_r(\hat\Pi_n)\convp\phi_r(\Pi_0)>0$ by the continuous mapping theorem and $\mathrm{rank}(\Pi_0)\equiv r_0>r$. Since the result \eqref{Eqn: rank estimator consistency, aux2} is true for any $r=0,\ldots,r_0-1$, we thus obtain
\begin{align}\label{Eqn: rank estimator consistency, aux3}
\limsup_{n\to\infty}P(\tilde r_n<r_0) =0~.
\end{align}

Next, note that $\tilde r_n>r_0$ if and only if $r_0\le k-1$ and either the relation \eqref{Eqn: rank estimator consistency, aux1} holds for some $r=r_0+1,\ldots, k-1$ or the following event occurs
\begin{align}\label{Eqn: rank estimator consistency, aux4}
\{\tilde r_n=k\}=A_{n,0}\cap \cdots\cap A_{n,k-1}\cap A_{n,k}~.
\end{align}
Hence, $\{\tilde r_n=r\}\subset A_{n,r_0}$ for all $r=r_0+1,\ldots,k$. Fix $r=\{r_0+1,\ldots,k\}$. We thus have
\begin{align}\label{Eqn: rank estimator consistency, aux5}
P(\tilde r_n=r) \le P(A_{n,r_0})=P(\tau_n^2\phi_{r_0}(\hat\Pi_n)>\hat c_{n,1-\alpha_n})~.
\end{align}
Fix $\epsilon\in(0,1)$ so that $c_{1-\epsilon}$ is a continuity point of the cdf $F$ of $\phi_{r,\Pi_0}''(\mathcal M)$. This can be done without loss of generality because the set of discontinuity points is countable. By Assumption \ref{Ass: alpha adjusted}(i), it holds that: for all $n$ sufficiently large,
\begin{align}\label{Eqn: rank estimator consistency, aux6}
F(c_{1-\epsilon})=1-\epsilon<1-\alpha_n~,
\end{align}
and hence $c_{1-\alpha_n}>c_{1-\epsilon}$. In turn, we obtain from \eqref{Eqn: rank estimator consistency, aux6} and Assumption \ref{Ass: alpha adjusted}(i) that there exists some $\delta>0$ satisfying: for all $n$ sufficiently large,
\begin{align}\label{Eqn: rank estimator consistency, aux7}
F(c_{1-\epsilon})+\delta <1-\alpha_n~.
\end{align}
Note that if $\hat c_{n,1-\alpha_n}\le c_{1-\epsilon}$, then we obtain from \eqref{Eqn: rank estimator consistency, aux7} that
\begin{align}\label{Eqn: rank estimator consistency, aux8}
F(c_{1-\epsilon})+\delta <1-\alpha_n\le \hat F_n(\hat c_{n,1-\alpha_n}) \le \hat F_n(c_{1-\epsilon})~.
\end{align}
By Lemma 10.11 in \citet{Kosorok2008}, we may thus conclude that
\begin{align}\label{Eqn: rank estimator consistency, aux9}
\limsup_{n\to\infty} P(\hat c_{n,1-\alpha_n}\le c_{1-\epsilon})\le \limsup_{n\to\infty} P(\hat F_n(c_{1-\epsilon})-F(c_{1-\epsilon})>\delta)=0~.
\end{align}
Combination of results \eqref{Eqn: rank estimator consistency, aux5} and \eqref{Eqn: rank estimator consistency, aux9}, together with Assumption \ref{Ass: Weaklimit: Pihat}, now yields
\begin{align}\label{Eqn: rank estimator consistency, aux10}
\limsup_{n\to\infty} P(\tilde r_n=r) \le \limsup_{n\to\infty}P(\tau_n^2\phi_{r_0}(\hat\Pi_n)>c_{1-\epsilon})=1-F(c_{1-\epsilon})\le \epsilon~.
\end{align}
Since $\epsilon>0$ and $r\in\{r_0+1,\ldots,k\}$ are both arbitrary, it follows from \eqref{Eqn: rank estimator consistency, aux10} that
\begin{align}\label{Eqn: rank estimator consistency, aux11}
\limsup_{n\to\infty} P(\tilde r_n>r_0) =0~.
\end{align}
The theorem now follows from results \eqref{Eqn: rank estimator consistency, aux3} and \eqref{Eqn: rank estimator consistency, aux11} since
\begin{align}
\liminf_{n\to\infty}P(\tilde r_n=r_0)\ge 1-\limsup_{n\to\infty} P(\tilde r_n<r_0)-\limsup_{n\to\infty}P(\tilde r_n>r_0)=1~. \tag*{$\qed$}
\end{align}

\section{Auxiliary Lemmas}\label{Sec: aux}
\renewcommand{\theequation}{D.\arabic{equation}}
\setcounter{equation}{0}

\begin{lem}\label{Lem: phi, diff lemma, ID}
Suppose $\Pi\in\mathbf{M}^{m\times k}$ with $\Pi\neq 0$ and $\text{rank}(\Pi)\leq r$. For $\epsilon>0$, let $\Psi(\Pi)_{1}^{\epsilon}$ be given as in the proof of Proposition \ref{Pro: phi, differentiability}. Let $\sigma^{+}_{\min}(\Pi)$ be the smallest positive singular value of $\Pi$. Then for all sufficiently small $\epsilon>0$, we have
\[
\min_{U\in \Psi(\Pi)_{1}^{\epsilon}}\|\Pi U\|\geq \frac{\sqrt{2}}{2}\sigma^{+}_{\min}(\Pi)\epsilon~.
\]
\end{lem}
\noindent{\sc Proof:} Let $\Pi=P\Sigma Q^\transpose$ be a singular value decomposition of $\Pi$, where $P\in\mathbb S^{m\times m}$, $Q\in\mathbb S^{k\times k}$, and $\Sigma\in\mathbf M^{m\times k}$ is diagonal with diagonal entries in descending order. Let $d\equiv k-r$ and $d_0\equiv k-r_0$ with $r_0\equiv \text{rank}(\Pi)$. For $U\in\mathbb S^{k\times d}$, let $U_Q\equiv Q^{\transpose} U$ and write $U_Q^\transpose = [U_Q^{(1)\transpose}, U_Q^{(2)\transpose}]$ such that $U_Q^{(1)}\in\mathbf{M}^{r_0\times d}$. Then we have that for $U\in\mathbb S^{k\times d}$,
\begin{align}\label{Eqn: phi, diff lemma, ID, aux1-1}
\|\Pi U\|=\|P\Sigma Q^\transpose  U\|=\|\Sigma U_Q\|\ge \sigma^{+}_{\min}(\Pi)\|U_{Q}^{(1)}\|~,
\end{align}
where the second equality follows by $P^{\transpose}P=I_{m}$, and the inequality follows by the fact that $\Sigma$ is diagonal with diagonal entries in descending order with $\sigma^{+}_{\min}(\Pi)=\sigma_{r_0}(\Pi)$ the smallest positive entry. Let $U_Q^{(2)}=P_U^{(2)}\Sigma_U^{(2)} Q_U^{(2)^\transpose }$ be a singular value decomposition of $U_Q^{(2)}$ where $Q_U^{(2)}\in\mathbb S^{d\times d}$, $P_U^{(2)}\in\mathbb S^{d_0\times d_0}$ and $\Sigma_U^{(2)}\in\mathbf M^{d_0\times  d}$. Since $r_0\le r$ and hence $d_0\geq d$, it follows that, for $U\in\mathbb S^{k\times d}$,
\begin{align}\label{Eqn: phi, diff lemma, ID, aux1-2}
\|U_Q^{(2)}\|^2 =\sum_{j=1}^{d}\sigma_{j}^{2}(U_Q^{(2)})\leq\sum_{j=1}^{d}\sigma_{j}(U_Q^{(2)}) = \text{tr}([I_{d},\mathbf{0}_{r-r_0}]\Sigma_{U}^{(2)})~,
\end{align}
where the inequality follows by the fact that $\sigma_j(U_Q^{(2)})\in[0, 1]$ as singular values of $U_Q^{(2)}$ due to $U_Q^{(2)\transpose}U_Q^{(2)}+U_Q^{(1)\transpose}U_Q^{(1)}=I_{d}$, and the second equality follows by noting that the diagonal entries of $\Sigma_{U}^{(2)}$ are singular values of $U_Q^{(2)}$. Since $\|U_Q^{(1)}\|^2+\|U_Q^{(2)}\|^2=\|U_{Q}\|^2=d$, thus combining \eqref{Eqn: phi, diff lemma, ID, aux1-1} and \eqref{Eqn: phi, diff lemma, ID, aux1-2} yields that for $U\in\mathbb S^{k\times d}$,
\begin{align}\label{Eqn: phi, diff lemma, ID, aux1-3}
\|\Pi U\|\ge \sigma^{+}_{\min}(\Pi) \sqrt{d - \text{tr}([I_{d},\mathbf{0}_{r-r_0}]\Sigma_{U}^{(2)})}~.
\end{align}
Since $\|U_Q^{(1)}\|^2+\|\Sigma_{U}^{(2)}\|^2=\|U_Q^{(1)}\|^2+\|U_Q^{(2)}\|^2=d$ and $\|[I_{d},\mathbf{0}_{r-r_0}]^\transpose\|^2=d$, then simple algebra yields that for $U\in\mathbb S^{k\times d}$,
\begin{align}\label{Eqn: phi, diff lemma, ID, aux1-4}
2(d - \text{tr}([I_{d},\mathbf{0}_{d-r_0}]\Sigma_{U}^{(2)})) = \|U_Q^{(1)}\|^2+\|\Sigma_U^{(2)}-[I_{d},\mathbf{0}_{r-r_0}]^\transpose\|^2~.
\end{align}
Write $Q=[Q_1,Q_2]$ such that $Q_1\in\mathbf{M}^{k\times r_0}$. Since $Q_{1}^{\transpose}Q_{1}=I_{r_0}$, $Q_{2}^{\transpose}Q_{2}=I_{d_0}$ and $Q_{1}^{\transpose}Q_{2}=0$ as well as $P_{U}^{(2)}$ and $Q_{U}^{(2)}$ are orthonormal, we then have that for $U\in\mathbb S^{k\times d}$,
\begin{align}\label{Eqn: phi, diff lemma, ID, aux1-5}
\hspace{-0.2cm}\|U_Q^{(1)}\|^2\hspace{-0.05cm}+\hspace{-0.05cm}\|\Sigma^{(2)}_U\hspace{-0.05cm}-\hspace{-0.05cm}[I_{d},\hspace{-0.1cm}\mathbf{0}_{r-r_0}]^\transpose\|^2\hspace{-0.05cm} =\hspace{-0.05cm} \|Q_1 U_Q^{(1)}\hspace{-0.05cm}+\hspace{-0.05cm}Q_2  P_U^{(2)}(\Sigma^{(2)}_U\hspace{-0.05cm}-\hspace{-0.05cm}[I_{d},\hspace{-0.1cm}\mathbf{0}_{r-r_0}]^\transpose)Q_U^{(2)\transpose}\|^2~.
\end{align}
Since $U_Q^{(1)}=Q_{1}^{\transpose}U$ and $U_Q^{(2)}=Q_{2}^{\transpose}U$ by construction and $Q_1Q_{1}^{\transpose}U+Q_2Q_{2}^{\transpose}U=U$ by $QQ^{\transpose}=I_{k}$, we then have that, for $U\in\mathbb S^{k\times d}$,
\begin{align}\label{Eqn: phi, diff lemma, ID, aux1-6}
\hspace{-0.2cm}|Q_1 U_Q^{(1)}+Q_2  P_U^{(2)}(\Sigma_2\hspace{-0.05cm}-\hspace{-0.05cm}[I_{d},\hspace{-0.1cm}\mathbf{0}_{r-r_0}]^\transpose)Q_U^{(2)\transpose}\|^2 =
\|U\hspace{-0.05cm}-\hspace{-0.05cm}Q_2 P_U^{(2)}[I_{d},\hspace{-0.1cm}\mathbf{0}_{r-r_0}]^\transpose Q_U^{(2)\transpose}\|^2~.
\end{align}
Noting that $Q_2 P_U^{(2)}[I_{d},\hspace{-0.1cm}\mathbf{0}_{r-r_0}]^\transpose Q_U^{(2)\transpose}\in \Psi(\Pi)$, we have by \eqref{Eqn: phi, diff lemma, ID, aux1-4}-\eqref{Eqn: phi, diff lemma, ID, aux1-6} that, for $U\in\mathbb S^{k\times d}$,
\begin{align}\label{Eqn: phi, diff lemma, ID, aux1-7}
2(d - \text{tr}([I_{d},\mathbf{0}_{r-r_0}]\Sigma_{U}^{(2)}))\geq\min_{U^{\prime}\in\Psi(\Pi)}\|U-U^{\prime}\|^{2}~.
\end{align}
Since $\Pi\neq 0$, then $\Psi(\Pi)_1^\epsilon\neq\emptyset$ for all sufficiently small $\epsilon>0$. Fix such an $\epsilon>0$. By the definition of $\Psi(\Pi)_{1}^{\epsilon}$, combining \eqref{Eqn: phi, diff lemma, ID, aux1-3} and \eqref{Eqn: phi, diff lemma, ID, aux1-7} yields that for all $U\in \Psi(\Pi)_{1}^{\epsilon}$,
\begin{align}\label{Eqn: phi, diff lemma, ID, aux1-8}
\|\Pi U\|\geq \frac{\sqrt{2}}{2}\sigma^{+}_{\min}(\Pi)\min_{U^{\prime}\in\Psi(\Pi)}\|U-U^{\prime}\| \geq \frac{\sqrt{2}}{2}\sigma^{+}_{\min}(\Pi)\epsilon~.
\end{align}
Then the lemma follows by applying minimum over $\Psi(\Pi)_{1}^{\epsilon}$ to both sides of \eqref{Eqn: phi, diff lemma, ID, aux1-8} and noting that the result continues to hold for all sufficiently small $\epsilon>0$. \qed

\begin{lem}\label{Lem: phi, auxlem, correspondence}
The correspondence $\varphi$ in the proof of Proposition \ref{Pro: phi, differentiability} is continuous at $0$.
\end{lem}
\noindent{\sc Proof:} Fix $U_0\in\Psi(\Pi)$, and define the correspondence $\bar\varphi: \mathbf{R}\twoheadrightarrow \Gamma^{\Delta}$ by $\bar\varphi(t)=\Gamma_{U_{0}}^{\Delta}(t)$, where $\Psi(\Pi)$, $\Gamma^\Delta$ and $\Gamma_{U_0}^\Delta(t)$ are given in the proof of Proposition \ref{Pro: phi, differentiability}. Let $d\equiv k-r$. For each $t_n$ and each $V_0\in\bar\varphi(0)$, define $f: \Gamma^\Delta\to\mathbf M^{k\times d}$ by
\begin{align*}
f(V)= V_0-\frac{t_n}{2}U_0V^\transpose V~.
\end{align*}
Since $f$ is continuous and $\Gamma^\Delta$ is compact, $f$ is a compact map in the sense of \citet{GranasDugunji2003Fixed}. By Theorem 0.2.3 in \citet{GranasDugunji2003Fixed}, one of the following two cases must happen: (i) $f$ has a fixed point $V_{1n}\in\Gamma^\Delta$, and (ii) there exists some $V_{2n}\in\Gamma^\Delta$ such that $\|V_{2n}\|=\Delta$ and $V_{2n}=\lambda_n f(V_{2n})$ with $\lambda_n\equiv \frac{\Delta}{\|f(V_{2n})\|}\in(0,1)$. In case (i), since $U_0\in\Psi(\Pi)$, $V_0\in\bar\varphi(0)$ and $f(V_{1n})=V_{1n}$, we have by simple algebra:
\begin{align}\label{Eqn: phi, auxlem, correspondence, aux1-1}
\hspace{-0.3cm}V_{1n}^\transpose U_0+U_0^\transpose V_{1n}=(V_0\hspace{-0.05cm}-\hspace{-0.05cm}\frac{t_n}{2}U_0V_{1n}^\transpose V_{1n})^\transpose U_0+U_0^\transpose (V_0\hspace{-0.05cm}-\hspace{-0.05cm}\frac{t_n}{2}U_0V_{1n}^\transpose V_{1n})=-t_nV_{1n}^\transpose V_{1n}~.
\end{align}
This together with $V_{1n}\in\Gamma^\Delta$ implies that $V_{1n}\in\bar\varphi(t_{n})$. Moreover, since $f(V_{1n})=V_{1n}$, $\|U_{0}\|=\sqrt{d}$ and $V_{1n}\in\Gamma^\Delta$, then by the sub-multiplicativity of Frobenius norm we have
\begin{align}\label{Eqn: phi, auxlem, correspondence, aux2}
\|V_{1n}-V_0\|=\|\frac{t_n}{2}U_0V_{1n}^\transpose V_{1n}\|\le \frac{t_n}{2}\sqrt{d}\Delta^2~.
\end{align}
In case (ii), since $U_0\in\Psi(\Pi)$, $\lambda_n^2 V_0\in\bar\varphi(0)$ and $\lambda_nV_{2n}=\lambda_n^2 f(V_{2n})$, then by analogous calculations as in \eqref{Eqn: phi, auxlem, correspondence, aux1-1}, we have
\begin{align*}
(\lambda_nV_{2n})^\transpose U_0+U_0^\transpose(\lambda_n V_{2n})=-t_n(\lambda_nV_{2n})^\transpose (\lambda_nV_{2n})~.
\end{align*}
This together with $\lambda_n V_{2n}\in\Gamma^\Delta$ due to $\lambda_n\in(0,1)$ and $V_{2n}\in\Gamma^\Delta$ implies that $\lambda_n V_{2n}\in\bar\varphi(t_{n})$. Moreover, since $\lambda_nV_{2n}=\lambda_n^2 f(V_{2n})$, similar to \eqref{Eqn: phi, auxlem, correspondence, aux2} we have:
\begin{align}\label{Eqn: phi, auxlem, correspondence, aux3}
&\|\lambda_nV_{2n}-V_0\|\le \|\lambda_n^{2}f(V_{2n})-\lambda_n^2 V_0\|+|\lambda_n^2-1|\|V_0\|\le \frac{t_n}{2}\sqrt{d}\Delta^2+|\lambda_n^2-1|\Delta~,
\end{align}
where the first inequality follows the triangle inequality and the second inequality follows since $\lambda_{n}\in(0,1)$. Now, for each $n\in\mathbf N$, define $V_n^*$ to be $V_{1n}$ if case (i) happens and $\lambda_n V_{2n}$ otherwise. Let $\delta_{n}\equiv1$ if case (i) happens and $\delta_{n}\equiv\lambda_{n}$ otherwise. Then $V_n^*\in\Gamma_{U_0}^\Delta(t_n)$ for all $n\in\mathbf N$, and combination of \eqref{Eqn: phi, auxlem, correspondence, aux2} and \eqref{Eqn: phi, auxlem, correspondence, aux3} yields
\begin{align*}
\|V_n^*-V_0\|\le \frac{t_n}{2}\sqrt{d}\Delta^2+|\delta_n^{2}-1|\Delta\to 0~,
\end{align*}
where we exploited the fact that if $V_{2n}$ exists infinitely often, $\delta_n=\lambda_n=\frac{\Delta}{\|f(V_{2n})\|}\to1$ due to $f(V_{2n})\to V_{0}$ as $n\to\infty$ and $\|V_{0}\|\leq \Delta$,  and $t_{n}\to0$ as $n\to\infty$. It follows that $\bar\varphi(t)$ is lower hemicontinuous at $t=0$ by Theorem 17.21 in \citet{AliprantisandBorder2006}.

The lower hemicontinuity of $\varphi(t)$ at $t=0$ follows easily from that of $\bar\varphi(t)$ again by Theorem 17.21 in \citet{AliprantisandBorder2006}. To see this, let $t_n\to 0$ and $(U_0,V_0)\in\varphi(0)$. Define $(U_n^*,V_n^*)$ to be $U_n^*=U_0$ and $V_n^*$ be as in previous construction for all $n\in\mathbf N$.  Clearly, $(U_n^*,V_n^*)\to (U_0,V_0)$, implying that $\varphi(t)$ is lower hemicontinuous at $t=0$. Since $\varphi(t)$ is contained in the compact set $\mathbb S^{k\times d}\times\Gamma^\Delta$ for all $t$, $\varphi(t)$ is upper hemicontinuous at $t=0$ by Theorem 17.20 in \citet{AliprantisandBorder2006}. We have therefore showed that $\varphi(t)$ is continuous at $t=0$. \qed

\begin{lem}\label{Lem: Rankcon, projection}
Suppose $\Pi\in\mathbf{M}^{m\times k}$ with $\text{rank}(\Pi)\leq r$, and $M\in\mathbf{M}^{m\times k}$ with $M\neq 0$. Let $\Psi(\Pi)=\amin_{U\in\mathbb S^{k\times (k-r)}}\|\Pi U\|^2$, and for $U\in\Psi(\Pi)$ and $\Delta>0$ let $\Gamma_U^\Delta(0)$ be as in the proof of Proposition \ref{Pro: phi, differentiability}.  For $\Delta$ sufficiently large, it follows that for all $U\in\Psi(\Pi)$,
\[
\min_{V\in \Gamma_U^{\Delta}(0)}\Vert\Pi V+MU\Vert^{2}=\min_{V\in\mathbf{M}^{k\times (k-r)}}\Vert\Pi V+MU\Vert^{2}~.
\]
\end{lem}
\noindent{\sc Proof:} The conclusion is trivial if $\Pi= 0$. Suppose that $\Pi\neq 0$ and let $d\equiv k-r$. Let $r_0=\mathrm{rank}(\Pi)$ and $\Pi=P\Sigma Q^\transpose $ be a singular value decomposition of $\Pi$, where $P\in\mathbb S^{m\times m}$, $Q\in\mathbb S^{k\times k}$, and $\Sigma\in\mathbf M^{m\times k}$ is diagonal with diagonal entries in descending order. Since $\Pi\neq 0$ and $r_0\le r$, we may write $\Sigma=[\Sigma_{1},0]$ with $\Sigma_{1}\in\mathbf{M}^{m\times r_0}$ of full rank so that
\begin{align}\label{Eqn: Rankcon, projection, aux1}
\min_{V\in\mathbf{M}^{k\times d}}\Vert\Pi V+MU\Vert^{2}=\min_{V\in\mathbf{M}^{r_0\times d}}\Vert [P\Sigma_1 V+MU\Vert^{2}~.
\end{align}
By the projection theorem, the minimum on the right hand side of \eqref{Eqn: Rankcon, projection, aux1} is attained at some point, say $V_{1}^{\ast}\in\mathbf{M}^{r_0\times d}$. Moreover, $V_{1}^{\ast}$ is uniformly bounded over $U\in\Psi(\Pi)$. Let $V^{\ast}\equiv Q[V_{1}^{\ast\transpose},0]^{\transpose}\in\mathbf{M}^{k\times d}$, then the minimum on the left hand side of \eqref{Eqn: Rankcon, projection, aux1} is attained at $V^{\ast}$. Decompose $Q$ as $Q=[Q_1, Q_2]$, where $Q_1\in\mathbf M^{k\times r_0}$. Then $V^{\ast} = Q_1V_{1}^{\ast}\in \Gamma_U^{\Delta}(0)$ for all $U\in\Psi(\Pi)$, when $\Delta$ is sufficiently large. It implies that the minimum on the right hand side of \eqref{Eqn: Rankcon, projection, aux1} is attained within $\Gamma_U^{\Delta}(0)$ as well for all $U\in\Psi(\Pi)$, when $\Delta$ is sufficiently large. This implies that when $\Delta$ is sufficiently large,
\begin{align*}
\min_{V\in \Gamma_U^{\Delta}(0)}\Vert\Pi V+MU\Vert^{2}\leq \min_{V\in\mathbf{M}^{k\times d}}\Vert\Pi V+MU\Vert^{2}
\end{align*}
for all $U\in\Psi(\Pi)$. The reverse inequality is simply true since $\Gamma_U^{\Delta}(0)\subset \mathbf{M}^{k\times d}$ all $U\in\Psi(\Pi)$ and all $\Delta>0$. This completes the proof of the lemma. \qed

\begin{lem}\label{Lem: phi, derivative, reprentation}
If $r_0\equiv \mathrm{rank}(\Pi)\le r$, then for any $M\in\mathbf M^{m\times k}$,
\begin{align}
\min_{U\in \Psi(\Pi)}\min_{V\in\mathbf{M}^{k\times d}}\Vert\Pi V+MU\Vert^{2}=\sum_{j=r-r_0+1}^{k-r_0}\sigma^{2}_j(P_2^\transpose MQ_2)~,
\end{align}
where $\Psi(\Pi)=\amin_{U\in\mathbb S^{k\times(k-r)}}\|\Pi U\|^2$.
\end{lem}
\noindent{\sc Proof:} Let $d\equiv k-r$ and $d_0\equiv k-r_0$. Noting that the column vectors in $Q_2$ form a orthonormal basis for the null space of $\Pi_{0}$, we may rewrite $\Psi(\Pi)$ as $\Psi(\Pi)=\{Q_2V: V\in\mathbb S^{d_0\times d}\}$. This, together with the projection theorem, implies
\begin{align}\label{Eqn: phi, derivative, reprentation, aux2}
\phi_{r,\Pi}^{\prime\prime}(M)=\min_{V\in\mathbb{S}^{d_0\times d}}\Vert (I-\Pi(\Pi^\transpose\Pi)^{-}\Pi^\transpose)MQ_2V\Vert^{2}~,
\end{align}
where $A^{-}$ denotes the Moore-Penrose inverse of a generic matrix $A$. By the singular value decomposition of $\Pi$, we have
\begin{multline}\label{Eqn: phi, derivative, reprentation, aux3}
(I-\Pi(\Pi^\transpose\Pi)^{-}\Pi^\transpose)P=P-P\Sigma Q^\transpose(Q\Sigma^\transpose P^\transpose P\Sigma Q^\transpose)^{-}Q\Sigma^\transpose P^\transpose P \\
 =P-P\Sigma Q^\transpose Q(\Sigma^\transpose P^\transpose P \Sigma )^{-}Q^\transpose Q\Sigma^\transpose P^\transpose P=P-P\Sigma(\Sigma^\transpose  \Sigma )^{-}\Sigma^\transpose=[0,P_2]~,
\end{multline}
where the second equality exploited Theorem 20.5.6 in \citet{Harville2008matrix}, the third equality follows from $P$ and $Q$ being orthonormal, and the fourth equality is obtained by carrying out the Moore-Penrose inverse by Exercise 2.7.4 in \citet{MagnusNeudecker2007Matrix} and noting that $\Sigma$ is diagonal. In view of \eqref{Eqn: phi, derivative, reprentation, aux3}, we have
\begin{multline}\label{Eqn: phi, derivative, reprentation, aux4}
\min_{V\in\mathbb{S}^{d_0\times d}}\Vert(I-\Pi(\Pi^\transpose\Pi)^{-}\Pi^\transpose)MQ_2V\Vert^{2}= \min_{V\in\mathbb{S}^{d_0\times d}}\Vert [0,P_2]P^\transpose MQ_2V\Vert^{2} \\
=\min_{V\in\mathbb{S}^{d_0\times d}}\Vert P_2P_2^\transpose MQ_2V\Vert^{2}=\min_{V\in\mathbb{S}^{d_0\times d}}\Vert P_2^\transpose MQ_2V\Vert^{2}=\sum_{j=r-r_0+1}^{k-r_0}\sigma^{2}_j(P_2^\transpose MQ_2)~,
\end{multline}
where the third equality follows from $P_{2}^{\transpose}P_{2}=I_{m-r_0}$ and the final equality follows from Lemma \ref{Lem: phi, representation}. Combining \eqref{Eqn: phi, derivative, reprentation, aux2} and \eqref{Eqn: phi, derivative, reprentation, aux4} concludes the proof of the lemma. \qed

\begin{lem}\label{Lem: phi, socHD}
Suppose $\text{rank}(\Pi)\leq r$ and let $\phi_{r,\Pi}^{\prime\prime}: \mathbf M^{m\times k}\to\mathbf R$ be given as in Proposition \ref{Pro: phi, differentiability}.
If $\text{rank}(\Pi)=r$, there exists a bilinear map $\Phi_{r,\Pi}'':\mathbf{M}^{m\times k}\times \mathbf{M}^{m\times k}\to \mathbf{R}$ such that $\phi_{r,\Pi}^{\prime\prime}(M)=\Phi_{r,\Pi}''(M,M)$ for all $M\in\mathbf{M}^{m\times k}$; if $\text{rank}(\Pi)<r$, such a $\Phi_{r,\Pi}''$ does not exist.
\end{lem}
\noindent{\sc Proof:} Let $\Pi=P\Sigma Q^\transpose$ is a singular value decomposition of $\Pi$, where $P\in\mathbb S^{m\times m}$ whose last $m-r$ columns constitutes $P_2$, $Q\in\mathbb S^{k\times k}$ whose last $k-r$ columns constitutes $Q_2$,, and $\Sigma\in\mathbf M^{m\times k}$ is diagonal with diagonal entries in descending order. Let $d\equiv k-r$. If $\text{rank}(\Pi)=r$, then Lemma \ref{Lem: phi, derivative, reprentation} and Lemma \ref{Lem: phi, representation} imply
\begin{align*}
\phi_{r,\Pi}^{\prime\prime}(M)=\min_{V\in\mathbb{S}^{d\times d}}\Vert P_2^\transpose MQ_2 V\Vert^{2} = \Vert P_2^\transpose MQ_2 \Vert^{2}~,
\end{align*}
for all $M\in\mathbf{M}^{m\times k}$, which is a quadratic form corresponding to the bilinear form $\Phi_{r,\Pi}''(M_{1},M_{2})\equiv \text{tr}(Q_2^\transpose M_{1}^{\transpose}P_2P_2^\transpose M_{2}Q_2)$ for $M_{1}\in\mathbf{M}^{m\times k}$ and $M_{2}\in\mathbf{M}^{m\times k}$.

Next, assume that $\text{rank}(\Pi)<r$. Suppose for the sake of a contradiction that there exists a bilinear map $\Phi_{r,\Pi}''$ corresponding to $\phi_{r,\Pi}''$. Bilinearity of $\Phi_{r,\Pi}''$ then implies that
\begin{align}\label{Eqn: phi, socHD, aux2}
\phi_{r,\Pi}^{\prime\prime}(M_{1})+\phi_{r,\Pi}^{\prime\prime}(M_{2})&=\frac{\phi_{r,\Pi}^{\prime\prime}(M_{1}+M_{2})+\phi_{r,\Pi}^{\prime\prime}(M_{1}-M_{2})}{2}
\end{align}
for all $M_{1}\in\mathbf{M}^{m\times k}$ and $M_{2}\in\mathbf{M}^{m\times k}$. Let $r_0 \equiv \text{rank}(\Pi)$ and $d_0\equiv k-r_0$. If $M=P_{2}HQ_{2}^\transpose$ for some $H\in\mathbf{M}^{(m-r_0)\times d_0}$, then Lemma \ref{Lem: phi, derivative, reprentation} and Lemma \ref{Lem: phi, representation} imply
\begin{align}\label{Eqn: phi, socHD, aux3}
\phi_{r,\Pi}^{\prime\prime}(M)= \sigma_{r-r_0+1}^{2}(H)+ \cdots + \sigma_{d_0}^{2}(H)~.
\end{align}
Now, let $H_{1}\in\mathbf{M}^{(m-r_0)\times d_0}$ be diagonal with the $(j,j)$th entry equal to $1$ for $j=1,\ldots,d_0$ and $H_{2}\in\mathbf{M}^{(m-r_0)\times d_0}$ be diagonal with the $(j,j)$th entry equal to $-1$ for $j=1$ and $1$ for $j=2,\ldots,d_0$. Set $M_{i}=P_{2}H_{i}Q_{2}^\transpose$ for $i=1,2$, the result in \eqref{Eqn: phi, socHD, aux3} implies $\phi_{r,\Pi}^{\prime\prime}(M_{1})= \phi_{r,\Pi}^{\prime\prime}(M_{2}) = k -r$, $\phi_{r,\Pi}^{\prime\prime}(M_{1}+M_{2}) = 4(k-r)-4$ and $\phi_{r,\Pi}^{\prime\prime}(M_{1}-M_{2}) = 0$. It follows that
\begin{align*}
2(k-r) = \phi_{r,\Pi}^{\prime\prime}(M_{1})+\phi_{r,\Pi}^{\prime\prime}(M_{2}) \neq \frac{\phi_{r,\Pi}^{\prime\prime}(M_{1}+M_{2})+\phi_{r,\Pi}^{\prime\prime}(M_{1}-M_{2})}{2}=2(k-r)-2~,
\end{align*}
which contradicts the result \eqref{Eqn: phi, socHD, aux2}. Thus, the second result of the lemma follows.\qed

\begin{lem}\label{Lem: Consistency: DerivEst}
Suppose Assumption \ref{Ass: Weaklimit: Pihat} holds. Let $\hat{\phi}^{\prime\prime}_{r,n}$ be the analytic estimator given by \eqref{Eqn: StructDevEst}. If $\hat r_n\convp r_0\equiv \mathrm{rank}(\Pi_0)$ and $r_0\le r<k$, then condition \eqref{Eqn: derivative consistency} holds.
\end{lem}

\noindent{\sc Proof:} For notational simplicity, let $d\equiv k-r$ and $\hat{d}_{n}\equiv k-\hat{r}_{n}$. Fix a sequence $\{M_n\}$ such that $M_{n}\to M$ as $n\to\infty$. By Lemma \ref{Lem: phi, representation}, we have:
\begin{multline}\label{Aux: Eqn: Lem: Consistency: DerivEst: 1}
|\hat\phi_{r,n}^{\prime\prime}(M_{n})-\hat\phi_{r,n}^{\prime\prime}(M)|= \big|\min_{U\in\mathbb{S}^{\hat{d}_{n}\times d}}\|\hat{P}_{2,n}^{\transpose}M_{n}\hat{Q}_{2,n}U\|-\min_{U\in\mathbb{S}^{\hat{d}_{n}\times d}}\|\hat{P}_{2,n}^{\transpose}M\hat{Q}_{2,n}U\|\big|\\
\times \big(\min_{U\in\mathbb{S}^{\hat{d}_{n}\times d}}\|\hat{P}_{2,n}^{\transpose}M_{n}\hat{Q}_{2,n}U\|+\min_{U\in\mathbb{S}^{\hat{d}_{n}\times d}}\|\hat{P}_{2,n}^{\transpose}M\hat{Q}_{2,n}U\|\big)~,
\end{multline}
where the inequality follows by the formula $(a^2-b^2)=(a+b)(a-b)$. For the first term on the right hand side of \eqref{Aux: Eqn: Lem: Consistency: DerivEst: 1}, we have
\begin{multline}\label{Aux: Eqn: Lem: Consistency: DerivEst: 2}
\big|\min_{U\in\mathbb{S}^{\hat{d}_{n}\times d}}\|\hat{P}_{2,n}^{\transpose}M_{n}\hat{Q}_{2,n}U\|-\min_{U\in\mathbb{S}^{\hat{d}_{n}\times d}}\|\hat{P}_{2,n}^{\transpose}M\hat{Q}_{2,n}U\|\big|\\
\leq \min_{U\in\mathbb{S}^{\hat{d}_{n}\times d}}\|\hat{P}_{2,n}^{\transpose}(M_{n}-M)\hat{Q}_{2,n}U\|\le \sqrt{kmd}\|M_{n}-M\|=o_{p}(1)~,
\end{multline}
where the first inequality follows by the Lipschitz continuity of the min operator and the triangle inequality, the second inequality holds by the submultiplicativity of Frobenius norm, $\|\hat{P}_{2,n}\|\leq \sqrt{m}$, $\|\hat{Q}_{2,n}\|\leq \sqrt{k}$ and $\|U\|=\sqrt{r}$ for all $U\in\mathbb{S}^{\hat{d}_{n}\times d}$, and the equality is because $M_{n}\to M$.  For the second term on the right hand side of \eqref{Aux: Eqn: Lem: Consistency: DerivEst: 1}, once again exploiting the sub-multiplicability of the Frobenius norm, $\|\hat{P}_{2,n}\|\leq \sqrt{m}$, $\|\hat{Q}_{2,n}\|\leq \sqrt{k}$, $\|U\|=\sqrt{r}$ for all $U\in\mathbb{S}^{\hat{d}_{n}\times d}$ and $M_n\to M$, we have that
\begin{multline}\label{Aux: Eqn: Lem: Consistency: DerivEst: 3}
\min_{U\in\mathbb{S}^{\hat{d}_{n}\times d}}\|\hat{P}_{2,n}^{\transpose}M_{n}\hat{Q}_{2,n}U\|+\min_{U\in\mathbb{S}^{\hat{d}_{n}\times d}}\|\hat{P}_{2,n}^{\transpose}M\hat{Q}_{2,n}U\| \\ \le \sqrt{kmd}\|M_n\|+\sqrt{kmd}\|M\|=O(1)~.
\end{multline}
Combining results \eqref{Aux: Eqn: Lem: Consistency: DerivEst: 1}-\eqref{Aux: Eqn: Lem: Consistency: DerivEst: 3}, then we obtain
\begin{align}\label{Aux: Eqn: Lem: Consistency: DerivEst: 4}
|\hat\phi_{r,n}^{\prime\prime}(M_{n})-\hat\phi_{r,n}^{\prime\prime}(M)|=o_{p}(1)~.
\end{align}

In view of \eqref{Aux: Eqn: Lem: Consistency: DerivEst: 4}, it thus suffices to show that
\begin{multline}\label{Aux: Eqn: Lem: Consistency: DerivEst: 5}
|\hat\phi_{r,n}^{\prime\prime}(M)-\phi_{r,\Pi_{0}}^{\prime\prime}(M)|\equiv \\
|\sum_{j=r-\hat{r}_{n}+1}^{k-\hat{r}_{n}}\sigma^{2}_j(\hat{P}_{2,n}^\transpose M \hat{Q}_{2,n})-\sum_{j=r-r_{0}+1}^{k-r_{0}}\sigma^{2}_j(P_{0,2}^\transpose MQ_{0,2})|=o_p(1)~.
\end{multline}
Let $\hat{q}_{j}$ be the $j$th column of $\hat{Q}_{2,n}$. Since $Q_{0}\in\mathbb{S}^{k\times k}$, we may write $\hat{q}_{j}=Q_{0}\hat u_{j}$ for some (random) $\hat u_{j}\in\mathbb{S}^{k\times 1}$. Noting that $\hat q_j$ is an eigenvector of $\hat\Pi_n^\transpose\hat\Pi_n$ associated with the eigenvalue $\sigma_{r_{0}+j}^2(\hat\Pi_n)$ when $\hat r_n=r_0$ and that $P(\hat r_n=r_0)\to 1$ as given, we have
\begin{multline}\label{Aux: Eqn: Lem: Consistency: DerivEst: 6}
[\hat{\Pi}_{n}^\transpose\hat{\Pi}_{n}-\Pi_{0}^\transpose\Pi_{0}-(\sigma_{r_{0}+j}^{2}(\hat{\Pi}_{n})-\sigma_{r_{0}+j}^{2}(\Pi_{0}))I_k+\Pi_{0}^\transpose\Pi_{0}-\sigma_{r_{0}+j}^{2}(\Pi_{0})I_k]Q_{0}\hat u_{j} \\
=[\hat{\Pi}_{n}^\transpose\hat{\Pi}_{n}-\sigma_{r_{0}+j}^{2}(\hat{\Pi}_{n})I_k]\hat{q}_{j}=o_p(1)~.
\end{multline}
Observe that $\|\hat{\Pi}_{n}^\transpose\hat{\Pi}_{n}-\Pi_{0}^\transpose\Pi_{0}\|=o_{p}(1)$ and $|\sigma_{r_{0}+j}^{2}(\hat{\Pi}_{n})-\sigma_{r_{0}+j}^{2}(\Pi_{0})|=o_{p}(1)$ by the continuous mapping theorem, the Weyl inequality \citep[Exercise 1.3.22(iv)]{Tao2012Matrix} and Assumption \ref{Ass: Weaklimit: Pihat}, we then conclude from \eqref{Aux: Eqn: Lem: Consistency: DerivEst: 6} that
\begin{align}\label{Aux: Eqn: Lem: Consistency: DerivEst: 7}
o_p(1)=[\Pi_{0}^\transpose\Pi_{0}-\sigma_{r_{0}+j}^{2}(\Pi_{0})I_k]Q_{0}\hat u_{j}=Q_0\Sigma_0^\transpose \Sigma_0 \hat u_j~,
\end{align}
where we exploited the singular value decomposition $\Pi_0=P_0\Sigma_0 Q_0^\transpose$, and the fact that $\sigma_{r_{0}+j}^{2}(\Pi_0)=0$. Since the first $r_{0}$ diagonal elements of $\Sigma_0^\transpose \Sigma_0$ are positive and $Q_0$ is nonsingular, we may conclude from result \eqref{Aux: Eqn: Lem: Consistency: DerivEst: 7} that the first $r_{0}$ elements of $\hat u_j$ are $o_p(1)$ and moreover by the definition of $\hat q_j$ that for some random $U_2\in \mathbb S^{(k-r_{0})\times (k-r_{0})}$,
\begin{align}\label{Aux: Eqn: Lem: Consistency: DerivEst: 8}
\hat{Q}_{2,n}=Q_{0,2} U_2+o_{p}(1)~,
\end{align}
By an analogous argument, we have that for some random $V_2\in\mathbb{S}^{(m-r_{0})\times(m-r_{0})}$,
\begin{align}\label{Aux: Eqn: Lem: Consistency: DerivEst: 9}
\hat{P}_{2,n}=P_{0,2}V_2+o_{p}(1)~.
\end{align}
Combining results \eqref{Aux: Eqn: Lem: Consistency: DerivEst: 8} and \eqref{Aux: Eqn: Lem: Consistency: DerivEst: 9} and the continuous mapping theorem yields
\begin{align}\label{Aux: Eqn: Lem: Consistency: DerivEst: 10}
\|\hat{P}_{2,n}^\transpose M\hat{Q}_{2,n}-V_2^{\transpose}P_{0,2}^\transpose MQ_{0,2}U_2\|=o_{p}(1)~.
\end{align}
Thus, \eqref{Aux: Eqn: Lem: Consistency: DerivEst: 5} follows from \eqref{Aux: Eqn: Lem: Consistency: DerivEst: 10}, the continuous mapping theorem and the fact that the singular values of $V_2^{\transpose}P_{0,2}^\transpose MQ_{0,2}U_2$ are equal to those of $P_{0,2}^\transpose MQ_{0,2}$. \qed

\begin{lem}\label{Lem: Consistency: RankEst}
Suppose Assumption \ref{Ass: Weaklimit: Pihat} holds. Let $\hat r_n$ be the maximal $j\in\{1,\ldots,k\}$ such that $\sigma_j(\hat\Pi_n)\ge\kappa_n$ if such a $j$ exists and $\hat r_n=0$ otherwise. If $\kappa_{n}\downarrow 0$ and $\tau_{n}\kappa_{n}\to\infty$, then it follows that
\[\lim_{n\to\infty} P(\hat{r}_{n}=r_{0})=1~.\]
\end{lem}
\noindent{\sc Proof:} On the one hand, note that if $\hat{r}_{n}> r_{0}$, then we must have $r_0\le k-1$, $\sigma_{r_{0}+1}(\hat{\Pi}_{n})\geq \kappa_{n}$ and $\sigma_{r_{0}+1}(\Pi_{0})=0$. In turn, it follows that
\begin{multline}\label{Aux: Eqn: Lem: Consistency: RankEst: 1}
\limsup_{n\to\infty}P(\hat{r}_{n}> r_{0})\le\limsup_{n\to\infty}P(|\sigma_{r_{0}+1}(\hat{\Pi}_{n})-\sigma_{r_{0}+1}(\Pi_{0})|\geq \kappa_{n})\\
\leq \limsup_{n\to\infty}P( \|\tau_{n}\{\hat{\Pi}_{n}-\Pi_{0}\}\|\geq \tau_{n}\kappa_{n})=0~,
\end{multline}
where the second inequality is by the Weyl inequality \citep[Exercise 1.3.22(iv)]{Tao2012Matrix}, and the equality follows from $\|\tau_{n}\{\hat{\Pi}_{n}-\Pi_{0}\}\|=O_p(1)$ by Assumption \ref{Ass: Weaklimit: Pihat} and $\tau_{n}\kappa_{n}\to\infty$ as given. On the other hand, if $\hat{r}_{n}<r_{0}$, then $r_0>0$ and $\sigma_{r_{0}}(\hat{\Pi}_{n})< \kappa_{n}$. Hence,
\begin{align}\label{Aux: Eqn: Lem: Consistency: RankEst: 2}
&\limsup_{n\to\infty}P(\hat{r}_{n}<r_{0})\le\limsup_{n\to\infty}P(|\sigma_{r_{0}}(\hat{\Pi}_{n})-\sigma_{r_{0}}(\Pi_{0})|>-\kappa_{n}+\sigma_{r_{0}}(\Pi_{0}))\notag\\
&\hspace{1.5cm}\leq \limsup_{n\to\infty}P( \|\tau_{n}\{\hat{\Pi}_{n}-\Pi_{0}\}\|\geq \tau_{n}\sigma_{r_{0}}(\Pi_{0})(1-\kappa_{n}/\sigma_{r_{0}}(\Pi_{0}))=0~,
\end{align}
where the first inequality exploited $\kappa_n<\sigma_{r_0}(\Pi_0)$ for all $n$ sufficiently large by $\kappa_n\downarrow 0$, the second inequality again follows by the Weyl inequality \citep[Exercise 1.3.22(iv)]{Tao2012Matrix} and also $\sigma_{r_0}(\Pi_0)>0$, and the equality is because $\|\tau_{n}\{\hat{\Pi}_{n}-\Pi_{0}\}\|=O_p(1)$ by Assumption \ref{Ass: Weaklimit: Pihat}, $\tau_{n}\to\infty$ and $\kappa_{n}\downarrow 0$. Combining \eqref{Aux: Eqn: Lem: Consistency: RankEst: 1} and \eqref{Aux: Eqn: Lem: Consistency: RankEst: 2} yields
\begin{align*}
\limsup_{n\to\infty}P(\hat{r}_{n}\neq r_{0})\leq \limsup_{n\to\infty}P(\hat{r}_{n}<r_{0})+\limsup_{n\to\infty}P(\hat{r}_{n}>r_{0})=0~.
\end{align*}
This completes the proof of the lemma. \qed

\begin{lem}\label{Lem: concentration inequality}
Let $\mathbb G\in\mathbf R^{k}$ follow $N(\mu, \Omega_0)$ and $g: \mathbf R^k\to\mathbf R$ be a Lipschitz map with Lipschitz constant $L$. Then, for $M$ the median of $g(\mathbb G)$ and any $x> 0$
\begin{align}
P(g(\mathbb G)-M>x)\le \frac{1}{2}\exp\{-\frac{1}{2}\frac{x^2}{C^2}\}
\end{align}
for some $C> 0$ depending on $L$ and $\|\Omega_0\|$.
\end{lem}
\noindent{\sc Proof:} This is a mild extension of Lemma A.2.2 in \citet{Vaart1996}, and we include a proof here only for completeness. Since $\mathbb G\sim N(\mu,\Omega_0)$, we may write $\mathbb G\overset{d}{=}\Omega_0^{1/2}Z+\mu$ for some $Z\sim N(0,I_k)$. Define a map $h: \mathbf R^k\to\mathbf R$ by $h(z)=g(\Omega_0^{1/2}z+\mu)$ for any $z\in\mathbf R^k$. Then by Lipschitz continuity of $g$ we have: for any $z_1,z_2\in\mathbf R^k$,
\begin{multline}
|h(z_1)-h(z_2)|=|g(\Omega_0^{1/2}z_1+\mu)-g(\Omega_0^{1/2}z_2+\mu)|\le L\|\Omega_0^{1/2}z_1-\Omega_0^{1/2}z_2\|\\
\le L\|\Omega_0^{1/2}\|\|z_1-z_2\| \le L\|\Omega_0\|^{1/2}\|z_1-z_2\|~,
\end{multline}
where the fact $\|\Omega_0^{1/2}\|\le \|\Omega_0\|^{1/2}$ follows from Theorem X.1.1 in \citet{Bhatia1997Matrix}. By replacing $L\|\Omega_0\|^{1/2}$ with $(L\|\Omega_0\|^{1/2})\vee 1$ if necessary, we may assume $C\equiv L\|\Omega_0\|^{1/2}>0$ without loss of generality. Since $M$ is the median of $g(\mathbb G)$ and hence also of $h(Z)$, we conclude that $M/C$ is the median of $h(Z)/C$. It follows from Lemma A.2.2 in \citet{Vaart1996} that: for any $x> 0$,
\begin{align}
P(g(\mathbb G)-M> x)=P(\frac{h(Z)}{C}-\frac{M}{C}> \frac{x}{C}) \le \frac{1}{2}\exp\{-\frac{1}{2}\frac{x^2}{C^2}\}~.
\end{align}
This completes the proof of the lemma. \qed

For the next two lemmas, we let $\mathrm{BL}_1(\mathbf R)$ be the set of real-valued Lipschitz functions on $\mathbf R$ with levels and Lipschitz constants both bounded by one.

\begin{lem}\label{Lem: alpha adjustment}
Let $T_n^*: \{X_i,W_{ni}\}_{i=1}^n\to\mathbf R$ be a bootstrap estimator for the distribution of $g(\mathbb G)$ such that $\mathbb G\in\mathbf R^k$ is Gaussian, $g: \mathbf R^k\to\mathbf R$ is a Lipschitz map, and,
\begin{align}
\sup_{f\in\mathrm{BL}_1(\mathbf R)}|E_W[f(T_n^*)]-E[f(g(\mathbb G))]|=o_p(1)~.
\end{align}
Suppose Assumption \ref{Ass: alpha adjusted} holds. Let $\hat c_{n,1-\alpha_n}$ be $(1-\alpha_n)$ conditional quantiles of $T_n^*$ given the data. If the cdf of $g(\mathbb G)$ is continuous and strictly increasing on $[r_0,\infty)$ for some $r_0\in\mathbf R$, then $\hat c_{n,1-\alpha_n}/\tau_n\convp 0$.
\end{lem}
% Check if the strict monotonicity condition can be weakened.
\noindent{\sc Proof:} Let $\hat F_n$ be the conditional cdf of $T_n^*$ given $\{X_i\}_{i=1}^n$, and $F$ be the cdf of $g(\mathbb G)$. By Lemma 10.11 in \citet{Kosorok2008}, we have
\begin{align}\label{Eqn: alpha adjustment, aux1}
\sup_{t\in[r_0,\infty)}|\hat F_n(t)-F(t)|=o_p(1)~.
\end{align}
By the definition of quantiles, we thus obtain from \eqref{Eqn: alpha adjustment, aux1} that, for any $r\in[r_0,\infty)$,
\begin{multline}\label{Eqn: alpha adjustment, aux2}
\limsup_{n\to\infty}P(\hat c_{n,1-\alpha_n} \le r )\le \limsup_{n\to\infty}P(\hat F_n( r)\ge 1-\alpha_n )\\
=\limsup_{n\to\infty}P(o_p(1)+F(r)\ge 1-\alpha_n )=0~.
\end{multline}
where we exploited the facts that $F(r)<1$ by strict monotonicity of $F$ on $[r_0,\infty)$ and that $\alpha_n\downarrow 0$. Next, fix $\epsilon>0$. Combination of \eqref{Eqn: alpha adjustment, aux2} and Lemma \ref{Lem: concentration inequality} yields
\begin{multline}\label{Eqn: alpha adjustment, aux3}
\alpha_n < 1-\hat F_n (\hat c_{n,1-\alpha_n}-\epsilon) =P(g(\mathbb G)> \hat c_{n,1-\alpha_n}-\epsilon)+o_p(1) \\
\le  \frac{1}{2}\exp\{-\frac{1}{2}\frac{(\hat c_{n,1-\alpha_n}-\epsilon-c_{0.5})^2}{C^2}\}+o_p(1)~,
\end{multline}
for some $C>0$ and $c_{0.5}$ the $0.5$-quantile of $g(\mathbb G)$. It follows from \eqref{Eqn: alpha adjustment, aux3} that
\begin{align}\label{Eqn: alpha adjustment, aux4}
(\frac{\hat c_{n,1-\alpha_n}}{\tau_n}-\frac{\epsilon}{\tau_n}-\frac{c_{0.5}}{\tau_n})^2\le 2C^2 (-\frac{\log\alpha_n}{\tau_n^2}+\frac{\log o_p(1)}{\tau_n^2}-\frac{\log 2}{\tau_n^2})~.
\end{align}
By Assumption \ref{Ass: alpha adjusted}(ii), $\tau_n\uparrow\infty$ and $\log o_p(1)\convp-\infty$ as $n\to\infty$, we may then conclude the proof of the lemma from result \eqref{Eqn: alpha adjustment, aux4}. \qed

\begin{lem}\label{Lem: alpha adjustment2}
Suppose Assumptions \ref{Ass: Weaklimit: Pihat}, \ref{Ass: Boostrap: Pihat} and \ref{Ass: alpha adjusted} hold. Let $\hat c_{n,1-\alpha}$ be defined by \eqref{Eqn: Quantile} for $\alpha\in(0,1)$ where $\kappa_n\to 0$ and $\tau_n\kappa_n\to\infty$ if $\hat\phi_{r,n}''$ is defined by \eqref{Eqn: NumericalDevEst} but no restrictions on $\hat r_n$ if $\hat\phi_{r,n}''$ is defined by \eqref{Eqn: StructDevEst}. If $\mathcal M$ is Gaussian but not constant, then $\hat c_{n,1-\alpha_n}/\tau_n^2\convp 0$.
\end{lem}
% Gaussiannity is not important; the concentration is the key.
% Check Chen and Santos see if the starting point of the cdf r0=0 perhaps quantile bigger than zero if enough
\noindent{\sc Proof:} Consider first the case when $\hat c_{n,1-\alpha}$ is defined by the analytic derivative estimator. By Lemma \ref{Lem: phi, representation} and simple manipulations, we have
\begin{align}\label{Eqn: alpha adjustment2, aux1}
\hat\phi_{r,n}''(\hat{\mathcal M}_n^*)^{1/2}\le \|\hat P_{2,n}^\transpose \hat{\mathcal M}_n^* \hat Q_{2,n}\|\le (mk)^{1/2} \| \hat{\mathcal M}_n^*\|~.
\end{align}
Let $\tilde c_{n,1-\alpha}$ be the $(1-\alpha)$th conditional quantile of $\| \hat{\mathcal M}_n^*\|$ for each $\alpha\in(0,1)$. Since $\mathcal M$ is Gaussian and the variance of $\mathrm{vec}(\mathcal M)$ is nonzero, $\|\mathcal M\|^2$ is equal in law to a weighted sum of independent $\chi^2(1)$ random variables. It follows that the cdf $\|\mathcal M\|$ is continuous and strictly increasing on $\mathbf R_+$. In turn, by Proposition 10.7 in \citet{Kosorok2008}, Assumptions \ref{Ass: Boostrap: Pihat} and \ref{Ass: alpha adjusted},  we obtain from Lemma \ref{Lem: alpha adjustment} that $\tilde c_{n,1-\alpha}/\tau_n\convp 0$. By result \eqref{Eqn: alpha adjustment2, aux1} and equivariance of quantiles to monotone transformations, we may then conclude that
\begin{align}
\frac{\hat c_{n,1-\alpha_n}}{\tau_n^2}\le \frac{\hat c_{n,1-\alpha_n}^2}{\tau_n^2}=o_p(1)~.
\end{align}

Next, turn to the case when $\hat c_{n,1-\alpha}$ is defined by the numerical derivative estimator. For each $\alpha\in(0,1)$, let $\bar c_{n,1-\alpha}$ be the conditional quantile (given the data) of
\begin{align}
\kappa_n \hat\phi_{r,n}''(\hat{\mathcal M}_n^*)= \frac{\phi_r(\hat\Pi_n+\kappa_n \hat{\mathcal M}_n^*)-\phi_r(\hat\Pi_n) }{\kappa_n}~.
\end{align}
By Assumptions \ref{Ass: Weaklimit: Pihat}, \ref{Ass: Boostrap: Pihat} and the rates conditions on $\kappa_n$ as given, we may employ Proposition \ref{Pro: phi, differentiability} and Theorem 3.3 in \citet{Chen_Fang2015FOD} to conclude that
\begin{align}
\sup_{f\in\mathrm{BL}_1(\mathbf R)} |E_W[f(\kappa_n \hat\phi_{r,n}''(\hat{\mathcal M}_n^*))]-E[f(\phi_{r,\Pi_0}'(\mathcal M))]|=o_p(1)~.
\end{align}
By simple algebra we may obtain that: for any $M_1,M_2\in\mathbf M^{m\times k}$,
\begin{multline}\label{Eqn: alpha adjustment2, aux2}
|\phi_{r,\Pi_0}'(M_1)-\phi_{r,\Pi_0}'(M_2)|=|\min_{U\in\Psi(\Pi)} 2\mathrm{tr}(U^\transpose\Pi^\transpose M_1U)-\min_{U\in\Psi(\Pi)} 2\mathrm{tr}(U^\transpose\Pi^\transpose M_2U)|\\
\le \max_{U\in\Psi(\Pi_0)} 2 \|\Pi_0 U\|\|(M_1-M_2) U\|\le 2 \sqrt k \|\Pi_0\| \|M_1-M_2\|~.
\end{multline}
By result \eqref{Eqn: alpha adjustment, aux2} and Lemma \ref{Lem: alpha adjustment}, we thus have $\bar c_{n,1-\alpha_n}/\tau_n=o_p(1)$ and hence
\begin{align}
\frac{\hat c_{n,1-\alpha_n}}{\tau_n^2}\le \frac{\bar c_{n,1-\alpha_n}}{\tau_n}\frac{1}{\tau_n\kappa_n}=o_p(1)~,
\end{align}
since $\tau_n\kappa_n\to\infty$. This completes the proof of the lemma. \qed

We next present lemmas that are relevant to Section \ref{Sec: mathching} and proceed by imposing:

\begin{ass}\label{Ass: saliency} % $p\in\Delta^I$ and $q\in\Delta^J$; (iii)
(i) The supports of $X$ and $Y$ are finite; (ii) the Jacobian matrix of $\mathrm{vec}(E_{\pi(A,p,q)}[XY^\transpose])$ with respect to $\mathrm{vec}(A)$ at $A_0$ is nonsingular.
\end{ass}

% Assumption \ref{Ass: saliency}(ii) states that the matching distribution uniquely solves the centralized problem \eqref{Eqn: saliency transport with entropy} with $\Phi$ specified as in \eqref{Eqn: saliency transport with entropy} -- see \citet{GalichonSalanie2010Tradeoff}, \citet{DeckerLiebMcCannStephens2013Unique} and \citet{DupuyGalichon2014Personality} for more discussions.

Assumption \ref{Ass: saliency}(i) formalizes the setup that the matching attributes are finitely valued.  Assumption \ref{Ass: saliency}(ii) is a technical condition, as implicitly imposed in \citet{GalichonSalanie2010Tradeoff} and \citet{DupuyGalichon2014Personality} who showed that the Jacobian coincides with the Fisher information matrix for $A_0$.

Next, let the supports $\mathcal X=\{x_1,\ldots,x_I\}$ and $\mathcal Y=\{y_1,\ldots,y_J\}$. Then, we may identify $p_0$ and $q_0$ as vectors in $(0,1)^I$ and $(0,1)^J$ respectively.

\begin{lem}\label{Lem: saliency, affinity diff}
If Assumption \ref{Ass: saliency} holds, then the implicit map $A: (0,1)^I\times (0,1)^J\times\mathbf M^{m\times k}\to\mathbf M^{m\times k}$ defined by \eqref{Eqn: saliency, mathcing moments}, i.e., $A(p_0,q_0, E[XY^\transpose])=A_0$, is Hadamard differentiable on some open neighborhood of the truth $(p_0,q_0,E[XY^\transpose])$.
\end{lem}
\noindent{\sc Proof:} First, note that $A$ is uniquely defined by Lemma 3 in \citet{DupuyGalichon2014Personality}. Next, define a map $\Psi: \mathbf M^{m\times k}\times (0,1)^I\times(0,1)^J\times \mathbf M^{m\times k}\to\mathbf R^{mk}$ by:
\begin{align}\label{Eqn: saliency, affinity diff, aux1}
\Psi(A,p,q,\Sigma)\equiv\mathrm{vec}(E_{\pi(A,p,q)}[X^\transpose Y]-\Sigma)~.
\end{align}
By Assumption \ref{Ass: saliency} and Lemma \ref{Lem: saliency, likelihood diff}, $\Psi$ is continuously differentiable on some open neighborhood of the truth $(A_0,p_0,q_0,E[XY^\transpose])$ -- note in particular that $X$ and $Y$ are finitely supported. In turn, Assumption \ref{Ass: saliency}(ii) allows us to invoke the implicit function theorem, see, for example, Theorem 9.28 in \citet{Rudin1976}, to conclude the proof. \qed

\iffalse
For notational simplicity, we write $a_0\equiv\mathrm{vec}(A_0)$ and $\theta_0\equiv (p^\transpose,q^\transpose,\mathrm{vec}(\Sigma_0)^\transpose)^\transpose$. Moreover, we define a map $\Psi: \mathbf R^{mk}\times (0,1)^I\times(0,1)^J\times \mathbf R^{mk}\to\mathbf R^{mk}$ by:
\begin{align}\label{Eqn: saliency, affinity diff, aux1}
\Psi(a,\theta)\equiv \mathrm{vec}(E_{\pi(A,p,q)}[X^\transpose Y]-\Sigma)~,
\end{align}
for any $A\in\mathbf M^{m\times k}$, $p\in(0,1)^I$, $q\in(0,1)^J$ and $\Sigma\in\mathbf M^{m\times k}$, with $a\equiv\mathrm{vec}(A)$ and $\theta\equiv(p^\transpose,q^\transpose,\mathrm{vec}(\Sigma)^\transpose)^\transpose$. We may thus identify \eqref{Eqn: saliency, mathcing moments} as a system of equations defined by
\begin{align}\label{Eqn: saliency, affinity diff, aux2}
\Psi(a_0,\theta_0)=0~.
\end{align}
It is now equivalent to showing that equation \eqref{Eqn: saliency, affinity diff, aux2} defines an implicit map $\theta_0\mapsto a(\theta_0)\equiv a_0$ that is Hadamard differentiable.
\fi

\begin{lem}\label{Lem: saliency, likelihood diff}
If Assumption \ref{Ass: saliency}(i) holds, then the map $(A_0,p_0,q_0)\mapsto \pi(A_0,p_0,q_0)(x,y)$ defined by \eqref{Eqn: saliency transport with entropy} where $\Phi$ is specified as in \eqref{Eqn: saliency, bilinear form} uniquely exists and is continuously differentiable on some open neighborhood of the truth $(A_0,p_0,q_0)$, for each $(x,y)\in\mathcal X\times\mathcal Y$.
\end{lem}
\noindent{\sc Proof:} First, we may rewrite the maximization problem \eqref{Eqn: saliency transport with entropy} as
\begin{align}\label{Eqn: saliency, likelihood diff, aux1}
\max_{\pi} \sum_{i=1}^{I}\sum_{j=1}^{J} \pi_{ij} x_i^\transpose A_0y_j-\sum_{i=1}^{I}\sum_{j=1}^{J}  \pi_{ij}\log\pi_{ij}~,
\end{align}
subject to: for all $i=1,\ldots,I$ and all $j=1,\ldots,J$,
\begin{align}
\sum_{j=1}^{J}\pi_{ij}=p_{0,i} ~, \quad \sum_{i=1}^{I}\pi_{ij}=q_{0,j}~, \label{Eqn: saliency, likelihood diff, aux3}
\end{align}
where $p_{0,i}=P(X=x_i)$ and $q_{0,j}=P(Y=y_j)$ for all $i=1,\ldots,I$ and all $j=1,\ldots,J$. By defining $x\log x=0$ if $x=0$, it is simple to see that the objective function in \eqref{Eqn: saliency, likelihood diff, aux1} is continuous. Since the constraints define a compact domain for $\pi$, it follows that an optimal matching distribution $\pi_0$ always exists. The uniqueness of $\pi_0$ follows from strict concavity of the objective function since $x\mapsto x\log x$ is strictly convex. Moreover, the right derivative of the objective function at 0 is infinite -- see equation \eqref{Eqn: saliency, likelihood diff, aux4} below or \citet[p.5]{GalichonSalanie2010Tradeoff}, implying that the optimal $\pi_0$ must satisfy $0<\pi_{0,ij}<1$ for all $i$ and $j$. Exploiting the constraints in \eqref{Eqn: saliency, likelihood diff, aux3}, together with the facts that $p_0,q_0$ and $\pi$ are pmfs, the constrained optimization can be converted into an unconstrained one in which the objective function in \eqref{Eqn: saliency, likelihood diff, aux1} is a function of $\{p_{0,i}\}_{i=1}^{I-1}$, $\{q_{0,j}\}_{j=1}^{J-1}$ and $\{\pi_{0,ij}\}_{i=1,j=1}^{I-1,J-1}$ only, with $\pi_{0,iJ}=p_{0,i}-\sum_{j=1}^{J-1}\pi_{0,ij} $, $\pi_{0,Ij}=q_{0,j}-\sum_{i=1}^{I-1}\pi_{0,ij}$ for all $i=1,\ldots,I-1$ and $j=1,\ldots,J-1$, and
\begin{multline}
\pi_{0,IJ}=1-\sum_{i=1}^{I-1}\sum_{j=1}^{J-1}\pi_{0,ij}-\sum_{i=1}^{I-1}\pi_{0,iJ}-\sum_{j=1}^{J-1}\pi_{0,Ij}\\ =1+\sum_{i=1}^{I-1}\sum_{j=1}^{J-1}\pi_{0,ij}-\sum_{i=1}^{I-1}p_{0,i}-\sum_{j=1}^{J-1}q_{0,j}~.
\end{multline}
It follows that the unique maximizer $\pi_0$ must satisfy the first order condition:
\begin{multline}
x_i^\transpose A_0y_j-x_I^\transpose A_0 y_j-x_i^\transpose A_0y_J+ x_I^\transpose A_0y_J -1-\log \pi_{0,ij}\\ +1+\log \pi_{0,Ij}+1+\log \pi_{0,iJ}-1-\log\pi_{0,IJ}=0~,
\end{multline}
or equivalently
\begin{multline}\label{Eqn: saliency, likelihood diff, aux4}
x_i^\transpose A_0y_j-x_I^\transpose A_0 y_j-x_i^\transpose A_0y_J+ x_I^\transpose A_0y_J\\ -\log \pi_{0,ij}+\log \pi_{0,Ij}+\log \pi_{0,iJ}-\log\pi_{0,IJ}=0~,
\end{multline}
for all $i=1,\ldots,I-1$ and $j=1,\ldots,J-1$, where $\pi_{0,iJ}$, $\pi_{0,Ij}$ and $\pi_{0,IJ}$ are functions of $\{p_{0,i}\}_{i=1}^{I-1}$, $\{q_{0,j}\}_{j=1}^{J-1}$ and $\{\pi_{0,ij}\}_{i=1,j=1}^{I-1,J-1}$ as defined previously.

Let us stack the equations in \eqref{Eqn: saliency, likelihood diff, aux4} along $i=1,\ldots,m$ sequentially for fixed $j=1,\ldots,k$, and let $d^*\equiv (I-1)(J-1)$. The left side of \eqref{Eqn: saliency, likelihood diff, aux4} is then a $\mathbf R^{d^*}$-valued function of $A_0$, $\{p_{0,i}\}_{i=1}^{I-1}$, $\{q_{0,j}\}_{j=1}^{J-1}$ and $\{\pi_{0,ij}\}_{i=1,j=1}^{I-1,J-1}$, which is obviously continuously differentiable. Moreover, the derivative of the left side in \eqref{Eqn: saliency, likelihood diff, aux4} with respect to $\mathrm{vec}(\{\pi_{0,ij}\}_{i=1,j=1}^{I-1,J-1})$ is then a matrix of size $d^*\times d^*$ which is given by: for $\mathbb J_d$ a generic $d\times d$ matrix of ones,
\begin{align}\label{Eqn: saliency, likelihood diff, aux5}
-\underline\pi_0-\underline\pi_{0,J} \otimes \mathbb J_{I-1}-\mathbb J_{J-1}\otimes \underline\pi_{0,I}-\pi_{0,IJ}^{-1} \mathbb J_{d^{*2}}~,
\end{align}
with $\underline\pi_{0,I}\equiv\mathrm{diag}(\{\pi_{0,iJ}^{-1}\}_{i=1}^{I-1})$, $\underline\pi_{0,J}\equiv\mathrm{diag}(\{\pi_{0,Ij}^{-1}\}_{j=1}^{J-1})$ and $\underline\pi_0\equiv\mathrm{diag}(\mathrm{vec}(\{\pi_{0,ij}^{-1}\}_{i=1,j=1}^{I-1,J-1}))$. Note that $\underline\pi_0$ is positive definite while $\underline\pi_{0,J} \otimes\mathbb  J_{I-1}$, $\mathbb J_{J-1}\otimes \underline\pi_{0,I}$ and $\pi_{0,IJ}^{-1}\otimes \mathbb J_{d^{*2}}$ are positive semidefinite, so the matrix in \eqref{Eqn: saliency, likelihood diff, aux5} is invertible. The conclusion now follows from the implicit function theorem -- see, for example, Theorem 9.28 in \citet{Rudin1976}. \qed

\section{Cointegration and Additional Examples}\label{Sec: B}
\renewcommand{\theequation}{E.\arabic{equation}}
\setcounter{equation}{0}

In this section, we present additional examples where knowledge on the rank of a matrix is of interest. We single out the treatment of inference on cointegration rank because (i) it is prominent in applied macroeconomics and (ii) Assumption \ref{Ass: Weaklimit: Pihat} may take a generalized form in this case where the convergence rates are heterogenous across entries of $\hat\Pi_n$ (but still falls within the scope of the Delta method).

\subsection{Inference on Cointegration Rank}\label{Sec: coint appendix}

Let $\{Y_t\}$ be a time series in $\mathbf R^k$ such that all its entries are unit root processes. For ease of exposition and to hight what is essential to our theory, we limit ourselves to processes without deterministic terms throughout. By the Granger representation theorem, the number $h_0$ of independent cointegrating vectors is precisely equal to $k-\mathrm{rank}(\Omega_0)$ with $\Omega_0$ the long run variance of $\Delta Y_t$.\footnote{By definition, $Y_t$ is said to be cointegrated if there is nonzero vector $\lambda\in\mathbf R^k$ such that $\lambda^\transpose Y_t$ is stationary, in which case $\lambda$ is called a cointegrating vector.} Within this (nonparametric) system framework, one may be interested in testing
\begin{align}\label{Eqn: coint test}
\mathrm H_0: h_0\ge h\qquad \text{v.s.}\qquad\mathrm H_1: h_0<h
\end{align}
for a given integer $h=1,\ldots,k-1$, which is equivalent to \eqref{Eqn: hypothesis, intro} with $\Pi_0=\Omega_0$ and $r=k-h$. The special case with $h=1$ is concerned with testing the null of cointegration. Problems of similar nature have been studied by \citet{StockWaston1988Trend}, \citet{Harris1997PCA}, \citet{Snell1999Coint} and \citet{NyblomHarvey2000Trend}, but they test the null $h_0=h$ instead. Note that the nonparametric tests of \citet{Bierens1997NPCoInt} and \citet{Shintani2001CoInt} are not directly applicable to \eqref{Eqn: coint test} because they test the null $h_0=h$ against $h_0>h$ (larger cointegration rank).

The testing problem \eqref{Eqn: coint test} is not only of interest in its own right \citep{Hayashi2000}, but also important as a complement to tests against larger cointegration rank especially in view of the potentially poor power of the latter tests, as forcefully argued by \citet{KwiatkowskiPhillipsSchmidtShin1992KPSS} and \citet{MaddalaKim1998Coint}. Nevertheless, through VAR or error-correction representations, our framework can accommodate the hypotheses
\begin{align}\label{Eqn: coint test2}
\mathrm H_0: h_0\le h\qquad \text{v.s.}\qquad\mathrm H_1: h_0>h~.
\end{align}
To see this, suppose that the error-correction representation of $\{Y_t\}$ is given by
\begin{align}\label{Eqn: coint, ECM}
\Delta Y_t=\Phi_0 Y_{t-1} + \sum_{j=1}^{p-1}\Phi_j\Delta Y_{t-j}+\epsilon_t~,
\end{align}
for some white noise $\{\epsilon_t\}$. Then $h_0=\mathrm{rank}(\Phi_0)$ by the Granger representation theorem, and hence \eqref{Eqn: coint test2} is equivalent to \eqref{Eqn: hypothesis, intro} with $\Pi_0=\Phi_0$ and $r=h$. The setup \eqref{Eqn: coint test2} is studied in the seminal work of \citet{Johansen1988CoInt,Johansen1991CoIntegration}, for which Johansen proposes the celebrated maximum eigenvalue test and the trace test, and derives their asymptotic distributions under $h_0=h$. The general limits under $h_0\le h$ are presented in \citet{Johansen1995likelihood} but no critical values are provided.

Below we study the problems \eqref{Eqn: coint test} and \eqref{Eqn: coint test2} separately as their treatments require different arguments, and proceed with the former.

\subsubsection*{Nonparametric Cointegration Test}

We start by estimating the long run variance $\Pi_0\equiv\Omega_0$ based on the periodogram. Specifically, for a kernel/density function $K: \mathbf R\to\mathbf R_+$, let
\begin{align}
\hat\Pi_n=\frac{2\pi}{n}\sum_{j=-\lfloor(n-1)/2\rfloor}^{\lfloor n/2\rfloor}  K_{b_n}(\omega_j) I_{\Delta Y,n}(\omega_j)~,
\end{align}
where $\lfloor a\rfloor$ is the integer part of $a\in\mathbf R$,  $K_{b_n}(\cdot)=K(\cdot/b_n)/b_n$, $\omega_j=2\pi j/n$ with $j=-\lfloor(n-1)/2\rfloor,\ldots,\lfloor n/2\rfloor$ are the natural frequencies, $b_n\to 0$ is a suitable bandwidth, and $\omega\mapsto I_{\Delta Y,n}(\omega)\in\mathbf M^{k\times k}$ is the periodogram of $\{\Delta Y_t\}_{t=1}^n$, i.e.,
\begin{align*}
I_{\Delta Y,n}(\omega) =\frac{1}{2\pi n}(\sum_{t=1}^{n} \Delta Y_t e^{-it\omega})(\sum_{t=1}^{n} \Delta Y_t e^{-it\omega})^{\mathrm{H}}
\end{align*}
for $M^{\mathrm{H}}$ denoting the Hermitian transpose of a generic complex matrix $M$. Then $\hat{\Pi}_{n}$ is asymptotically normal at the rate $\sqrt{nb_{n}}$ under regularity conditions -- see, for example, \citet[Theorem V.5.11]{Hannan1970TS} and \citet[Theorem 7.4.4]{Brillinger2001TS} for classical treatments, and \citet{PhillipsSunJin2006Spectral} and \citet{Politis2011Higher} for recent developments.

We construct the estimator $\hat{\mathcal M}_n^*$ employing the multivariate linear process bootstrap recently developed by \citet{JentschPolitis2015LPB}. Let $\hat\Gamma_{n,j}=\sum_{t=1}^{n-j}\Delta Y_t\Delta Y_{t+j}^\transpose/n$ for $j\ge 0$ and $\hat\Gamma_{n,j}=\hat\Gamma_{n,-j}^\transpose$ if $j<0$. Define $\hat V_n\in\mathbf M^{nk\times nk}$ to be a block matrix whose $(i,j)$th block is given by $\varrho((i-j)/l_n) \hat\Gamma_{n,i-j}$ for $i,j=1,\ldots,n$, where $\varrho:\mathbf R\to\mathbf R$ is a flat-top kernel and $l_n$ is a banding parameter \citep{Politis2001Flat,Politis2011Higher}. The matrix $\hat V_n$ serves as an estimator of the covariance matrix of $\Delta_n=[\Delta Y_1^\transpose,\ldots,\Delta Y_n^\transpose]^\transpose\in\mathbf R^{nk}$. One may modify $\hat V_n$ if necessary to ensure that it is positive definite \citep[p.1124]{JentschPolitis2015LPB}. Let $Z_n=L_n^{-1} \Delta_n\in\mathbf R^{nk}$ where $L_n$ is from the Cholesky decomposition $\hat V_n=L_nL_n^\transpose$, $Z_{n,i}$ the $i$th entry of $Z_n$, and $\bar Z_{n,i}=(Z_{n,i}-\bar Z_n)/\hat\sigma_n$ for $\bar Z_n=\sum_{i=1}^{nk}Z_{n,i}/(nk)$ and $\hat\sigma_n^2=\sum_{i=1}^{nk}(Z_{n,i}-\bar Z_n)^2/(nk)$. Now, draw an i.i.d.\ sample $\{Z_{n,i}^*\}_{i=1}^{nk}$ from $\{\bar Z_{n,i}\}_{i=1}^{nk}$ with replacement. Define $Z_n^*\in\mathbf R^{nk}$ whose $i$th entry is $Z_{n,i}^*$, and let $\Delta_n^*=L_n Z_n^*\in\mathbf R^{nk}$.

Finally, our bootstrap sample $\{\Delta Y_t^*\}_{t=1}^n$ is such that $\Delta_n^*=(\Delta Y_1^{*\transpose},\ldots,\Delta Y_n^{*\transpose})^\transpose$, and then the bootstrap estimator $\hat\Pi_n^*$ is defined analogously to $\hat\Pi_n$ but with $\{\Delta Y_{t}\}_{t=1}^n$ replaced by $\{\Delta Y_t^*\}_{t=1}^n$. In order to construct a bootstrap estimator $\hat{\mathcal M}_n^*$ that satisfies Assumption \ref{Ass: Boostrap: Pihat}, we need to properly center $\hat\Pi_n^*$, as is well understood for bootstrap in nonparametric settings \citep{Hall1992bootstrap}. To this end, define
\begin{align}
\tilde\Pi_n=\sum_{j=-(n-1)}^{n-1}\varrho(\frac{j}{l_n})\hat\Gamma_{n,j}~.
\end{align}
Under regularity conditions, the bootstrap consistency of $\hat{\mathcal M}_n^*\equiv\sqrt{nb_n}\{\hat\Pi_n^*-\tilde\Pi_n\}$ is formally established by Theorem 4.2 in \citet{JentschPolitis2015LPB}. We refer the reader to \citet{JentschKreiss2010Multiple}, \citet{PolitisRomano1993Field}, \citet{Politis_Romano1994subsample}, and \citet{BerkowitzDiebold1998Spectra} for alternative resampling schemes.

\subsubsection*{Cointegration Test in Error-Correction Models}

Now consider the error-correction model, and suppose that $\{\epsilon_t\}$ is a white noise having nonsingular covariance matrix $\Sigma_0$. Since $h_0=\mathrm{rank}(\Phi_0)$ under \eqref{Eqn: coint, ECM}, the problem \eqref{Eqn: coint test2} is equivalent to \eqref{Eqn: hypothesis, intro} by identifying $\Pi_0$ with $\Phi_0$ and $r$ with $h\in\{0,1,\ldots,k-1\}$. The special case $h=0$ reduces to a test of no cointegration against existence of cointegration. For ease and transparency of our exposition, suppose $p=1$ and hence there are no lagged variables $\Delta Y_{t-j}$ in \eqref{Eqn: coint, ECM}; the general case can be handled in a straightforward manner by combining our arguments below with Lemma A.6 in \citet{LiaoPhillips2015Automated}.

We proceed with some clarifications on notation. Let $\Pi_0=P_0\Sigma_0 Q_0^\transpose$ be a singular value decomposition of $\Pi_0$; write $P_0=[P_{0,1},P_{0,2}]$ and $Q_0=[Q_{0,1},Q_{0,2}]$ where $P_{0,1}\in\mathbb S^{k\times r_0}$ and $Q_{0,1}\in\mathbb S^{k\times r_0}$ with $r_0\equiv\mathrm{rank}(\Pi_0)$. On the other hand, it is more common to have $\Pi_0=\alpha_0\beta_0^\transpose$ where $\alpha_0\in\mathbf M^{k\times r_0}$ (whose columns are called adjustment coefficients) and $\beta_0\in\mathbf M^{k\times r_0}$ (whose columns are cointegrating vectors) both have full rank $r_0$ \citep{Johansen1995likelihood}. As pointed out by \citet{Johansen1988CoInt,Johansen1991CoIntegration}, $\alpha_0$ and $\beta_0$ are not identified, but their column spaces are. To flesh out the connections between these two sets of notation, let $\Sigma_{0,1}\in\mathbf M^{r_0\times r_0}$ be the left top block of $\Sigma_0$. By direct calculations, we obtain $\Pi_0=P_{0,1}\Sigma_{0,1}Q_{0,1}^\transpose$. Consequently, we may take $\alpha_0=P_{0,1}\Sigma_{0,1}$ and $\beta_0=Q_{0,1}$. In turn, the corresponding orthogonal complement versions  $\alpha_{0,\perp}\in\mathbf M^{k\times (k-r_0)}$ and $\beta_{0,\perp}\in\mathbf M^{k\times (k-r_0)}$ (both of full rank) can be taken to be $\alpha_{0,\perp}=P_{0,2}$ and $\beta_{0,\perp}=Q_{0,2}$, satisfying $\alpha_{0,\perp}^\transpose\alpha_0=\mathbf 0_{(k-r_0)\times r_0}$ and $\beta_{0,\perp}^\transpose\beta_0=\mathbf 0_{(k-r_0)\times r_0}$ as required. Finally, define
\begin{align*}
B_0\equiv \begin{bmatrix}
\beta_0^\transpose\\
\alpha_{0,\perp}^\transpose
\end{bmatrix}=\begin{bmatrix}
Q_{0,1}^\transpose\\
P_{0,2}^\transpose
\end{bmatrix}~.
\end{align*}

In applying our inferential framework, we need to construct a matrix estimator $\hat\Pi_n$ that converges weakly and a consistent bootstrap analog. In what follows, let $\{Y_t\}_{t=0}^n$ be a time series sample in $\mathbf R^k$ that is generated according to \eqref{Eqn: coint, ECM} (with $p=1$).

\noindent\underline{\sc Asymptotic Distributions:} For this, we employ the OLS estimator:
\begin{align}
\hat\Pi_n=(\sum_{t=1}^{n}\Delta Y_t Y_{t-1}^\transpose) (\sum_{t=1}^{n}Y_{t-1} Y_{t-1}^\transpose)^{-1}~.
\end{align}
Under standard regularity conditions, Lemma A.2 in \citet{LiaoPhillips2015Automated}, together with the continuous mapping theorem, implies that
\begin{align}\label{Eqn: chisquare: coint0}
\{\hat{\Pi}_{n}-\Pi_{0}\}B_{0}^{-1}D_{n}B_{0}\convl \mathcal{M}\equiv \mathcal{M}_{1}+\mathcal{M}_{2}~,
\end{align}
where $D_{n}\equiv \text{diag}(\sqrt{n}\mathbf{1}_{r_{0}},n \mathbf{1}_{k-r_{0}})$, $\text{vec}(\mathcal{M}_{1}^{\transpose})\sim N(0,\Sigma_0\otimes(Q_{0,1}\Sigma_{1}^{-1}Q^{\transpose}_{0,1}))$ with $\Sigma_{1}\equiv \text{Var}(Q_{0,1}^{\transpose}Y_{t})$, and $\mathcal M_2\in\mathbf M^{k\times k}$ is such that
\begin{align}
\mathcal{M}_{2}\sim \Sigma_0^{1/2}\int_{0}^{1}dB_{k}(t)B_{k}(t)^{\transpose}\Sigma_0^{1/2}P_{0,2}(P_{0,2}^{\transpose}\Sigma_0^{1/2}\int_{0}^{1}B_{k}(t)B_{k}(t)^{\transpose}dt\,\Sigma_0^{1/2}P_{0,2})^{-1}P_{0,2}^{\transpose}
\end{align}
with $B_{k}(\cdot)$ is a $k$-dimensional standard Brownian motion defined on the unit interval.

Inspecting result \eqref{Eqn: chisquare: coint0}, it seems that Assumption \ref{Ass: Weaklimit: Pihat} is being violated because the ``convergence rate'' $B_{0}^{-1}D_{n}B_{0}$ is not a scalar. However, this creates no conceptual difficulties if we interpret $\tau_n$ there as linear maps $\tau_n: \mathbf M^{m\times k}\to\mathbf M^{m\times k}$ -- such an insight has been noted in \citet[p.413]{Vaart1996}. In the current setup, we have $\tau_n: \mathbf M^{m\times k}\to\mathbf M^{m\times k}$ defined by: for any $M\in\mathbf M^{k\times k}$,
\begin{align}
\tau_nM\equiv\tau_n(M)= M B_{0}^{-1}D_{n}B_{0}~.
\end{align}
Therefore, in order to invoke the Delta method, the only question that remains is: is our map $\phi_r$ as defined in \eqref{Eqn: phi, defn} suitably differentiable with respect to these linear maps? The answer is affirmative, as shown by Proposition \ref{Pro: Derivative: Ex: Conintegration_VAR} stated at the end of this subsection. In particular, if $\mathrm{rank}(\Pi_0)\le r\equiv h$, then we have
\begin{align}\label{Eqn: coint, VAR derivative}
\lim_{n\rightarrow \infty}\frac{\phi_r(\Pi_0 +\tau_n^{-1}M_n)-\phi_r(\Pi_0)}{n^{-2}} = \phi_{r,\Pi_0}''(M)\equiv \sum_{j=h-h_{0}+1}^{k-h_{0}}\sigma^{2}_{j}(P_{0,2}^{\transpose} M Q_{0,2})~,
\end{align}
whenever $M_n\to M$ as $n\to\infty$. By a modification of the Delta method -- see Proposition \ref{Pro: Weaklimit: Ex: Conintegration_VAR}, we thus obtain from \eqref{Eqn: coint, VAR derivative} and \eqref{Eqn: chisquare: coint0} that, under $\mathrm{H}_0$ in \eqref{Eqn: coint test2},
\begin{align}\label{Eqn: coint, VAR weak limit}
n^2\phi_r(\hat\Pi_n)\convl \phi_{r,\Pi_0}''(\mathcal M)\equiv \sum_{j=h-h_{0}+1}^{k-h_{0}}\sigma^{2}_{j}(P_{0,2}^{\transpose} \mathcal M Q_{0,2})~.
\end{align}
The limit in \eqref{Eqn: coint, VAR weak limit} shows the importance of acknowledging the generic possibility that the true cointegration rank $h_0$ may be strictly less than the hypothesized value $h$.

\iffalse
This does not matter for testing the null of no cointegration (i.e., $h=0$) as $h_0$ must be equal to $h$ under $\mathrm H_0$ in this case.
\fi

In positioning our work in the literature, we note that existing tests are mainly based on the following standardized version of $\hat{\Pi}_{n}$ \citep{HubrichLutkepohlSaikkonen2001Coint,Sadoon2017Rank}:
\begin{align}
\hat{\Pi}_{s,n} = (\sum_{t=1}^{n}\Delta Y_{t}\Delta Y_{t}^{\transpose})^{-1/2} \hat{\Pi}_{n} (\sum_{t=1}^{n}Y_{t-1}Y_{t-1}^{\transpose})^{1/2}~.
\end{align}
For example, the classical trace statistic of \citet{Johansen1991CoIntegration,Johansen1988CoInt} is given by
\begin{align}
\mathrm{LR}_{n}(\hat{\Pi}_{s,n}) = -n\sum_{j=r+1}^{k}\log(1-\sigma^{2}_{j}(\hat{\Pi}_{s,n}))~.
\end{align}
whose asymptotic distribution under $\mathrm H_0': h_0=h$ is: for $d_0\equiv k-r_0$,
\begin{align}\label{Eqn: chisquare: coint6}
\mathrm{tr}\big(\int_{0}^{1}dB_{d_0}(t)B_{d_0}(t)^{\transpose}(\int_{0}^{1}B_{d_0}(t)B_{d_0}(t)^{\transpose}dt)^{-1}\int_{0}^{1}B_{d_0}(t)dB_{d_0}(t)^{\transpose}\big)~.
\end{align}
The asymptotic distribution under $\mathrm H_0: h_0\le h$, however, is different from \eqref{Eqn: chisquare: coint6} in general -- see \citet[p.157-8,168]{Johansen1995likelihood}. This may adversely affect the trace test through channels as discussed in the main text, and hence in turn provides an alternative explanation on why its finite sample performance can be poor, as documented in the literature \citep{MaddalaKim1998Coint,Johansen2002Correction}.

% As noted by \citet{Kleibergen_Paap2006rank}, the standardization is not necessary here because both $Y_t$ and $\Delta Y_t$ are different realizations of the same (economic) variables over time. So we rather focus on $\hat{\Pi}_{n}$ instead of $\hat{\Pi}_{s,n}$ in this paper.

%When $r_{0}=k-1$, \citet{Kleibergen_Paap2006rank} proved that a normalized version of $n^{2}\phi_r(\hat{\Pi}_{n})$ follows a pivotal but nonanalytical distribution under additional condition. When $r_{0}<k-1$, the asymptotic distribution can be more highly nonstandard.

% COMPARE TO KP and others

\noindent\underline{\sc Bootstrap Inference:} The limiting distribution in \eqref{Eqn: coint, VAR weak limit} is highly nonstandard, and in particular depends on the true rank $h_0$. In order to apply our bootstrap procedure, we need to estimate both the ``derivative'' $\phi_{r,\Pi_0}''$ and the limit $\mathcal M$. Estimation of $\phi_{r,\Pi_0}''$ is no more special than what we have discussed in Section \ref{Sec: 3-3}. For example, one may estimate $\phi_{r,\Pi_0}''$ by \eqref{Eqn: StructDevEst} with $\hat r_n$ given by \eqref{Eqn: bootstrap overview0}. Since $\sqrt n \{\hat\Pi_n-\Pi_0\}$ converges in distribution by \eqref{Eqn: coint, VAR weak limit}, we thus still have $\hat r_n\convp r_0= h_0$ provided $\kappa_n\to 0$ and $\sqrt n\kappa_n\to\infty$ by Lemma \ref{Lem: Consistency: RankEst}. Condition \eqref{Eqn: derivative consistency} in turn follows from Lemma \ref{Lem: Consistency: DerivEst}. In fact, one can show along the lines in the proof of Lemma \ref{Lem: Consistency: RankEst} that it suffices to have $\kappa_n\to 0$ and $n\kappa_n\to\infty$.

Given an estimator $\hat\phi_{r,n}''$ of $\phi_{r,\Pi_0}''$, we may thus approximate the law of $\phi_{r,\Pi_0}''(\mathcal M)$ in \eqref{Eqn: coint, VAR weak limit} by the conditional law (given the data) of $\hat\phi_{r,n}''(\hat{\mathcal M}_n^*)$ as long as $\hat{\mathcal M}_n^*$ is consistent for $\mathcal M$. To this end, we employ a residual-based bootstrap following \citet{Giersbergen1996BootTrace}, \citet{Swensen2006Bootstrap} and \cite{CavaliereRahbekTaylor2012BootstrapCoIn}, who study bootstrap cointegration tests for $\mathrm{H}_0': h_0=h$ based on error-correction models. Although these rank tests are potentially subject to the deficiencies illustrated in Sections \ref{Sec: 2} and \ref{Sec: 4-3}, their work show that the residual bootstrap procedure produces bootstrap samples that mimic the data well, a property we exploit directly. Moreover, in order to properly account for the possibility $h_0\equiv \mathrm{rank}(\Pi_0)<h$, we need a (preliminary) estimator $\hat r_n$ for $h_0$ that is consistent under both $\mathrm H_0$ and $\mathrm H_1$. For example, in view of Lemma \ref{Lem: Consistency: RankEst}, we may take
\begin{align}
\hat r_n=\max\{j=1,\ldots,k: \sigma_j(\hat\Pi_n)\ge \kappa_n\}~,
\end{align}
if the set is nonempty and $\hat r_n=0$ otherwise, where $\kappa_n\to 0$ and $n \kappa_n\to\infty$. The residual bootstrap now goes as follows.
\iffalse
\begin{enumerate}
  \item[]\underline{Step 1:} Given an estimator $\hat r_n$ that is consistent for $h_0$ under both $\mathrm H_0$ and $\mathrm H_1$, calculate the reduced rank estimator $\tilde\Pi_{n}$ following the maximum likelihood approach of \citet{Johansen1988CoInt,Johansen1991CoIntegration}, and obtain the residuals $\{\hat{\epsilon}_{t}\}$ as well as their centered versions $\{\bar{\epsilon}_{t}\}$, i.e., $\bar{\epsilon}_{t}\equiv \hat{\epsilon}_{t}-n^{-1}\sum_{t=1}^{n}\hat{\epsilon}_{t}$.
  \item[]\underline{Step 2:} Check if $|I_{k}-\lambda(\tilde\Pi_{n} + I_k)|=0$ has roots on or outside the unit circle, and if $\tilde{P}_{2,n}^{\transpose}\tilde{Q}_{2,n}$ has full rank, where the columns in $\tilde{P}_{2,n}$ and $\tilde{Q}_{2,n}$ are left and right singular vectors of $\tilde\Pi_{n}$ associated with the smallest $k-\hat r_n$ singular values. If so, proceed to the next step -- see Remark 1 in \citet{Swensen2006Bootstrap} for related discussions.
  \item[]\underline{Step 3:} Construct a bootstrap sample $\{Y_{t}^{\ast}\}_{t=1}^{n}$ recursively from
   \begin{align*}
   \Delta Y_t^*=\tilde\Pi_{n} Y_{t-1}^*+\epsilon_t^*~,
   \end{align*}
   with the initial value $Y_{0}$ and $\epsilon^{\ast}_{t}$ being generated from $\{\bar{\epsilon}_{t}\}_{t=1}^{n}$ by the nonparametric bootstrap. Calculate the bootstrap least square estimator
      \begin{align}
      \hat{\Pi}^{\ast}_{n} = \sum_{t=1}^{n}\Delta Y^{\ast}_{t}Y_{t-1}^{\ast\transpose}(\sum_{t=1}^{n}Y^{\ast}_{t-1}Y_{t-1}^{\ast\transpose})^{-1}~.
      \end{align}
\end{enumerate}
\fi

\par
\begingroup
\leftskip2em
\rightskip\leftskip

\noindent\underline{Step 1:} Given an estimator $\hat r_n$ that is consistent for $h_0$ under both $\mathrm H_0$ and $\mathrm H_1$, calculate the reduced rank estimator $\tilde\Pi_{n}$ following the maximum likelihood approach of \citet{Johansen1988CoInt,Johansen1991CoIntegration}, and obtain the residuals $\{\hat{\epsilon}_{t}\}$ as well as their centered versions $\{\bar{\epsilon}_{t}\}$, i.e., $\bar{\epsilon}_{t}\equiv \hat{\epsilon}_{t}-n^{-1}\sum_{t=1}^{n}\hat{\epsilon}_{t}$.

\noindent\underline{Step 2:} Check if $|I_{k}-\lambda(\tilde\Pi_{n} + I_k)|=0$ has roots on or outside the unit circle, and if $\tilde{P}_{2,n}^{\transpose}\tilde{Q}_{2,n}$ has full rank, where the columns in $\tilde{P}_{2,n}$ and $\tilde{Q}_{2,n}$ are left and right singular vectors of $\tilde\Pi_{n}$ associated with the smallest $k-\hat r_n$ singular values. If so, proceed to the next step -- see Remark 1 in \citet{Swensen2006Bootstrap} for discussions.

\noindent\underline{Step 3:} Construct a bootstrap sample $\{Y_{t}^{\ast}\}_{t=1}^{n}$ recursively from
   \begin{align*}
   \Delta Y_t^*=\tilde\Pi_{n} Y_{t-1}^*+\epsilon_t^*~,
   \end{align*}
   with the initial value $Y_{0}$ and $\epsilon^{\ast}_{t}$ being generated from $\{\bar{\epsilon}_{t}\}_{t=1}^{n}$ by the nonparametric bootstrap. Calculate the bootstrap least square estimator
      \begin{align}
      \hat{\Pi}^{\ast}_{n} = \sum_{t=1}^{n}\Delta Y^{\ast}_{t}Y_{t-1}^{\ast\transpose}(\sum_{t=1}^{n}Y^{\ast}_{t-1}Y_{t-1}^{\ast\transpose})^{-1}~.
      \end{align}
\endgroup
\par
\noindent Let $\hat{B}_{n}$ be the analog of $B_{0}$ based on $\tilde\Pi_{n}$, and $\hat{D}_{n}$ the analog of $D_{n}$ based on $\hat r_n$. Following the proof of Lemma A.2 of \citet{LiaoPhillips2015Automated}, we have: almost surely, %please see exercise 4.12 of Johansen (1995) for the currect Granger representation.
\begin{align}\label{Eqn: chisquare: coint3}
\hat{\mathcal M}_n^*\equiv \{\hat{\Pi}^{\ast}_{n}-\tilde\Pi_{n}\}\hat{B}_{n}^{-1}\hat{D}_{n}\hat{B}_{n}\overset{L^{\ast}}{\to}\mathcal{M}~.
\end{align}

Given an estimator $\hat\phi_{r,n}''$ satisfying \eqref{Eqn: derivative consistency} and the bootstrap estimator $\hat{\mathcal M}_n^*$ as in \eqref{Eqn: chisquare: coint3}, we may finally estimate the limit in \eqref{Eqn: coint, VAR weak limit} by the conditional law (given the data) of
\begin{align}\label{Eqn: chisquare: coint5}
\hat\phi_{r,n}''(\hat{\mathcal M}_n^*)\equiv \hat\phi_{r,n}^{\prime\prime}(\{\hat{\Pi}^{\ast}_{n}-\tilde\Pi_{n}\}\hat{B}_{n}^{-1}\hat{D}_{n}\hat{B}_{n})~.
\end{align}
Let $\hat c_{n,1-\alpha}$ be the conditional $(1-\alpha)$-quantile of $\hat\phi_{r,n}''(\hat{\mathcal M}_n^*)$ given the data. Then our test for \eqref{Eqn: coint test2} that rejects $\mathrm H_0$ if $n^2\phi_r(\hat\Pi_n)>\hat c_{n,1-\alpha}$ has asymptotic size control and is consistent, along the lines in Theorem \ref{Thm: SizePowerTest}.

To conclude, we present results that establish weak convergence of our statistic.

\begin{pro}\label{Pro: Derivative: Ex: Conintegration_VAR}
Let $\phi_r: \mathbf{M}^{k\times k}\to\mathbf{R}$ be defined as in \eqref{Eqn: phi, defn} with $m=k$ and $\Pi_0\in\mathbf{M}^{k\times k}$ satisfy $\phi_r(\Pi_0)=0$. Then, for $r_0\equiv\mathrm{rank}(\Pi_0)$ and $T_n\equiv \mathrm{diag}(t_{n}\mathbf{1}_{r_0},t_{n}^{2}\mathbf{1}_{k-r_0})$ with $t_n>0$,
\begin{align*}
\lim_{n\rightarrow \infty}\frac{\phi_r(\Pi_0 +M_nT_nB_0)}{t_n^4} = \sum_{j=r-r_0+1}^{k-r_0}\sigma^{2}_{j}(P_{0,2}^{\transpose} M Q_{0,2})~,
\end{align*}
whenever $t_n\downarrow 0$ and $\{M_n\}\subset\mathbf{M}^{k\times k}$ satisfies $M_nB_0\to M\in\mathbf{M}^{m\times k}$ as $n\to\infty$.
\end{pro}
\noindent{\sc Proof:} Let $\{M_n\}\subset\mathbf M^{k\times k}$ be such that $M_nB_0\to M$ and $t_n\downarrow 0$ as $n\to\infty$.  Thus we may write $M_{n}=[M_{n,1},M_{n,2}]$ and $M=M_1+M_2$ such that $M_{n,1}\in\mathbf{M}^{k\times r_0}$ and
\begin{align}\label{Eqn: Lem: Derivative: Ex: Conintegration_VAR, aux0}
M_{n,1}Q_{0,1}^{\transpose}\to M_{1}~,\quad M_{n,2}P_{0,2}^{\transpose}\to M_{2}~.
\end{align}
Clearly, $M_{1}U=0$ for all $U\in\Psi(\Pi_0)$. For $\epsilon>0$, let $\Psi(\Pi_0)^{\epsilon}$ and $\Psi(\Pi_0)_{1}^{\epsilon}$ be given in the proof of Proposition \ref{Pro: phi, differentiability}. In what follows we consider the nontrivial case when $\Pi_0\neq 0$ and $M_{2}\neq 0$. Let $d=k-r$. Then $\Psi(\Pi_0)\subsetneqq \mathbb S^{k\times d}$ and hence $\Psi(\Pi_0)_1^\epsilon\neq \emptyset$ for $\epsilon$ sufficiently small. Let $\sigma^{+}_{\min}(\Pi_0)$ be the smallest positive singular value of $\Pi_0$, which exists since $\Pi_0\neq 0$. Let $\Delta\equiv 5\sqrt{2}[\sigma_{\min}^{+}(\Pi_0)]^{-1}(\max_{U\in \mathbb S^{k\times d}}\Vert M_{2}U\Vert+\max_{U\in \mathbb S^{k\times d}}\Vert M_{1}U\Vert)>0$, which holds since $M_{2}\neq 0$. Then it follows that, for all $n$ sufficiently large,
\begin{align}\label{Eqn: Lem: Derivative: Ex: Conintegration_VAR, aux1}
\min_{U\in \Psi(\Pi_0)_{1}^{t_{n}\Delta}}&\Vert(\Pi_0 +M_nT_nB_0)U\Vert\ge \min_{U\in \Psi(\Pi_0)_{1}^{t_{n}\Delta}}\Vert\Pi_0 U\Vert-\max_{U\in\mathbb{S}^{k\times d}}\Vert M_nT_nB_0 U\Vert \notag\\
& \ge\frac{\sqrt 2}{2} t_{n}\sigma_{\min}^{+}(\Pi_0)\Delta -t_{n}\max_{U\in \mathbb S^{k\times d}}\Vert M_{n,1}Q_{0,1}^{\transpose}U\Vert -
t^{2}_{n}\max_{U\in \mathbb S^{k\times d}}\Vert M_{n,2}P_{0,2}^{\transpose}U\Vert \notag\\
& >t^{2}_{n}\max_{U\in \mathbb S^{k\times d}}\Vert M_{n,2}P_{0,2}^{\transpose}U\Vert \ge \min_{U\in \Psi(\Pi_0)}\|( \Pi_0 +M_nT_nB_0)U\|\notag\\
&\ge\sqrt{\phi_r(\Pi_0+M_nT_nB_0)}~,
\end{align}
where the first inequality follows by the Lipschitz continuity of the min operator, the triangle inequality and the fact that $\Psi(\Pi_0)_{1}^{t_{n}\Delta}\subset \mathbb{S}^{k\times d}$, the second inequality follows by Lemma \ref{Lem: phi, diff lemma, ID} and the triangle inequality, the third inequality follows by the definition of $\Delta$, $t_{n}\downarrow 0$, $M_{n,1}Q_{0,1}^{\transpose}\to M_{1}$, $M_{n,2}P_{0,2}^{\transpose}\to M_{2}$ as $n\to\infty$ and the simple fact that $2a-a_n>0$ for all $n$ large if $a_n\to a>0$, the fourth inequality holds by the facts that $\Pi_0 U=0$ and $Q_{0,1}^{\transpose}U=0$ for $U\in \Psi(\Pi_0)$, and the last by Lemma \ref{Lem: phi, representation}.

Next, let $\Gamma^{\Delta}$ and the correspondence $\varphi: \mathbf R\twoheadrightarrow\mathbb S^{k\times d}\times\Gamma^\Delta$ be given as in the proof of Proposition \ref{Pro: phi, differentiability} for $\Delta>0$. Then it follows that
\begin{align}\label{Eqn: Lem: Derivative: Ex: Conintegration_VAR, aux2}
\max_{U\in \Psi(\Pi_0)^{t_{n}\Delta}}\Vert M_nT_nB_0 U\Vert &\leq t_{n}\max_{(U,V)\in\varphi(t_{n})}\|(M_{n,1}Q_{0,1}^{\transpose})(U+t_{n}V)\| + t^{2}_{n}\max_{U\in \mathbb S^{k\times d}}\Vert M_{n,2}P_{0,2}^{\transpose}U\Vert\notag\\
&\leq t^{2}_{n}\max_{V\in\Gamma^{\Delta}}\|M_{n,1}Q_{0,1}^{\transpose}V\| + t^{2}_{n}\max_{U\in \mathbb S^{k\times d}}\Vert M_{n,2}P_{0,2}^{\transpose}U\|~,
\end{align}
where the first inequality follows by the triangle inequality, $M_n=[M_{n,1},M_{n,2}]$ and $\Psi(\Pi_0)^{t_{n}\Delta}\subset \mathbb{S}^{k\times d}$, and the second inequality follows from $Q_{0,1}^{\transpose}U=0$ for $U\in \Psi(\Pi_0)$ and $\varphi(t_{n})\subset \Psi(\Pi_0)\times \Gamma^{\Delta}$. By analogous arguments as in \eqref{Eqn: Lem: Derivative: Ex: Conintegration_VAR, aux1}, we have, for all $n$ large,
\begin{align}\label{Eqn: Lem: Derivative: Ex: Conintegration_VAR, aux3}
\min_{U\in \Psi(\Pi_0)_{1}^{t^{3/2}_{n}\Delta}\cap \Psi(\Pi_0)^{t_{n}\Delta}} & \hspace{-0.5cm}\Vert ( \Pi_0 +M_nT_nB_0)U\Vert\ge \min_{U\in \Psi(\Pi_0)_{1}^{t^{3/2}_{n}\Delta}}\Vert\Pi_0 U\Vert-\max_{U\in \Psi(\Pi_0)^{t_{n}\Delta}}\Vert M_n T_nB_0 U\Vert \notag\\
\ge &\frac{\sqrt 2}{2} t^{3/2}_{n}\sigma_{\min}^{+}(\Pi_0)\Delta -t^{2}_{n}\max_{V\in \Gamma^{\Delta}}\Vert M_{n,1}Q_{0,1}^{\transpose}V\Vert - t^{2}_{n}\max_{U\in \mathbb S^{k\times d}}\Vert M_{n,2}P_{0,2}^{\transpose}U\Vert \notag\\
> & t^{2}_{n}\max_{U\in \mathbb S^{k\times d}}\Vert M_{n,2}P_{0,2}^{\transpose}U\Vert \ge \min_{U\in \Psi(\Pi_0)}\|( \Pi_0 +M_nT_nB_0)U\|\notag\\
\ge&\sqrt{\phi_r(\Pi_0+M_n T_nB_0)}~,
\end{align}
where the first inequality follows by the Lipschitz continuity of the min operator, the triangle inequality, $\Psi(\Pi_0)_{1}^{t^{3/2}_{n}\Delta}\cap \Psi(\Pi_0)^{t_{n}\Delta} \subset \Psi(\Pi_0)_{1}^{t_{n}^{3/2}\Delta}$ and $\Psi(\Pi_0)_{1}^{t^{3/2}_{n}\Delta}\cap \Psi(\Pi_0)^{t_{n}\Delta}\subset \Psi(\Pi_0)^{t_{n}\Delta} $, the second inequality follows by \eqref{Eqn: Lem: Derivative: Ex: Conintegration_VAR, aux2} and Lemma \ref{Lem: phi, diff lemma, ID},
the third inequality follows by the definition of $\Delta$ and $\Gamma^{\Delta}$, $t_{n}\downarrow 0$, $M_{n,1}Q_{0,1}^{\transpose}\to M_{1}$ and $M_{n,2}P_{0,2}^{\transpose}\to M_{2}$ as $n\to\infty$, the fourth inequality holds by the facts that $\Pi_0 U=0$ and $Q_{0,1}^{\transpose}U=0$ for $U\in \Psi(\Pi_0)$. In turn, by analogous arguments, we have, for all $n$ sufficiently large,
\begin{align}\label{Eqn: Lem: Derivative: Ex: Conintegration_VAR, aux4}
&\min_{U\in \Psi(\Pi_0)_{1}^{t^{2}_{n}\Delta}\cap \Psi(\Pi_0)^{t_{n}^{3/2}\Delta}}\Vert ( \Pi_0 +M_nT_nB_0)U\Vert>\sqrt{\phi_r(\Pi_0+M_nT_nB_0)}~.
\end{align}
Combining \eqref{Eqn: Lem: Derivative: Ex: Conintegration_VAR, aux1}, \eqref{Eqn: Lem: Derivative: Ex: Conintegration_VAR, aux3}, \eqref{Eqn: Lem: Derivative: Ex: Conintegration_VAR, aux4} and Lemma \ref{Lem: phi, representation}, we thus obtain that, for all $n$ large,
\begin{align}\label{Eqn: Lem: Derivative: Ex: Conintegration_VAR, aux5}
\phi_r(\Pi_0 +M_nT_nB_0) = \min_{U\in \Psi(\Pi_0)^{t^{2}_{n}\Delta}}\Vert ( \Pi_0 + M_nT_nB_0)U\Vert^{2}~.
\end{align}

Now, for the right hand side of \eqref{Eqn: Lem: Derivative: Ex: Conintegration_VAR, aux5}, we have
\begin{align}\label{Eqn: Lem: Derivative: Ex: Conintegration_VAR, aux6}
&\big|\min_{U\in \Psi(\Pi_0)^{t^{2}_{n}\Delta}}\Vert ( \Pi_0 + M_nT_nB_0)U\Vert^{2}-\min_{U\in \Psi(\Pi_0)^{t^{2}_{n}\Delta}}\Vert ( \Pi_0 +t_{n}M_{1}+t_{n}^{2}M_{2} )U\Vert^{2}\big|\notag\\
&\leq (O(t^{2}_{n})+O(t^{2}_{n}))\max_{U\in \Psi(\Pi_0)^{t^{2}_{n}\Delta}}\Vert (t_{n}(M_{1,n}Q_{0,1}^{\transpose}-M_{1})+t^{2}_{n}(M_{2,n}P_{0,2}^{\transpose}-M_{2}))U\Vert~,
\end{align}
where the inequality follows by the formula $a^{2}-b^{2}=(a+b)(a-b)$, the Lipschitz inequality of the min operator, the triangle inequality, and the facts that  $\min_{U\in \Psi(\Pi_0)^{t^{2}_{n}\Delta}}\Vert ( \Pi_0 + M_nT_nB_0)U\Vert = O(t^{2}_{n})$ and $\min_{U\in \Psi(\Pi_0)^{t^{2}_{n}\Delta}}\Vert ( \Pi_0 + MT_nB_0)U\Vert = O(t^{2}_{n})$. For the second term on the right hand side of \eqref{Eqn: Lem: Derivative: Ex: Conintegration_VAR, aux6}, we have
\begin{align}\label{Eqn: Lem: Derivative: Ex: Conintegration_VAR, aux7}
&\max_{U\in \Psi(\Pi_0)^{t^{2}_{n}\Delta}}\Vert (t_{n}(M_{1,n}Q_{0,1}^{\transpose}-M_{1})+t^{2}_{n}(M_{2,n}P_{0,2}^{\transpose}-M_{2}))U\Vert\notag\\
&\leq t_{n}\max_{(U,V)\in \varphi(t_{n}^{2})} \|(M_{n,1}Q_{0,1}^{\transpose}-M_{1})(U+t_{n}^{2}V)\|+t_{n}^{2}\max_{U\in \Psi(\Pi_0)^{t^{2}_{n}\Delta}}\|(M_{n,2}P_{0,2}^{\transpose}-M_{2})U\|\notag\\
&\leq \max_{V\in \Gamma^{\Delta}} t_{n}^{3}\|(M_{n,1}Q_{0,1}^{\transpose}-M_{1})V\|+t_{n}^{2}\max_{U\in \Psi(\Pi_0)^{t^{2}_{n}\Delta}}\|(M_{n,2}P_{0,2}^{\transpose}-M_{2})U\| = o(t_{n}^{2})~,
\end{align}
where the first inequality follows by the triangle inequality and the definition of $\varphi(t_{n}^{2})$, the second inequality follows by the fact that $Q_{0,1}^{\transpose}U=0$ and $M_{1}U=0$ for $U\in \Psi(\Pi_0)$ and $\varphi(t^{2}_{n})\subset \Psi(\Pi_0)\times \Gamma^{\Delta}$, and the equality follows by applying the sub-multiplicativity of Frobenius norm and the facts that $M_{n,1}Q_{0,1}^{\transpose}\to M_{1}$ and $M_{n,2}P_{0,2}^{\transpose}\to M_{2}$ as $n\to \infty$. Combining results \eqref{Eqn: Lem: Derivative: Ex: Conintegration_VAR, aux5}, \eqref{Eqn: Lem: Derivative: Ex: Conintegration_VAR, aux6} and \eqref{Eqn: Lem: Derivative: Ex: Conintegration_VAR, aux7}, we then obtain
\begin{align}\label{Eqn: Lem: Derivative: Ex: Conintegration_VAR, aux8}
\phi_r(\Pi_0 +M_nT_nB_0) = \min_{U\in \Psi(\Pi_0)^{t^{2}_{n}\Delta}}\Vert ( \Pi_0 + t_{n}M_{1} + t_{n}^{2}M_{2})U\Vert^{2} + o(t^{4}_{n})~.
\end{align}

Next, the first term on the right hand side of \eqref{Eqn: Lem: Derivative: Ex: Conintegration_VAR, aux8} can be written as
\begin{align}\label{Eqn: Lem: Derivative: Ex: Conintegration_VAR, aux9}
\min_{U\in \Psi(\Pi_0)^{t^{2}_{n}\Delta}}\Vert ( \Pi_0 + t_{n}M_{1} + t_{n}^{2}M_{2})U\Vert^{2} &= \min_{(U,V)\in\varphi(t_{n}^{2})}\Vert( \Pi_0 + t_{n}M_{1} + t_{n}^{2}M_{2})(U+t_{n}^{2}V)\Vert^{2}\notag\\
&= t_{n}^{4} \min_{(U,V)\in\varphi(t_{n}^{2})}\Vert \Pi_0 V + M U\Vert^{2} + o(t_{n}^{4})~,
\end{align}
where the second equality follows by the facts that $\Pi_0 U=0$ and $M_{1}U=0$ for $U\in \Psi(\Pi_0)$, and $\|V\|\leq \Delta $ for all $V\in\Gamma^{\Delta}$. By analogous arguments in \eqref{Eqn: Rankcon HDD, aux8}, we have
\begin{align}\label{Eqn: Lem: Derivative: Ex: Conintegration_VAR, aux10}
\min_{(U,V)\in\varphi(t_{n}^{2})}\Vert \Pi_0 V +M U\Vert^{2} = \min_{U\in\Psi(\Pi_0)}\min_{V\in\mathbf{M}^{k\times d}}\|\Pi_0 V+ M U\|^{2}+o(1)~.
\end{align}
Combining \eqref{Eqn: Lem: Derivative: Ex: Conintegration_VAR, aux8}, \eqref{Eqn: Lem: Derivative: Ex: Conintegration_VAR, aux9} and \eqref{Eqn: Lem: Derivative: Ex: Conintegration_VAR, aux10}, we may conclude that
\begin{multline}\label{Eqn: Lem: Derivative: Ex: Conintegration_VAR, aux11}
\lim_{n\to\infty}\frac{\phi_r(\Pi_0 +M_nT_nB_0)}{t_n^4} = \min_{U\in\Psi(\Pi_0)}\min_{V\in\mathbf{M}^{k\times d}}\|\Pi_0 V+ M U\|\\
=\sum_{j=r-r_0+1}^{k-r_0}\sigma^{2}_{j}(P_{0,2}^{\transpose} M Q_{2})~,
\end{multline}
where the second equality follows by Lemma \ref{Lem: phi, derivative, reprentation}, as desired. \qed

\begin{pro}\label{Pro: Weaklimit: Ex: Conintegration_VAR}
Suppose that there is an estimator $\hat{\Pi}_{n}: \{X_{i}\}_{i=1}^{n}\to\mathbf{M}^{k\times k}$ for $\Pi_{0}\in\mathbf{M}^{k\times k}$ such that $\{\hat{\Pi}_{n}-\Pi_{0}\}B_{0}^{-1}D_{n}B_{0}\overset{L}{\rightarrow}\mathcal{M}$ for some $\tau_{n}\uparrow \infty$ and random matrix $\mathcal{M}\in\mathbf{M}^{k\times k}$, where $D_{n}\equiv \text{diag}(\tau_{n}\mathbf{1}_{r_{0}},\tau_{n}^{2}\mathbf{1}_{k-r_{0}})$. If $\mathrm{rank}(\Pi_0)\le r$, then we have
\[\tau_{n}^{4}\phi_r(\hat{\Pi}_{n}) \convl \sum_{j=r-r_{0}+1}^{k-r_{0}}\sigma^{2}_{j}(P_{0,2}^{\transpose}\mathcal{M}Q_{0,2})~.\]
\end{pro}
\noindent{\sc Proof:} For each $n\in\mathbf{N}$, define $g_{n}: \mathbf{M}^{k\times k }\to\mathbf{R}$ by
\begin{align}\label{Eqn: Pro: Weaklimit: Ex: Conintegration_VAR, aux11}
g_{n}(M)\equiv \tau_{n}^{4}\phi_r(\Pi_{0} + M D_{n}^{-1}B_{0})~.
\end{align}
By Proposition \ref{Pro: Derivative: Ex: Conintegration_VAR}, $g_{n}(M_{n})\to \sum_{j=r-r_{0}+1}^{k-r_{0}}\sigma^{2}_{j}(P_{0,2}^{\transpose} M Q_{0,2})$ whenever $M_{n}B_0\to M$ as $n\to\infty$. In turn, since $\tau_{n}^{4}\phi_r(\hat{\Pi}_{n}) = g_{n}((\hat{\Pi}_{n}-\Pi_{0})B_{0}^{-1}D_{n})$, the proposition follows by Theorem 1.11.1(i) in \citet{Vaart1996}. \qed

\subsection{Additional Examples}\label{Sec: examples more}

Our first example in this section arises in finite mixture models of dynamic discrete choices where a problem of both theoretical and practical importance is inference on the number of types \citep{McLachlanPeel2004Mixture,KasaharaShimotsu2009NPIDmixture}. It is also related to incomplete information games with multiple equiliria studied in \citet{Xiao2018MultiEquili}.

\begin{ex}[Finite Mixtures, Discrete Choices and Multiple Equilibria]\label{Ex: game}
Consider an individual with characteristic $Z_t\in\mathcal Z\equiv\{z_1,\ldots,z_d\}$ who makes a choice $S_t\in\mathcal S\equiv\{0,1\}$ depending on his/her unknown (to econometricians) type, at time $t=1, 2$. Suppose that there are $\gamma_0$ (finite) types. Under regularity conditions, \citet{KasaharaShimotsu2009NPIDmixture} establish a lower bound for $\gamma_0$, i.e., $\gamma_0\ge \mathrm{rank}(\Pi_0)$ with
\begin{align}\label{Eqn: mixture, Pi0}
\Pi_0=\begin{bmatrix}
1 & \tilde p_{X_2}(1,z_1) & \cdots & \tilde p_{X_2}(1,z_d) \\
\tilde p_{X_1}(1,z_1) & \tilde p_{X_1,X_2}(1,z_1;1,z_1) & \cdots & \tilde p_{X_1,X_2}(1,z_1;1,z_d)\\
\vdots & \vdots & \ddots & \vdots\\
\tilde p_{X_1}(1,z_d) & \tilde p_{X_1,X_2}(1,z_d;1,z_1) & \cdots & \tilde p_{X_1,X_2}(1,z_d;1,z_d)
\end{bmatrix}~,
\end{align}
where $X_t\equiv (Z_t,S_t)\in \mathcal X\equiv\mathcal Z\times\mathcal S$ for $t=1,2$, $\tilde p_{X_1}(1,z)\equiv\sum_{x_2\in\mathcal X} \tilde p_{X_1,X_2}(1,z;x_2)$, $\tilde p_{X_2}(1,z)\equiv\sum_{x_1\in\mathcal X} \tilde p_{X_1,X_2}(x_1;1,z)$, and, for any $x=(z,s)$ and $x'=(z',s')$ in $\mathcal X$,
\begin{align*}
\tilde p_{X_1,X_2}(z,s;z',s')\equiv \frac{p_{X_1,X_2}(z,s;z',s')}{p_{Z_2|X_1}(z';z,s)}
\end{align*}
with $p_{X_1,X_2}$ the probability mass function (pmf) of $(X_1,X_2)$ and $p_{Z_2|X_1}$ the conditional pmf of $Z_2$ given $X_1$. Under additional conditions, \citet{KasaharaShimotsu2009NPIDmixture} show in fact $\gamma_0= \mathrm{rank}(\Pi_0)$. We focus on discrete variables and two periods for ease of exposition, but the results extend to more general cases by Remark 2(iv) and Propositions 3 and 8 in \citet{KasaharaShimotsu2009NPIDmixture}. The number of types is crucial for the specification of mixture distributions, and yet inference on $\gamma_0$ without parametric assumptions on the component distributions has been understood to be challenging \citep{KasaharaShimotsu2009NPIDmixture,KasaharaShimotsu2014EstimationMixture,BonhommeJochmansRobin2016Mixture}. By further restricting each component distribution to have independent marginals, \citet{KasaharaShimotsu2014EstimationMixture} and \citet{BonhommeJochmansRobin2016Mixture} accomplish nonparametric estimation of $\gamma_0$ based on the rank test of \citet{Kleibergen_Paap2006rank}. Interestingly, \citet{Xiao2018MultiEquili} derives a similar nonparametric identification result for the number of equilibria in incomplete information games and obtains a consistent estimator based on the rank test of \citet{Robin_Smith2000rank}. \qed
\end{ex}

\iffalse
the baseline model studied by \citet{KasaharaShimotsu2009NPIDmixture} yields the following joint density of the observables: for $X_t\equiv (Z_t,S_t)$,
\begin{multline}
f_{X_1,\ldots,X_T}(x_1,\ldots,x_T)\\
=\sum_{j=1}^{J} \pi^j f_{Z_1,S_1}^j(z_1,s_1)\prod_{t=2}^{T} f_{Z_t|Z_{t-1},S_{t-1}}(z_t;z_{t-1},s_{t-1}) f_{S_t|Z_t}^j(s_t;z_t)~,
\end{multline}
where $\pi_j$ is the probability of belonging to type $j$, $f_{Z_1,S_1}^j$ is the density of the initial observable $X_1$ if the individual belongs to type $j$, $f_{Z_t|Z_{t-1},S_{t-1}}$ is the transition density, and $f_{S_t|Z_t}^j$ is the conditional choice probability if the individual belongs to type $j$ and chooses $S_t$.
\fi

Our second example pertains to the existence of general common features \citep{Engle_Kozicki1993CF}, which conceptually includes cointegration as a special case.

\begin{ex}[Common Features]\label{Ex: CommonFeatures}
Let $\{Y_t\}$ be a $k\times 1$ time series. According to \citet{Engle_Kozicki1993CF}, a feature that is present in each component of $Y_t$ is said to be common to $Y_t$ if there exists a nonzero linear combination of $Y_t$ that fails to have the feature. To fix ideas, suppose that $\{Y_t\}$ is generated according to
\begin{align}\label{Eqn: ex, CF}
Y_t=\Gamma_0^\transpose Z_t+\Xi_0^\transpose W_t+u_t~,
\end{align}
where $W_t$ can be thought of as control variables, and $Z_t$ is an $m\times 1$ vector reflecting the feature under consideration with $m\geq k$. For example, testing for the existence of common serial correlation would set $Z_t$ to be lags of $Y_t$, and testing for the existence of common conditionally heteroskedastic factors would set $Z_t$ to be relevant factors. We refer to \citet{Engle_Kozicki1993CF}, \citet{Engle_Susmel1993CommonVol} and \citet{Dovonon_Renault2013testing} for details of these and other examples. By the specification of \eqref{Eqn: ex, CF}, existence of common features means that $\Gamma_0$ is not of full rank. Thus, testing for the existence of common features reduces to examining the hypotheses in \eqref{Eqn: hypothesis, intro} with
\begin{align}
\Pi_{0}= \Gamma_{0}\text{ and } r=k-1~.
\end{align}
Since the number of common features is generally unknown \textit{a priori}, the assumption $\text{rank}(\Pi_{0})\geq k-1$ that underlies the hypotheses in \eqref{Eqn: hypothesis, literature, intro} may again be unrealistic. \qed
\end{ex}

Our next example involves estimation of the rank of demand systems, a notion developed by \citet{Gorman1981Engle} for exactly aggregable demand systems and generalized by \citet{Lewbel1991Rank} to all demand systems.

\begin{ex}[Consumer Demand]\label{Ex: ConsumerDemand}
An Engel curve is the function describing the allocation of an individual's consumption expenditures with the prices of all goods fixed, and the rank of a demand system is the dimension of the space spanned by the Engel curves of the system \citep{Lewbel1991Rank}. Suppose that there are $k$ goods in the system and that the Engel curve is given by
\begin{align}
Y = \Gamma_{0}G(Z) + u~,
\end{align}
where $Y$ is a $k\times 1$ vector of budget shares on the $k$ goods, $Z$ is the total expenditure, $G(\cdot)$ is a $r_{0}\times 1$ vector of unknown function with $r_{0}\leq k$, and $u$ is an error term. The rank of the demand system ie precisely $r_0$, and in fact also equal to the rank of
\begin{align}\label{Eqn: ex, CD}
\Pi_{0}=E[Q(Z)Y^{\transpose}]~,
\end{align}
where $Q(\cdot)$ is an $m\times1$ vector of known functions with $m\geq k$, under suitable conditions. Estimation of the rank of the demand system is important because it provides evidence on consistency of consumer behaviors with utility maximization, and has implications for welfare comparisons and aggregation across goods and across consumers \citep{Lewbel1991Rank,Lewbel2006Engel,Barnett_Serletis2009Measuring}.\qed
\end{ex}

Our fourth example shows the importance of rank estimation in identifying the number of factors in factor models \citep{Anderson2003multivariate, LamYao2012Factor}.

\begin{ex}[Factor Analysis]\label{Ex: FactorAnalysis}
Let $Y\in\mathbf{R}^{d}$ be generated by the following model
\begin{eqnarray}\label{Eqn: Ex: FactorAnalysis: 1}
Y = \mu_{0} +\Lambda_{0}F + u ~,
\end{eqnarray}
where $F$ is a $r_{0}\times 1$ vector of unobserved common factors with $r_{0}\leq d$, and $u$ is an error term. Partition $Y=[Y_{1}^{\transpose},Y_{2}^{\transpose},Y_{3}^{\transpose}]^{\transpose}$ for $Y_{1}\in\mathbf{R}^{m}$ and $Y_{2}\in\mathbf{R}^{k}$ with some $r_{0}\leq k\leq m< d$ and $m+k\leq d$, and also $\Lambda_{0}=[\Lambda_{0,1}^{\transpose}, \Lambda_{0,2}^{\transpose}, \Lambda_{0,3}^{\transpose}]^{\transpose}$ with $\Lambda_{0,1}$ and $\Lambda_{0,2}$ having $m$ and $k$ rows. Then under appropriate restrictions, the rank of $\text{Var}(F)$ is equal to the rank of
\begin{align}\label{Eqn: ex, FA}
\Pi_{0}=\text{Cov}(Y_{1},Y_{2})~.
\end{align}
Thus, determining the number $r_{0}$ of the common factors reduces to estimation of the rank of $\Pi_{0}$. Such a problem also arises in the interbattery factor analysis \citep{Gill_Lewbel1992rank}, the dynamic analysis of time series \citep{LamYao2012Factor}, and finance and macroeconomics \citep{Bai_Ng2002NumFac,Bai_Ng2007PrimiShocks}. \qed
\end{ex}

Our final example is taken from \citet{Gill_Lewbel1992rank}, and manifests how matrix rank determination is useful in model selection in time series models.

\begin{ex}[Model Selection]\label{Ex: ModelSelection}
Let $\{Y_t\}$ be a $p\times 1$ weakly stationary time series, which has the following state space representation:
\begin{align}
Y_{t}=\Gamma_0 Z_t+u_t~,\,Z_t=\Lambda_0 Z_{t-1}+\epsilon_{t}~,
\end{align}
where $Z_t$ is a $r_{0}\times 1$ vector of state variables, and $u_t$ and $\epsilon_t$ are error terms. It turns out that the number $r_{0}$ of state variables is equal to the rank of the Hankel matrix
\begin{align}\label{Eqn: ex, MS}
\Pi_{0}= E(\begin{bmatrix}
Y_{t+1}\\
\vdots\\
Y_{t+b}
\end{bmatrix}\begin{bmatrix}
Y_{t}^\transpose & \cdots & Y_{t-b+1}^\transpose
\end{bmatrix})~,
\end{align}
for $b$ sufficiently large \citep[p.52]{Aoki1990State}. Consequently, determining the number of state variables $r_{0}$ to model $Y_{t}$ reduces to determining the rank of $\Pi_{0}$. When $Y_{t}$ is a scalar and follows an ARMA$(p_{1},p_{2})$ model, then $Y_{t}$ has a state space representation with the number $r_{0}$ of state variables equal to $\max(p_{1},p_{2})$ \citep{Aoki1990State}. Thus, determining the rank of the Hankel matrix is crucial for model specification in these contexts.\qed
\end{ex}

For simplicity, we verify the main assumptions only for Example \ref{Ex: game}. Let $\{X_{it}\}_{i=1}^n$ be a sample generated by the mixture model with $X_{it}=(Z_{it},S_{it})$ for $t=1,2$. Then we estimate $\Pi_0$ by its empirical analog:
\begin{align}
\hat\Pi_n=\begin{bmatrix}
1 & \tilde p_{X_2,n}(1,z_1) & \cdots & \tilde p_{X_2,n}(1,z_d) \\
\tilde p_{X_1,n}(1,z_1) & \tilde p_{X_1,X_2,n}(1,z_1;1,z_1) & \cdots & \tilde p_{X_1,X_2,n}(1,z_1;1,z_d)\\
\vdots & \vdots & \ddots & \vdots\\
\tilde p_{X_1,n}(1,z_d) & \tilde p_{X_1,X_2,n}(1,z_d;1,z_1) & \cdots & \tilde p_{X_1,X_2,n}(1,z_d;1,z_d)
\end{bmatrix}~,
\end{align}
where $\tilde p_{X_1,n}(1,z)\equiv\sum_{x_2\in\mathcal X} \tilde p_{X_1,X_2,n}(1,z;x_2)$, $\tilde p_{X_2,n}(1,z)\equiv\sum_{x_1\in\mathcal X} \tilde p_{X_1,X_2,n}(x_1;1,z)$, and, for any $x=(z,s)$ and $x'=(z',s')$ in $\mathcal X$,
\begin{align*}
\tilde p_{X_1,X_2,n}(z,s;z',s')\equiv \frac{\hat p_{X_1,X_2,n}(z,s;z',s')}{\hat p_{Z_2|X_1,n}(z';z,s)}
\end{align*}
with $\hat p_{X_1,X_2,n}$ the empirical pmf of $\{(X_{i1},X_{i2})\}_{i=1}^n$ and $\hat p_{Z_2|X_1,n}(\cdot\,; z,s)$ the empirical conditional pmf of $\{Z_{i2}\}$ given $X_1=(z,s)$. Since sums and ratios are differentiable maps (at nonzero denominators as assumed in \citet{KasaharaShimotsu2009NPIDmixture}), a simple application of the Delta method shows that $\hat\Pi_n$ satisfies Assumption \ref{Ass: Weaklimit: Pihat} with $\tau_n=\sqrt n$ and $\mathcal M$ some centered Gaussian matrix, under standard regularity conditions.

Next, suppose that $\{(X_{i1},X_{i2})\}_{i=1}^n$ are i.i.d.\ across $i$ for ease of exposition. Let $\{(X_{i1}^*,X_{i2}^*)\}_{i=1}^n$ be an i.i.d.\ sample drawn with replacement from $\{(X_{i1},X_{i2})\}_{i=1}^n$. Then we propose the bootstrap estimator $\hat\Pi_n^*$ as follows:
\begin{align}
\hat\Pi_n^*=\begin{bmatrix}
1 & \tilde p_{X_2,n}^*(1,z_1) & \cdots & \tilde p_{X_2,n}^*(1,z_d) \\
\tilde p_{X_1,n}^*(1,z_1) & \tilde p_{X_1,X_2,n}^*(1,z_1;1,z_1) & \cdots & \tilde p_{X_1,X_2,n}^*(1,z_1;1,z_d)\\
\vdots & \vdots & \ddots & \vdots\\
\tilde p_{X_1,n}^*(1,z_d) & \tilde p_{X_1,X_2,n}^*(1,z_d;1,z_1) & \cdots & \tilde p_{X_1,X_2,n}^*(1,z_d;1,z_d)
\end{bmatrix}~,
\end{align}
where $\tilde p_{X_1,n}^*(1,z)\equiv\sum_{x_2\in\mathcal X} \tilde p_{X_1,X_2,n}^*(1,z;x_2)$, $\tilde p_{X_2,n}^*(1,z)\equiv\sum_{x_1\in\mathcal X} \tilde p_{X_1,X_2,n}^*(x_1;1,z)$, and, for any $x=(z,s)$ and $x'=(z',s')$ in $\mathcal X$,
\begin{align*}
\tilde p_{X_1,X_2,n}^*(z,s;z',s')\equiv \frac{\hat p_{X_1,X_2,n}^*(z,s;z',s')}{\hat p_{Z_2|X_1,n}^*(z';z,s)}
\end{align*}
with $\hat p_{X_1,X_2,n}^*$ and $\hat p_{Z_2|X_1,n}^*(\cdot\,; z,s)$ the bootstrap analogs of $\hat p_{X_1,X_2,n}$ and $\hat p_{Z_2|X_1,n}(\cdot\,; z,s)$ respectively based on  $\{(X_{i1}^*,X_{i2}^*)\}_{i=1}^n$. Assumption \ref{Ass: Boostrap: Pihat} now follows from the Delta method for bootstrap, as a result of the same differentiability mentioned previously -- see, for example, Theorem 23.5 in \citet{Vaart1998}.

\end{appendices}

%\phantomsection
\addcontentsline{toc}{section}{References}
\putbib
\end{bibunit}

%\bibliographystyle{ims}
%\phantomsection
%\addcontentsline{toc}{section}{References}
%\bibliography{mybibliography}
\end{document}